\documentclass[journal]{IEEEtran}
\usepackage{cite}

\ifCLASSINFOpdf
\else
   \usepackage[dvips]{graphicx}
\fi
\usepackage{amsmath}
\usepackage{amssymb}
\usepackage{balance}
\usepackage{algorithm}
\usepackage{algorithmicx}
\usepackage{algpseudocode}

\newtheorem{example}{Example}
\newtheorem{remark}{Remark}
\newtheorem{theorem}{Theorem}
\newtheorem{assumption}{Assumption}
\newtheorem{lemma}{Lemma}
\newtheorem{proposition}{Proposition}
\newtheorem{corollary}{Corollary}

\newcommand{\ato}{\overset{\mathrm{a.s.}}{\to}}
\newcommand{\aeq}{\overset{\mathrm{a.s.}}{=}}

\begin{document}
%
\title{On the Convergence of Orthogonal/Vector AMP: Long-Memory Message-Passing Strategy}
%
%
%

\author{Keigo~Takeuchi,~\IEEEmembership{Member,~IEEE}
\thanks{K.~Takeuchi is with the Department of Electrical and Electronic Information Engineering, Toyohashi University of Technology, Toyohashi 441-8580, Japan (e-mail: takeuchi@ee.tut.ac.jp).}
\thanks{
The author was in part supported by the Grant-in-Aid 
for Scientific Research~(B) (JSPS KAKENHI Grant Numbers 21H01326), Japan. 
The material in this paper will be presented in part at 2022 IEEE International Symposium on Information Theory.
}
}

%
%

\markboth{IEEE transactions on information theory}%
{Takeuchi: On the Convergence of Orthogonal/Vector AMP: Long-Memory Message-Passing Strategy}
%

\IEEEpubid{0000--0000/00\$00.00~\copyright~2015 IEEE}


\maketitle

\begin{abstract}
Orthogonal/vector approximate message-passing (AMP) is a powerful 
message-passing (MP) algorithm for signal reconstruction in compressed sensing. 
This paper proves the convergence of Bayes-optimal orthogonal/vector AMP in 
the large system limit. The proof strategy is based on a novel long-memory 
(LM) MP approach: A first step is a construction of LM-MP that 
is guaranteed to converge systematically. A second step is a large-system 
analysis of LM-MP via an existing framework of state evolution. 
A third step is to prove the convergence of state evolution recursions for 
Bayes-optimal LM-MP via a new statistical interpretation of existing LM 
damping. The last is an exact reduction of the state evolution recursions for 
Bayes-optimal LM-MP to those for Bayes-optimal orthogonal/vector AMP. 
The convergence of the state evolution recursions for Bayes-optimal LM-MP 
implies that for Bayes-optimal orthogonal/vector AMP. 
Numerical simulations are presented to show the verification of state 
evolution results for damped orthogonal/vector AMP and a negative aspect 
of LM-MP in finite-sized systems. 
\end{abstract}

\begin{IEEEkeywords}
Compressed sensing, orthogonal/vector approximate message-passing (AMP), 
long memory, damping, state evolution, convergence analysis. 
\end{IEEEkeywords}

%
\IEEEpeerreviewmaketitle

\section{Introduction}
\subsection{Motivation}
\IEEEPARstart{C}{onsider} a reconstruction of an $N$-dimensional 
sparse signal vector $\boldsymbol{x}\in\mathbb{R}^{N}$ with 
$N^{-1}\mathbb{E}[\|\boldsymbol{x}\|^{2}]=1$ from 
compressed, noisy, and linear measurements $\boldsymbol{y}
\in\mathbb{R}^{M}$~\cite{Donoho06,Candes061} with $M\leq N$, given by 
\begin{equation} \label{model}
\boldsymbol{y} = \boldsymbol{A}\boldsymbol{x} + \boldsymbol{w}. 
\end{equation}
In (\ref{model}), $\boldsymbol{A}\in\mathbb{R}^{M\times N}$ is a known 
sensing matrix. The vector $\boldsymbol{w}\in\mathbb{R}^{M}$ denotes an 
additive noise vector with $\sigma^{2}
=M^{-1}\mathbb{E}[\|\boldsymbol{w}\|^{2}]$. A goal of compressed 
sensing is to reconstruct the signal vector $\boldsymbol{x}$ from the 
knowledge on the sensing matrix $\boldsymbol{A}$ and the 
measurement vector $\boldsymbol{y}$.  

Compressed sensing has been applied to several practical issues, such as 
magnetic resonance imaging~\cite{Lustig07}, radar~\cite{Herman09}, image 
super-resolution~\cite{Yang10}, channel estimation~\cite{Bajwa10},  
spectrum sensing~\cite{Zeng11}, error-correcting codes~\cite{Joseph12}, and 
multiuser detection~\cite{Shim12}. The prior distribution of the sparse 
signal vector is known in specific instances, such as error-correcting codes 
or multiuser detection, while it is unknown in many signal processing 
instances. This paper is more relevant to such specific instances---the prior 
distribution is known---since the knowledge on the prior distribution 
is utilized in the theoretical analysis of iterative reconstruction algorithms. 

Approximate message-passing (AMP)~\cite{Donoho09,Rangan11} is a powerful 
message-passing (MP) algorithm for signal reconstruction. AMP with 
Bayes-optimal denoiser---called Bayes-optimal AMP---can be regarded as an 
exact approximation of loopy belief propagation in the large system 
limit~\cite{Kabashima03}, where both $M$ and $N$ tend to infinity while the 
compression ratio $\delta=M/N\in(0,1]$ is kept constant. Furthermore, 
Bayes-optimal AMP was proved to achieve the Bayes-optimal 
performance~\cite{Tanaka02,Guo05,Reeves19,Barbier19} for zero-mean, 
independent and identically distributed (i.i.d.), and  
sub-Gaussian sensing matrices~\cite{Bayati11,Bayati15} if the fixed-point (FP) 
equations describing the Bayes-optimal performance have a unique solution.    
Spatial coupling~\cite{Kudekar11} is required when the FP equations have 
multiple solutions. See \cite{Krzakala12,Donoho13,Javanmard13,Takeuchi15} 
for the details.  

The main weakness of AMP is that AMP does not converge for sensing matrices 
beyond zero-mean i.i.d.\ matrices, such as non-zero mean 
matrices~\cite{Caltagirone14} or ill-conditioned matrices~\cite{Rangan191}. 
To tackle this convergence issue, several modified MP 
algorithms~\cite{Kabashima14,Vila15,Manoel15,Rangan17,Ma17,Rangan192} 
were proposed. In particular, orthogonal AMP (OAMP)~\cite{Ma17} or 
equivalently vector AMP (VAMP)~\cite{Rangan192} is a promising algorithm 
to solve the weakness of AMP. 
Since they are equivalent to each other in the linear measurement 
system~(\ref{model}), the two MP algorithms are referred 
to as OAMP in this paper.

The main feature of OAMP is the use of \emph{extrinsic} messages to eliminate 
intractable dependencies between the current message and all preceding 
messages. The extrinsic messages were originally used in a single loop 
algorithm to solve a FP of the expectation consistency (EC) free 
energy~\cite{Opper05}, inspired by expectation propagation~\cite{Minka01}. 
In fact, Bayes-optimal OAMP can be regarded as an exact approximation 
of EP in the large system limit~\cite{Cespedes14,Takeuchi201}.  

The asymptotic performance of OAMP can be characterized via rigorous 
state evolution~\cite{Rangan192,Takeuchi201}. 
Bayes-optimal OAMP was proved to achieve the Bayes-optimal 
performance~\cite{Takeda06,Tulino13,Barbier18} for all orthogonally 
invariant sensing matrices if the FP equations describing the 
Bayes-optimal performance have a unique solution. The class of orthogonally 
invariant matrices is a general class beyond zero-mean i.i.d.\ sensing 
matrices.    

\IEEEpubidadjcol

Strictly speaking, state 
evolution~\cite{Bayati11,Bayati15,Rangan192,Takeuchi201} does not guarantee 
the convergence\footnote{
Throughout this paper, we only consider the convergence of state evolution 
recursions for MP algorithms in the large system limit, which may be phrased 
as the convergence of MP algorithms in the large system limit or more simply 
as the convergence of MP algorithms. Thus, convergence for finite-sized 
systems is out of the scope of this paper. 
} of MP algorithms for AMP or OAMP while it describes 
the asymptotic dynamics of these MP algorithms rigorously. As a result, we 
need to investigate the convergence of state evolution recursions to a FP for 
individual problems, e.g.\ \cite{Liu16,Gerbelot20,Mondelli21}.

For finite-sized systems, damping~\cite{Rangan191} is used to improve 
the convergence property of MP algorithms. Conventional state evolution 
cannot describe the dynamics of damped AMP or OAMP anymore. To the best of 
author's knowledge, theoretical analysis for damped MP algorithms are 
very limited~\cite{Rangan191,Mimura19}. 

This paper addresses these two issues by proposing a novel damped MP 
algorithm that realizes the convergence \emph{in principle} and the exact 
description for its dynamics. 
The damping issue has been already solved in \cite{Liu21} 
via the framework of long-memory (LM) MP~\cite{Takeuchi211}. LM-MP utilizes 
all preceding messages in computing the current message while conventional 
MP uses messages only in the latest iteration to update the current message. 
LM-MP was originally used to solve the Thouless-Anderson-Palmer (TAP) 
equation for Ising models with orthogonally invariant spin interaction via 
non-rigorous dynamical functional theory~\cite{Opper16}. In this direction, 
Fan~\cite{Fan22} proposed an LM-MP algorithm and analyzed the dynamics of the 
LM-MP algorithm rigorously. See \cite{Venkataramanan21} for a generalization 
of \cite{Fan22}. 

The state evolution framework for LM-MP in \cite{Takeuchi211} is fully general, 
so that rigorous state evolution results for LM-MP can be obtained just by 
confirming whether the model of estimation errors for the LM-MP is included 
into a general error model proposed in \cite{Takeuchi211}. In this direction, 
convolutional AMP (CAMP)~\cite{Takeuchi202,Takeuchi211} was proposed as 
an LM generalization of AMP. The framework was used to construct VAMP with 
warm-started conjugate gradient (WS-CG) in \cite{Skuratovs20}.   
To solve an convergence issue in CAMP for sensing 
matrices with high condition numbers~\cite{Takeuchi22}, memory AMP 
(MAMP)~\cite{Liu21} was proposed. 
State evolution results in \cite{Liu21} can describe the exact dynamics of 
MAMP even if damping is used. 
Since the main purpose of these LM-MP algorithms is to 
reduce the complexity of OAMP, such LM-MP 
algorithms are out of the scope of this paper. 

The ``convergence in principle'' is the main idea to prove the convergence 
of OAMP. It means that the convergence of an iterative algorithm is 
intuitively trivial while the convergence of conventional MP algorithms is 
non-trivial and therefore needs to be proved for individual problems. 
The current message in LM-MP can be regarded as an additional  
measurement for the signal vector that is correlated with all preceding 
messages. When all measurements are utilized in a \emph{Bayes-optimal} manner 
to estimate the signal vector, using the additional 
measurements never degrades the estimation performance. As a result, the 
performance of the obtained Bayes-optimal LM-MP is monotonically convergent 
in principle as the iteration proceeds. 

To design the Bayes-optimal estimation of the signal vector in each 
iteration, we start with an LM generalization of OAMP---called LM-OAMP. 
The estimation errors in LM-OAMP are proved to be jointly 
Gaussian-distributed for all iterations in the large system limit 
via the unified framework of state evolution~\cite{Takeuchi211}. Thus, 
the design issue of Bayes-optimal 
LM-OAMP reduces to a solvable issue, i.e.\ the Bayes-optimal estimation of 
the signal vector from correlated additive white Gaussian noise (AWGN) 
measurements. The convergence of Bayes-optimal OAMP is guaranteed by 
proving the equivalence between Bayes-optimal LM-OAMP and OAMP. 

\subsection{Contributions}
The main contributions of this paper are threefold. In particular, 
the first two contributions were presented in \cite{Takeuchi222}. 

A first contribution is a statistical interpretation for the optimization of 
damping in \cite{Liu21}. This paper derives the optimized damping in terms 
of a sufficient statistic, without using state evolution recursions, while 
it was obtained via an optimization problem for state evolution 
recursions~\cite{Liu21}. This statistical interpretation is a key technical 
tool to prove the convergence of Bayes-optimal LM-OAMP rigorously.  

A second contribution is a derivation of state evolution recursions 
for LM-OAMP (Theorems~\ref{THEOREM_GAUSSIANITY} and \ref{THEOREM_SE}) and 
a rigorous convergence analysis of state evolution recursions for 
Bayes-optimal LM-OAMP (Theorem~\ref{THEOREM_SE_BAYES}), as well as an exact 
reduction of Bayes-optimal LM-OAMP to conventional Bayes-optimal 
OAMP~\cite{Ma17}. These results provide a rigorous proof for the convergence 
of conventional Bayes-optimal OAMP in the large system limit.  

The last contribution is a negative aspect of LM-OAMP. Numerical simulations 
show that LM-OAMP requires larger systems than conventional OAMP for state 
evolution to provide a good approximation for finite-sized systems. As a 
result, damped LM-OAMP is slightly inferior to OAMP with heuristic damping 
for small-to-moderate system sizes while it is consistent with state evolution 
for large systems. This observation explains why MAMP with the optimized LM 
damping~\cite{Liu21} needs large systems to realize good convergence 
properties.

\subsection{Organization}
After summarizing the notation used in this paper, Section~\ref{sec2} 
presents the statistical interpretation and technical results in the first 
contribution as preliminaries to propose LM-OAMP in Section~\ref{sec3}. 
Section~\ref{sec4} presents state evolution results 
for LM-OAMP. The convergence of state evolution recursions for 
Bayes-optimal LM-OAMP is analyzed in 
Section~\ref{sec5}. Numerical simulations for damped LM-OAMP are presented in 
Section~\ref{sec6}. Section~\ref{sec7} concludes this paper. 
Several theorems and lemmas are proved in the appendices. 

\subsection{Notation} \label{sec1_D}
Throughout this paper, $\boldsymbol{M}^\mathrm{T}$, 
$\mathrm{Tr}(\boldsymbol{M})$, and $\det\boldsymbol{M}$ denote the 
transpose, trace, and determinant of a matrix $\boldsymbol{M}$, respectively.  
The notation $[\boldsymbol{M}]_{t',t}$ represents the $(t', t)$ element of 
$\boldsymbol{M}$. 

For a vector $\boldsymbol{v}_{t}$ with a subscript $t$, 
the $n$th element of $\boldsymbol{v}_{t}$ is written as $v_{n,t}$. 
The notation $\|\cdots\|$ denotes the Euclidean norm. The vector 
$\boldsymbol{1}$ represents a vector whose elements are all one while 
$\boldsymbol{e}_{n}$ is the $n$th column of the identity matrix 
$\boldsymbol{I}$. The notations $\otimes$ and $\delta_{i,j}$ denote the 
Kronecker product and delta, respectively. 

The Gaussian distribution with mean $\boldsymbol{\mu}$ 
and covariance $\boldsymbol{\Sigma}$ is written as 
$\mathcal{N}(\boldsymbol{\mu},\boldsymbol{\Sigma})$. The notations  
$\ato$ and $\aeq$ represent the almost sure convergence and equivalence, 
respectively. 

For a function $f:\mathbb{R}^{t}\to\mathbb{R}$ of $t$ variables, the notation 
$f(\boldsymbol{x}_{1},\ldots,\boldsymbol{x}_{t})$ means the element-wise 
application of $f$, i.e.\ $[f(\boldsymbol{x}_{1},\ldots,\boldsymbol{x}_{t})]_{n}
=f([\boldsymbol{x}_{1}]_{n},\ldots,[\boldsymbol{x}_{t}]_{n})$. For a vector 
$\boldsymbol{v}\in\mathbb{R}^{N}$, the arithmetic mean of $\boldsymbol{v}$ is 
written as $\langle\boldsymbol{v}\rangle=N^{-1}\sum_{n=1}^{N}v_{n}$. 
Combining these notations, we find that 
$\langle f(\boldsymbol{x},\boldsymbol{y})\rangle$ means 
$N^{-1}\sum_{n=1}^{N}f(x_{n}, y_{n})$ for $\boldsymbol{x},\boldsymbol{y}
\in\mathbb{R}^{N}$. 

For any finite set $\mathcal{T}$ of non-negative integers, we define 
the one-to-one mapping $i=I_{\mathcal{T}}(t)\in\{1,\ldots,|\mathcal{T}|\}$ to 
represent that $t\in\mathcal{T}$ is the $i$th minimum element in 
$\mathcal{T}$, e.g.\ for $\mathcal{T}=\{2, 3, 4\}$ we have 
$I_{\mathcal{T}}(2)=1$, $I_{\mathcal{T}}(3)=2$, and $I_{\mathcal{T}}(4)=3$.

\section{Preliminaries}
\label{sec2}
\subsection{Measurement Model}
The purpose of Section~\ref{sec2} is to present the technical background 
of LM-OAMP proposed in Section~\ref{sec3}. This section is separated 
from the other sections in terms of the notation. 

Consider correlated $(t+1)$ AWGN measurements 
$\{Y_{\tau}\in\mathbb{R}\}_{\tau=0}^{t}$ for a signal $X\in\mathbb{R}$. 
For a subset $\mathcal{T}_{\tau}\subset\{0,\ldots,\tau\}$, we reconstruct the 
signal $X$ from partial measurements 
$\boldsymbol{Y}_{\tau}\in\mathbb{R}^{1\times|\mathcal{T}_{\tau}|}$ that 
consist of $\{Y_{t'}: t'\in\mathcal{T}_{\tau}\}$, given by  
\begin{equation} \label{correlated_measurement} 
\boldsymbol{Y}_{\tau} = X\boldsymbol{1}^{\mathrm{T}} + \boldsymbol{W}_{\tau}, 
\end{equation}
where $\boldsymbol{W}_{\tau}=\{W_{t'}: t'\in\mathcal{T}_{\tau}\}
\sim\mathcal{N}(\boldsymbol{0},\boldsymbol{\Sigma}_{\tau})$ denotes 
a zero-mean Gaussian row vector with 
covariance $\mathbb{E}[\boldsymbol{W}_{\tau}^{\mathrm{T}}\boldsymbol{W}_{\tau}]
=\boldsymbol{\Sigma}_{\tau}\in\mathbb{R}^{|\mathcal{T}_{\tau}|\times 
|\mathcal{T}_{\tau}|}$. The covariance matrix $\boldsymbol{\Sigma}_{\tau}$ 
is assumed to be positive definite. When all measurements are used, i.e.\ 
$\mathcal{T}_{\tau}=\{0,\ldots,\tau\}$, $\boldsymbol{\Sigma}_{\tau}$ 
is the $(\tau+1)\times(\tau+1)$ upper-left submatrix in  
$\boldsymbol{\Sigma}_{t}$ for all $\tau<t$.  
The goal is to estimate the signal $X$ from the knowledge 
on the measurement vector $\boldsymbol{Y}_{\tau}$. 

We know that the posterior mean estimator 
$\mathbb{E}[X|\boldsymbol{Y}_{\tau}]$ is the best among 
all possible estimators $f(\boldsymbol{Y}_{\tau})$ in terms of the 
mean-square error (MSE) $\mathbb{E}[\{X-f(\boldsymbol{Y}_{\tau})\}^{2}]$. 
In computing the posterior mean estimator, we use a two-step approach: 
A first step is the derivation of a sufficient statistic $S_{\tau}$ for 
estimation of $X$ based on the measurements $\boldsymbol{Y}_{\tau}$. 
The second step is to compute the conditional mean of $X$ given the 
sufficient statistic $S_{\tau}$, instead of the original measurements 
$\boldsymbol{Y}_{\tau}$. This two-step approach is useful in proving 
technical lemmas to establish the main results 
while it is equivalent to direct computation of the posterior mean estimator, 
i.e.\ $\mathbb{E}[X|\boldsymbol{Y}_{\tau}]=\mathbb{E}[X|S_{\tau}]$.

\subsection{Sufficient Statistic}
\label{appen_AB}
We evaluate a sufficient statistic for estimation of $X$ based on the 
measurements $\boldsymbol{Y}_{\tau}$. By definition, the probability 
density function (pdf) of $\boldsymbol{Y}_{\tau}$ given $X$ is defined as 
\begin{equation}
p(\boldsymbol{Y}_{\tau} | X) 
= \frac{\exp\{-(\boldsymbol{Y}_{\tau}-X\boldsymbol{1}^{\mathrm{T}})
\boldsymbol{\Sigma}_{\tau}^{-1}
(\boldsymbol{Y}_{\tau}-X\boldsymbol{1}^{\mathrm{T}})^{\mathrm{T}}/2\}}
{\{(2\pi)^{|\mathcal{T}_{\tau}|}\det\boldsymbol{\Sigma}_{\tau}\}^{1/2}}.  
\end{equation} 
Evaluating the exponent in the pdf yields 
\begin{IEEEeqnarray}{rl}
&(\boldsymbol{Y}_{\tau}-X\boldsymbol{1}^{\mathrm{T}})
\boldsymbol{\Sigma}_{\tau}^{-1}
(\boldsymbol{Y}_{\tau}-X\boldsymbol{1}^{\mathrm{T}})^{\mathrm{T}} 
\nonumber \\
=& \boldsymbol{Y}_{\tau}\boldsymbol{\Sigma}_{\tau}^{-1}
\boldsymbol{Y}_{\tau}^{\mathrm{T}} 
- 2X\boldsymbol{Y}_{\tau}\boldsymbol{\Sigma}_{\tau}^{-1}\boldsymbol{1}
+ X^{2}\boldsymbol{1}^{\mathrm{T}}\boldsymbol{\Sigma}_{\tau}^{-1}\boldsymbol{1},
\end{IEEEeqnarray}
which implies that $\boldsymbol{Y}_{\tau}\boldsymbol{\Sigma}_{\tau}^{-1}
\boldsymbol{1}$ is a sufficient statistic for estimation of $X$ given  
$\boldsymbol{Y}_{\tau}$. Normalizing this sufficient 
statistic, we arrive at 
\begin{equation} \label{sufficient_statistic}
S_{\tau}
=\frac{\boldsymbol{Y}_{\tau}\boldsymbol{\Sigma}_{\tau}^{-1}\boldsymbol{1}}
{\boldsymbol{1}^{\mathrm{T}}\boldsymbol{\Sigma}_{\tau}^{-1}\boldsymbol{1}}
= X + \tilde{W}_{\tau}, 
\end{equation}
with 
\begin{equation} \label{W_tilde}
\tilde{W}_{\tau}
= \frac{\boldsymbol{W}_{\tau}\boldsymbol{\Sigma}_{\tau}^{-1}\boldsymbol{1}}
{\boldsymbol{1}^{\mathrm{T}}\boldsymbol{\Sigma}_{\tau}^{-1}\boldsymbol{1}}.
\end{equation} 

It is straightforward to confirm that $\tilde{W}_{\tau}$ is a zero-mean 
Gaussian random variable with variance $\mathbb{E}[\tilde{W}_{\tau}^{2}]$, 
given by 
\begin{equation} \label{var_W_tilde}
\mathbb{E}[\tilde{W}_{\tau}^{2}]
=\frac{1}{\boldsymbol{1}^{\mathrm{T}}\boldsymbol{\Sigma}_{\tau}^{-1}\boldsymbol{1}}.
\end{equation} 
Furthermore, the correlation $\mathbb{E}[\tilde{W}_{\tau'}\tilde{W}_{\tau}]$ 
is given by 
\begin{equation} \label{correlation_W}
\mathbb{E}[\tilde{W}_{\tau'}\tilde{W}_{\tau}] 
= \frac{\boldsymbol{1}^{\mathrm{T}}\boldsymbol{\Sigma}_{\tau'}^{-1}
\mathbb{E}[\boldsymbol{W}_{\tau'}^{\mathrm{T}}\boldsymbol{W}_{\tau}]
\boldsymbol{\Sigma}_{\tau}^{-1}\boldsymbol{1}}
{\boldsymbol{1}^{\mathrm{T}}\boldsymbol{\Sigma}_{\tau'}^{-1}\boldsymbol{1}
\boldsymbol{1}^{\mathrm{T}}\boldsymbol{\Sigma}_{\tau}^{-1}\boldsymbol{1}}. 
\end{equation}

When all measurements are used, i.e.\ $\mathcal{T}_{\tau}=\{0,\ldots,\tau\}$,  
the correlation~(\ref{correlation_W}) reduces to 
\begin{equation} \label{correlation_W_reduction}
\mathbb{E}[\tilde{W}_{\tau'}\tilde{W}_{\tau}] 
=\frac{1}{\boldsymbol{1}^{\mathrm{T}}\boldsymbol{\Sigma}_{\tau}^{-1}\boldsymbol{1}}
= \mathbb{E}[\tilde{W}_{\tau}^{2}] 
\end{equation} 
for all $\tau'\leq\tau$, since $\mathbb{E}[\boldsymbol{W}_{\tau'}^{\mathrm{T}}
\boldsymbol{W}_{\tau}]=(\boldsymbol{I}_{\tau'+1}, \boldsymbol{O})
\boldsymbol{\Sigma}_{\tau}$ holds for 
$\boldsymbol{W}_{\tau}=(W_{0},\ldots,W_{\tau})$. This property is used to prove 
the convergence of the error covariance in Bayes-optimal LM-OAMP via that 
of its diagonal elements, which are monotonically decreasing MSEs  
when we use the posterior mean estimator of the signal vector 
given all preceding messages.

Finally, we investigate the impact of small eigenvalues for 
$\boldsymbol{\Sigma}_{\tau}$ on the variance $\mathbb{E}[\tilde{W}_{\tau}^{2}]$. 
\begin{proposition} \label{proposition_eigenvalue}
Suppose that $\mathbb{E}[\tilde{W}_{\tau}^{2}]>\sigma_{0}^{2}$ holds for some 
$\sigma_{0}^{2}>0$. For the eigen-decomposition $\boldsymbol{\Sigma}_{\tau}
=\sum_{\tau'}\lambda_{\tau'}\boldsymbol{u}_{\tau'}
\boldsymbol{u}_{\tau'}^{\mathrm{T}}$, we have 
\begin{equation} \label{eigenvector}
\boldsymbol{1}^{\mathrm{T}}\boldsymbol{u}_{\tau'}
= {\cal O}(\sqrt{\lambda_{\tau'}})
\quad \hbox{as $\lambda_{\tau'}\downarrow0$.} 
\end{equation}
\end{proposition}
\begin{IEEEproof}
Applying the eigen-decomposition of $\boldsymbol{\Sigma}_{\tau}^{-1}$ 
to the definition of $\mathbb{E}[\tilde{W}_{\tau}^{2}]$ in (\ref{var_W_tilde}), 
we obtain   
\begin{equation}
\mathbb{E}[\tilde{W}_{\tau}^{2}] 
= \left\{
 \sum_{\tau'}\frac{ (\boldsymbol{1}^{\mathrm{T}}\boldsymbol{u}_{\tau'})^{2}}
 {\lambda_{\tau'}}
\right\}^{-1}>\sigma_{0}^{2}, 
\end{equation}
which implies $(\boldsymbol{1}^{\mathrm{T}}\boldsymbol{u}_{\tau'})^{2}
/\lambda_{\tau'}<\sigma_{0}^{-2}$ for all $\tau'$. Thus, 
we arrive at (\ref{eigenvector}). 
\end{IEEEproof}

The quantity $\sigma_{0}^{2}$ in Proposition~\ref{proposition_eigenvalue} 
may be obtained via the Bayes-optimal performance for the original 
compressed sensing problem~(\ref{model}). There exists 
$\sigma_{0}^{2}>0$ as long as noisy measurements are considered.  

Proposition~\ref{proposition_eigenvalue} implies that the inner product 
$\boldsymbol{1}^{\mathrm{T}}\boldsymbol{u}_{\tau'}$ must tend to zero 
as some eigenvalue $\lambda_{\tau'}>0$ goes to zero. In other words,  
all eigenvectors associated with small eigenvalues have a tendency to 
be orthogonal to the vector $\boldsymbol{1}$. Unless this asymptotic 
orthogonality holds, error-free estimation of $X$ is possible via the matched 
filter (MF) $\boldsymbol{Y}_{\tau}\boldsymbol{u}_{\tau'}
/\boldsymbol{1}^{\mathrm{T}}\boldsymbol{u}_{\tau'}\to X$ in probability 
as $\lambda_{\tau'}\downarrow0$.

\subsection{Extrinsic Denoiser} \label{sec2_C} 
The Bayes-optimal estimator is defined as the posterior mean estimator 
$f_{\mathrm{opt}}(S_{t})=\mathbb{E}[X|S_{t}]$ of $X$ given the sufficient 
statistic $S_{t}$. By definition, we have $\mathbb{E}[X|S_{t}]
=\mathbb{E}[X|\boldsymbol{Y}_{t}]$. To realize asymptotic Gaussianity 
in LM-OAMP,  
we define the extrinsic denoiser $f_{t}^{\mathrm{ext}}:\mathbb{R}^{|\mathcal{T}_{t}|}
\to\mathbb{R}$ as
\begin{equation} \label{extrinsic_denoiser}
f_{t}^{\mathrm{ext}}(\boldsymbol{Y}_{t}) = \frac{f_{\mathrm{opt}}(S_{t}) 
- \sum_{\tau\in\mathcal{T}_{t}}\xi_{\tau,t}Y_{\tau}}
{1 - \sum_{\tau\in\mathcal{T}_{t}}\xi_{\tau,t}}, 
\end{equation}
where $\xi_{\tau,t}$ for $\tau\in\mathcal{T}_{t}$ denotes the averaged 
partial derivative of $f_{\mathrm{opt}}(S_{t})$ with respect to $Y_{\tau}$, given by 
\begin{equation}  \label{xi}
\xi_{\tau,t} 
= \mathbb{E}[f_{\mathrm{opt}}'(S_{t})]
\frac{\boldsymbol{e}_{I_{\mathcal{T}_{t}}(\tau)}^{\mathrm{T}}
\boldsymbol{\Sigma}_{t}^{-1}\boldsymbol{1}}
{\boldsymbol{1}^{\mathrm{T}}\boldsymbol{\Sigma}_{t}^{-1}\boldsymbol{1}}.  
\end{equation}
See Section~\ref{sec1_D} for the definition of the mapping 
$I_{\mathcal{T}_{t}}(\cdot)$. 
By definition, we have $\sum_{\tau\in\mathcal{T}_{t}}\xi_{\tau,t}
=\mathbb{E}[f_{\mathrm{opt}}'(S_{t})]$.   

The numerator in the extrinsic denoiser~(\ref{extrinsic_denoiser}) is the 
Onsager correction of the Bayes-optimal estimator $f_{\mathrm{opt}}(S_{t})$, 
which originates from a general error model in \cite{Takeuchi211}. 
The significance of the denominator is presented in the following proposition:
\begin{proposition} \label{proposition_denoiser}
Among all possible Lipschitz-continuous extrinsic denoisers 
$\psi_{t}:\mathbb{R}^{|\mathcal{T}_{t}|}
\to\mathbb{R}$ satisfying $\mathbb{E}[\partial\psi(\boldsymbol{Y}_{t})
/\partial Y_{\tau}]=0$ 
for all $\tau\in\mathcal{T}_{t}$, the extrinsic denoiser $f_{t}^{\mathrm{ext}}$ 
configured from the Bayes-optimal estimator 
minimizes $\mathbb{E}[\{\psi_{t}(\boldsymbol{Y}_{t}) - X\}^{2}]$.  
\end{proposition}
\begin{IEEEproof}
See Appendix~\ref{appen_AA}. 
\end{IEEEproof} 
 
Proposition~\ref{proposition_denoiser} implies that the denominator is the 
best option in terms of the MSE, as long as the Bayes-optimal denoiser is 
used. 

\subsection{Estimator for Error Covariance} \label{sec2_D}
For denoisers $f_{t'}$ and $f_{t}$, LM-OAMP needs a consistent estimator 
for the error covariance $\mathbb{E}[\{X-f_{t'}(S_{t'})\}
\{X-f_{t}(S_{t})\}]$ under the assumption of the known signal power 
$\mathbb{E}[X^{2}]$ and covariance $\mathbb{E}[\tilde{W}_{t'}\tilde{W}_{t}]$.  
When the Bayes-optimal estimators $f_{t'}=f_{\mathrm{opt}}(S_{t'})$ and 
$f_{t}=f_{\mathrm{opt}}(S_{t})$ are used, we can use the posterior covariance 
$C(S_{t'},S_{t})$ as a consistent estimator, given by  
\begin{equation} \label{posterior_covariance}
C(S_{t'},S_{t}) 
= \mathbb{E}[\{X-f_{\mathrm{opt}}(S_{t'})\}\{X-f_{\mathrm{opt}}(S_{t})\}
| S_{t'},S_{t}]. 
\end{equation}
In particular, $C(S_{t},S_{t})$ reduces to the posterior variance  
$\mathbb{E}[\{X-f_{\mathrm{opt}}(S_{t})\}^{2}|S_{t}]$. However, it is not 
straightforward to construct a consistent estimator for general denoisers.

We derive a consistent estimator of the error covariance for 
general $f_{t'}$ and $f_{t}$. 
For any $t'$ and $t$, let $\{S_{t',n}, S_{t,n}\}_{n=1}^{N}$ denote 
independent samples that follow the same joint distribution as for the two 
sufficient statistics $(S_{t'}, S_{t})$, i.e.\ 
\begin{equation}
S_{t,n} = X_{n} + \tilde{W}_{t,n}, 
\end{equation} 
where $\{X_{n}\}$ are independent random variables satisfying $X_{n}\sim X$, 
while $\{\tilde{W}_{t',n}, \tilde{W}_{t,n}\}_{n=1}^{N}$ are independent zero-mean 
Gaussian random variables with covariance 
$\mathbb{E}[\tilde{W}_{t',n}\tilde{W}_{t,n}]
=\mathbb{E}[\tilde{W}_{t'}\tilde{W}_{t}]$. 

\begin{lemma} \label{LEMMA_CONSISTENT} 
Suppose that $f_{t'}$ and $f_{t}$ are Lipschitz-continuous denoisers and let 
\begin{IEEEeqnarray}{r}
\hat{C}(S_{t',n},S_{t,n}) 
= \mathbb{E}[X^{2}]  
+ f_{t'}(S_{t',n})f_{t}(S_{t,n})
\nonumber \\
+ \mathbb{E}[\tilde{W}_{t'}\tilde{W}_{t}]f_{t}'(S_{t,n})
- S_{t',n}f_{t}(S_{t,n})
\nonumber \\ 
+ \mathbb{E}[\tilde{W}_{t'}\tilde{W}_{t}]f_{t'}'(S_{t',n})
- S_{t,n}f_{t'}(S_{t',n}).  \label{consistent_estimator}
\end{IEEEeqnarray}
Then, the sample average of $\{\hat{C}(S_{t',n},S_{t,n})\}_{n=1}^{N}$ is 
a consistent estimator: $N^{-1}\sum_{n=1}^{N}\hat{C}(S_{t',n},S_{t,n})$ converges 
almost surely to the error covariance $\mathbb{E}[\{X-f_{t'}(S_{t'})\}
\{X-f_{t}(S_{t})\}]$ as $N\to\infty$. 
\end{lemma}
\begin{IEEEproof}
See Appendix~\ref{proof_LEMMA_CONSISTENT}. 
\end{IEEEproof}

\begin{corollary}
Suppose that $f_{t}$ is a Lipschitz-continuous denoiser and let 
\begin{equation}
\hat{C}(S_{t,n}) 
= \mathbb{E}[X^{2}]  
+ \mathbb{E}[\tilde{W}_{t}^{2}]f_{t}'(S_{t,n})
- S_{t,n}f_{t}(S_{t,n}).  
\end{equation}
Then, the sample average $N^{-1}\sum_{n=1}^{N}\hat{C}(S_{t,n})$ is a consistent 
estimator for $\mathbb{E}[X\{X-f_{t}(S_{t})\}]$ as $N\to\infty$. 
\end{corollary}
\begin{IEEEproof}
Use Lemma~\ref{LEMMA_CONSISTENT} for $S_{t'}=S_{t}$ and $f_{t'}=0$. 
\end{IEEEproof}

\subsection{Properties of Bayes-Optimal Estimator} 
We investigate properties of the Bayes-optimal estimator 
$f_{\mathrm{opt}}(S_{t})=\mathbb{E}[X|S_{t}]$.  
We have the following technical tools to analyze Bayes-optimal LM-OAMP: 
 
\begin{lemma} \label{LEMMA_BAYES}
Let $\mathcal{T}_{\tau}=\{0,\ldots,\tau\}$ for all $\tau$.  
\begin{itemize}
\item The monotonicity $\mathbb{E}[\{X-f_{\mathrm{opt}}(S_{t'})\}^{2}]
\geq\mathbb{E}[\{X-f_{\mathrm{opt}}(S_{t})\}^{2}]$ 
and $\mathbb{E}[\tilde{W}_{t'}^{2}]\geq\mathbb{E}[\tilde{W}_{t}^{2}]$ holds for 
all $t'<t$. 
\item The identity $C(S_{t'},S_{t})=C(S_{t},S_{t})$ for 
the posterior covariance~(\ref{posterior_covariance}) holds for $t'\leq t$ 
if $\mathbb{E}[\tilde{W}_{t'}\tilde{W}_{t}]=\mathbb{E}[\tilde{W}_{t}^{2}]$ 
is satisfied. 
\end{itemize}
\end{lemma}
\begin{IEEEproof}
See Appendix~\ref{proof_LEMMA_BAYES}. 
\end{IEEEproof}

\begin{lemma} \label{LEMMA_MONOTONICITY} 
Let $\mathcal{T}_{\tau}=\{0,\ldots,\tau\}$ for all $\tau$ and suppose that 
the covariance $\boldsymbol{\Sigma}_{t}$ for the noise vector 
$\boldsymbol{W}_{t}$ in (\ref{correlated_measurement}) satisfies 
$[\boldsymbol{\Sigma}_{t}]_{\tau',\tau}=[\boldsymbol{\Sigma}_{t}]_{\tau,\tau'}
=[\boldsymbol{\Sigma}_{t}]_{\tau,\tau}$ for all $\tau'<\tau$. 
If $\{\boldsymbol{\Sigma}_{\tau}\}_{\tau=1}^{t}$ are positive definite, 
the following properties hold: 
\begin{itemize}
\item The monotonicity $[\boldsymbol{\Sigma}_{t}]_{\tau,\tau}
>[\boldsymbol{\Sigma}_{t}]_{\tau+1,\tau+1}$ holds for all 
$\tau\in\{0,\ldots,t-1\}$. 
\item The identity $\mathbb{E}[\tilde{W}_{t}^{2}]
=\mathbb{E}[W_{t}^{2}]$ holds for $\tilde{W}_{t}$ given 
in (\ref{W_tilde}).
\item The identities $S_{t}=Y_{t}$ and $\xi_{\tau,t}=0$ hold for all 
$\tau\neq t$.  
\end{itemize}
\end{lemma}
\begin{IEEEproof}
See Appendix~\ref{proof_LEMMA_MONOTONICITY}. 
\end{IEEEproof}

The former property in Lemma~\ref{LEMMA_BAYES} is used to prove the 
convergence of state evolution recursions for Bayes-optimal LM-OAMP. 
The latter property and 
Lemma~\ref{LEMMA_MONOTONICITY} are useful in proving an exact reduction of 
Bayes-optimal LM-OAMP to conventional Bayes-optimal OAMP.

\section{Long-Memory OAMP} \label{sec3}
\subsection{Outline}
LM-OAMP is a generalization of conventional OAMP~\cite{Ma17,Rangan192}. 
OAMP consists of two modules---called modules A and B in this paper.  
Module~A uses a linear filter to compute a posterior message of the 
signal vector $\boldsymbol{x}$ while module~B utilizes a separable denoiser to 
refine the posterior message for each signal element. A crucial step in each 
module is the so-called Onsager correction to obtain extrinsic messages of 
$\boldsymbol{x}$. This step realizes asymptotic Gaussianity for the 
estimation errors in the two modules~\cite{Rangan192,Takeuchi201}. 
The obtained messages may be damped in a heuristic manner to improve the 
convergence property of OAMP for finite-sized systems.  

\begin{figure}[t]
\begin{center}
\includegraphics[width=\hsize]{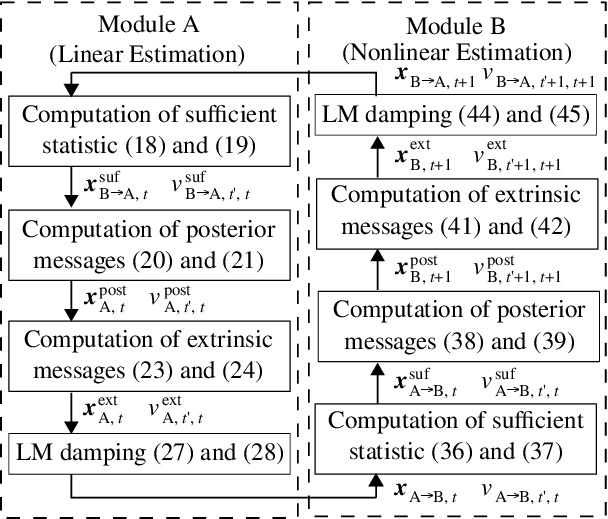}
\caption{
Diagram of LM-OAMP. 
}
\label{fig1} 
\end{center}
\end{figure}

LM-OAMP consists of two modules in Fig.~\ref{fig1}, 
of which each has four steps. In the first two steps, 
each module utilizes all preceding messages to compute a posterior message 
of $\boldsymbol{x}$ while OAMP only uses those in the latest iteration. 
A first step in computing the posterior message is computation of a 
sufficient statistic for estimation of 
$\boldsymbol{x}$ given all preceding messages, as presented in 
Section~\ref{sec2}. The second step computes a posterior message of 
$\boldsymbol{x}$ based on the sufficient statistic. This step corresponds to 
the computation of the posterior messages in conventional OAMP. 
While the first two steps are equivalent to direct computation of the posterior 
message given all preceding messages, the two-step approach simplifies 
the formulation of the Onsager correction in a third step. 

To realize asymptotic Gaussianity in LM-OAMP, each module computes the 
Onsager correction of each posterior message. This third step is heavily 
based on a general framework of state evolution~\cite{Takeuchi211}, which can 
evaluate asymptotic correlations between the current message and all 
preceding messages rigorously. 

The last step is optional: LM damping~\cite{Liu21} is employed to improve 
the convergence property of LM-OAMP for finite-sized systems. As long as 
the Onsager-corrected messages in the third step are damped, LM damping 
does not break asymptotic Gaussianity for the estimation errors while 
heuristic damping of posterior messages breaks asymptotic 
Gaussianity~\cite{Takeuchi211}.  

\subsection{Notation}
The notations used in LM-OAMP are summarized. 
Let $\boldsymbol{x}_{\mathrm{A}\to \mathrm{B},t}\in\mathbb{R}^{N}$ and 
$\{v_{\mathrm{A}\to \mathrm{B},t',t}\in\mathbb{R}\}_{t'=0}^{t}$ denote mean and 
covariance messages for the signal vector $\boldsymbol{x}$, respectively,  
passed from module~A to module~B in iteration~$t$. The covariance 
$v_{\mathrm{A}\to \mathrm{B},\tau',\tau}$ corresponds 
to an estimate of the true error covariance 
$N^{-1}\mathbb{E}[(\boldsymbol{x}_{\mathrm{A}\to \mathrm{B},\tau'}
-\boldsymbol{x})^{\mathrm{T}}
(\boldsymbol{x}_{\mathrm{A}\to \mathrm{B},\tau}-\boldsymbol{x})]$ for all 
iterations $\tau',\tau\in\{0,\ldots,t\}$. For notational convenience, 
we define the infinite-dimensional real 
symmetric matrix $\boldsymbol{V}_{\mathrm{A}\to \mathrm{B}}$ satisfying 
$[\boldsymbol{V}_{\mathrm{A}\to \mathrm{B}}]_{\tau',\tau}
=[\boldsymbol{V}_{\mathrm{A}\to \mathrm{B}}]_{\tau,\tau'}
=v_{\mathrm{A}\to \mathrm{B},\tau',\tau}$. 
Messages passed in the opposite direction are written as 
$\boldsymbol{x}_{\mathrm{B}\to \mathrm{A},t}\in\mathbb{R}^{N}$, 
$\{v_{\mathrm{B}\to \mathrm{A},t',t}\in\mathbb{R}\}_{t'=0}^{t}$, and 
$\boldsymbol{V}_{\mathrm{B}\to \mathrm{A}}$. 

The two modules utilize part of preceding messages in each iteration~$t$. 
Let $\mathcal{T}_{\mathrm{A},t}\subset\{0,\ldots,t\}$ and 
$\mathcal{T}_{\mathrm{B},t}\subset\{0,\ldots,t\}$ denote subsets of indices that 
represent preceding messages used in iteration~$t$ for modules A and B, 
respectively. We assume that both subsets $\mathcal{T}_{\mathrm{A},t}$ and 
$\mathcal{T}_{\mathrm{B},t}$ contain the current index $t$. 

We write the message matrix obtained by arranging the preceding messages 
$\{\boldsymbol{x}_{\mathrm{A}\to \mathrm{B},t'}:t'\in\mathcal{T}_{\mathrm{B},t}\}$ as 
$\boldsymbol{X}_{\mathrm{A}\to \mathrm{B},t}\in
\mathbb{R}^{N\times|\mathcal{T}_{\mathrm{B},t}|}$.
More precisely, $\boldsymbol{x}_{\mathrm{A}\to \mathrm{B},t'}$ for 
$t'\in\mathcal{T}_{\mathrm{B},t}$ 
is the $i$th column in $\boldsymbol{X}_{\mathrm{A}\to \mathrm{B},t}$ with 
$i=I_{\mathcal{T}_{\mathrm{B},t}}(t')$, defined in Section~\ref{sec1_D}. 
Furthermore, we define $\boldsymbol{V}_{\mathrm{A}\to \mathrm{B},t',t}\in
\mathbb{R}^{|\mathcal{T}_{\mathrm{B},t'}|\times|\mathcal{T}_{\mathrm{B},t}|}$ 
as the matrix obtained by extracting the $(\tau', \tau)$ elements 
from $\boldsymbol{V}_{\mathrm{A}\to \mathrm{B}}$ for all $(\tau', \tau)\in
\mathcal{T}_{\mathrm{B},t'}\times\mathcal{T}_{\mathrm{B},t}$. 

Similarly, we define the message matrices  
$\boldsymbol{X}_{\mathrm{B}\to \mathrm{A},t}$ and 
$\boldsymbol{V}_{\mathrm{B}\to \mathrm{A},t',t}\in
\mathbb{R}^{|\mathcal{T}_{\mathrm{A},t'}|\times|\mathcal{T}_{\mathrm{A},t}|}$ 
passed in the opposite direction via the preceding messages 
$\{\boldsymbol{x}_{\mathrm{B}\to \mathrm{A},t'}:t'\in\mathcal{T}_{\mathrm{A},t}\}$ and  
$\boldsymbol{V}_{\mathrm{B}\to \mathrm{A}}$. In particular, LM-OAMP reduces to 
conventional OAMP~\cite{Ma17} when the subsets 
$\mathcal{T}_{\mathrm{A},t}=\mathcal{T}_{\mathrm{B},t}=\{t\}$ 
are used, while $\mathcal{T}_{\mathrm{A},t}=\mathcal{T}_{\mathrm{B},t}
=\{0,\ldots,t\}$ is the best in terms of the estimation performance. 
Interestingly, this paper proves that Bayes-optimal LM-OAMP reduces to 
Bayes-optimal OAMP for 
$\mathcal{T}_{\mathrm{A},t}=\mathcal{T}_{\mathrm{B},t}=\{0,\ldots,t\}$.

Update rules in LM-OAMP are defined so as to realize   
asymptotic Gaussianity for $\{\boldsymbol{x}_{\mathrm{A}\to \mathrm{B},t'}\}$. 
While a mathematically rigorous definition for asymptotic Gaussianity is 
presented shortly, a rough intuition is to regard the estimation errors  
$\boldsymbol{h}_{t} = \boldsymbol{x}_{\mathrm{A}\to \mathrm{B},t} - \boldsymbol{x}$ 
and $\boldsymbol{b}_{t}= \boldsymbol{V}^{\mathrm{T}}
(\boldsymbol{x}_{\mathrm{B}\to \mathrm{A},t}-\boldsymbol{x})$ 
for the singular-value decomposition (SVD) $\boldsymbol{A}
=\boldsymbol{U}\boldsymbol{\Sigma}\boldsymbol{V}^{\mathrm{T}}$ 
as zero-mean Gaussian vectors with covariance 
$\mathbb{E}[\boldsymbol{h}_{\tau}\boldsymbol{h}_{\tau'}^{\mathrm{T}}]
=[\boldsymbol{V}_{\mathrm{A}\to \mathrm{B}}]_{\tau',\tau}\boldsymbol{I}_{N}$ and 
$\mathbb{E}[\boldsymbol{b}_{\tau}\boldsymbol{b}_{\tau'}^{\mathrm{T}}]
=[\boldsymbol{V}_{\mathrm{B}\to \mathrm{A}}]_{\tau',\tau}\boldsymbol{I}_{N}$, 
respectively. These properties help us understand LM-OAMP intuitively while 
they are too strong to justify. Asymptotic Gaussianity proved in 
Theorem~\ref{THEOREM_GAUSSIANITY} is enough strong to prove the 
Bayes-optimality of LM-OAMP while it is weaker than the properties mentioned 
above.

\subsection{Module~A (Linear Estimation)}
Module~A consists of four steps in each iteration~$t$. In the first step, 
module~A uses the preceding message matrix 
$\boldsymbol{X}_{\mathrm{B}\to \mathrm{A},t}$ and the covariance messages  
$\{v_{\mathrm{B}\to \mathrm{A},\tau',\tau}: \tau', \tau\in\{0,\ldots,t\}\}$ to 
compute a sufficient statistic 
$\boldsymbol{x}_{\mathrm{B}\to \mathrm{A},t}^{\mathrm{suf}}\in\mathbb{R}^{N}$ 
for estimation of $\boldsymbol{x}$ and the corresponding covariance 
$\{v_{\mathrm{B}\to \mathrm{A},t',t}^{\mathrm{suf}}\}_{t'=0}^{t}$, given by 
\begin{equation} \label{statistic_mean_A}
\boldsymbol{x}_{\mathrm{B}\to \mathrm{A},t}^{\mathrm{suf}} 
= \frac{\boldsymbol{X}_{\mathrm{B}\to \mathrm{A},t}
\boldsymbol{V}_{\mathrm{B}\to \mathrm{A},t,t}^{-1}\boldsymbol{1}}
{\boldsymbol{1}^{\mathrm{T}}\boldsymbol{V}_{\mathrm{B}\to \mathrm{A},t,t}^{-1}
\boldsymbol{1}},  
\end{equation} 
\begin{equation} \label{statistic_covariance_A} 
v_{\mathrm{B}\to \mathrm{A},t',t}^{\mathrm{suf}}
= \frac{\boldsymbol{1}^{\mathrm{T}}\boldsymbol{V}_{\mathrm{B}\to \mathrm{A},t',t'}^{-1}
\boldsymbol{V}_{\mathrm{B}\to \mathrm{A},t',t}
\boldsymbol{V}_{\mathrm{B}\to \mathrm{A},t,t}^{-1}
\boldsymbol{1}}
{\boldsymbol{1}^{\mathrm{T}}\boldsymbol{V}_{\mathrm{B}\to \mathrm{A},t',t'}^{-1}
\boldsymbol{1}\boldsymbol{1}^{\mathrm{T}}
\boldsymbol{V}_{\mathrm{B}\to \mathrm{A},t,t}^{-1}\boldsymbol{1}}. 
\end{equation}
For the initial iteration $t=0$, we use $\boldsymbol{x}_{\mathrm{B}\to \mathrm{A},0}
=\boldsymbol{0}$ and $v_{\mathrm{B}\to \mathrm{A},0,0}
=\mathbb{E}[\|\boldsymbol{x}\|^{2}]/N$.  
These update rules are obtained by regarding the estimation errors 
$\{\boldsymbol{x}_{\mathrm{A}\to \mathrm{B},t'}-\boldsymbol{x}\}$ as zero-mean 
Gaussian vectors with covariance 
$\mathbb{E}[(\boldsymbol{x}_{\mathrm{A}\to \mathrm{B},\tau}-\boldsymbol{x})
(\boldsymbol{x}_{\mathrm{A}\to \mathrm{B},\tau'}-\boldsymbol{x})^{\mathrm{T}}]
=[\boldsymbol{V}_{\mathrm{B}\to \mathrm{A},t}]_{\tau',\tau}\boldsymbol{I}_{N}$. 
See Section~\ref{sec2} for the derivation.  

The covariance matrix $\boldsymbol{V}_{\mathrm{B}\to \mathrm{A},t,t}$ is guaranteed 
to be asymptotically positive definite via state evolution as long as $t$ is 
finite. However, $\boldsymbol{V}_{\mathrm{B}\to \mathrm{A},t,t}$ has small positive 
eigenvalues that converge to zero as $t\to\infty$. Nonetheless, such small 
eigenvalues causes no numerical issues in computing 
$\boldsymbol{V}_{\mathrm{B}\to \mathrm{A},t,t}^{-1}\boldsymbol{1}$ 
since the eigenvectors associated with such small eigenvalues 
should be orthogonal to the vector $\boldsymbol{1}$, as proved in 
Proposition~\ref{proposition_eigenvalue}. 

\begin{remark}
The sufficient statistic~(\ref{statistic_mean_A}) is a normalized linear 
combination of $\{\boldsymbol{x}_{\mathrm{B}\to \mathrm{A},t'}\}_{t'=0}^{t}$ while 
it might not be a convex combination. Thus, (\ref{statistic_mean_A}) can be 
regarded as a damped estimate in module~B. In fact, (\ref{statistic_mean_A}) 
was originally obtained via the optimization of damping~\cite{Liu21}. 
Nonetheless, the interpretation in terms of a sufficient statistic 
is important in analyzing the convergence of LM-OAMP as a technical tool. 
\end{remark}

The second step is computation of the posterior mean  
$\boldsymbol{x}_{\mathrm{A},t}^{\mathrm{post}}\in\mathbb{R}^{N}$ and 
covariance $\{v_{\mathrm{A},t',t}^{\mathrm{post}}\in\mathbb{R}\}_{t'=0}^{t}$. 
A linear filter $\boldsymbol{W}_{t}\in\mathbb{R}^{M\times N}$ is applied 
to the residual after interference subtraction based on 
the sufficient statistic~(\ref{statistic_mean_A}). 
\begin{equation} \label{post_mean_A}
\boldsymbol{x}_{\mathrm{A},t}^{\mathrm{post}} 
= \boldsymbol{x}_{\mathrm{B}\to \mathrm{A},t}^{\mathrm{suf}} 
+ \boldsymbol{W}_{t}^{\mathrm{T}}(\boldsymbol{y} 
- \boldsymbol{A}\boldsymbol{x}_{\mathrm{B}\to \mathrm{A},t}^{\mathrm{suf}}), 
\end{equation}
\begin{equation} \label{post_covariance_A}
v_{\mathrm{A},t',t}^{\mathrm{post}} 
= \gamma_{t',t}v_{\mathrm{B}\to \mathrm{A},t',t}^{\mathrm{suf}}
+ \frac{\sigma^{2}}{N}\mathrm{Tr}\left(
 \boldsymbol{W}_{t'}\boldsymbol{W}_{t}^{\mathrm{T}}
\right),
\end{equation}
with 
\begin{equation} \label{gamma_t't}
\gamma_{t',t} 
= \frac{1}{N}\mathrm{Tr}\left\{
 \left(
  \boldsymbol{I}_{N} - \boldsymbol{W}_{t'}^{\mathrm{T}}\boldsymbol{A}
 \right)^{\mathrm{T}}
 \left(
  \boldsymbol{I}_{N} - \boldsymbol{W}_{t}^{\mathrm{T}}\boldsymbol{A}
 \right)
\right\}. 
\end{equation}
The formula~(\ref{post_covariance_A}) for the posterior covariance is 
justified in the large system limit via state evolution while the posterior 
mean~(\ref{post_mean_A}) is equivalent to that in conventional 
OAMP~\cite{Ma17}.   
  
The third step is the Onsager correction to realize asymptotic 
Gaussianity. Module~A computes the extrinsic mean 
$\boldsymbol{x}_{\mathrm{A},t}^{\mathrm{ext}}\in\mathbb{R}^{N}$ and covariance 
$\{v_{\mathrm{A},t',t}^{\mathrm{ext}}\}_{t'=0}^{t}$, given by  
\begin{equation} \label{ext_mean_A} 
\boldsymbol{x}_{\mathrm{A},t}^{\mathrm{ext}} 
= \frac{\boldsymbol{x}_{\mathrm{A},t}^{\mathrm{post}} 
- \sum_{t'\in\mathcal{T}_{\mathrm{A},t}}\xi_{\mathrm{A},t',t}
\boldsymbol{x}_{\mathrm{B}\to \mathrm{A},t'}}
{1 - \xi_{\mathrm{A},t}},
\end{equation}
\begin{equation} \label{ext_covariance_A}
v_{\mathrm{A},t',t}^{\mathrm{ext}} 
= \frac{v_{\mathrm{A},t',t}^{\mathrm{post}} 
- \xi_{\mathrm{A},t'}\xi_{\mathrm{A},t}v_{\mathrm{B}\to \mathrm{A},t',t}^{\mathrm{suf}}}
{(1-\xi_{\mathrm{A},t'})(1-\xi_{\mathrm{A},t})}, 
\end{equation}
where $\xi_{\mathrm{A},t',t}\in\mathbb{R}$ is given by 
\begin{equation} \label{xi_A} 
\xi_{\mathrm{A},t',t} 
= \xi_{\mathrm{A},t}\frac{\boldsymbol{e}_{I_{\mathcal{T}_{\mathrm{A},t}}(t')}^{\mathrm{T}}
\boldsymbol{V}_{\mathrm{B}\to \mathrm{A},t,t}^{-1}\boldsymbol{1}}
{\boldsymbol{1}^{\mathrm{T}}\boldsymbol{V}_{\mathrm{B}\to \mathrm{A},t,t}^{-1}
\boldsymbol{1}} 
\end{equation}
for $t'\in\mathcal{T}_{\mathrm{A},t}$, with 
\begin{equation} \label{gamma_t}
\xi_{\mathrm{A},t} 
= \frac{1}{N}\mathrm{Tr}\left(
 \boldsymbol{I}_{N} - \boldsymbol{W}_{t}^{\mathrm{T}}\boldsymbol{A}
\right). 
\end{equation}
The factor $\xi_{\mathrm{A},t',t}$ is the empirical average for the 
partial derivatives of the posterior mean 
$\boldsymbol{x}_{\mathrm{A},t}^{\mathrm{post}}$ 
with respect to the elements in $\boldsymbol{x}_{\mathrm{B}\to \mathrm{A},t'}$. 
The formula~(\ref{ext_covariance_A}) for the covariance is 
justified via state evolution. 

The numerator in (\ref{ext_mean_A}) is the Onsager correction of 
the posterior message $\boldsymbol{x}_{\mathrm{A},t}^{\mathrm{post}}$ to realize 
asymptotic Gaussianity, and originates from a general error model 
in \cite{Takeuchi211}. 
The denominator in (\ref{ext_mean_A}) has been selected to simplify state 
evolution, as considered in VAMP~\cite{Rangan192}. In particular,    
this selection is the best when the true posterior mean estimator is used. 
See Section~\ref{sec2_C} for the details. 

The last step is LM damping~\cite{Liu21} of the extrinsic messages.   
Module~A feeds the damped messages $\boldsymbol{x}_{\mathrm{A}\to \mathrm{B},t}$ and 
$\{v_{\mathrm{A}\to \mathrm{B},t',t}\}_{t'=0}^{t}$ forward to module~B. 
\begin{equation} \label{mean_AB} 
\boldsymbol{x}_{\mathrm{A}\to \mathrm{B},t} 
= \sum_{\tau=0}^{t}\theta_{\mathrm{A},\tau,t}
\boldsymbol{x}_{\mathrm{A},\tau}^{\mathrm{ext}}, 
\end{equation}
\begin{equation} \label{covariance_AB}
v_{\mathrm{A}\to \mathrm{B},t',t} 
= \sum_{\tau'=0}^{t'}\sum_{\tau=0}^{t}\theta_{\mathrm{A},\tau',t'}\theta_{\mathrm{A},\tau,t} 
v_{\mathrm{A},\tau',\tau}^{\mathrm{ext}}, 
\end{equation}
where the damping factors $\{\theta_{\mathrm{A},\tau,t}\}$ satisfy the following 
normalization condition: 
\begin{equation} \label{damping_A}
\sum_{\tau=0}^{t}\theta_{\mathrm{A},\tau,t} = 1, 
\quad \theta_{\mathrm{A},t,t}\neq0 
\end{equation}
for all $t$, where the condition $\theta_{\mathrm{A},t,t}\neq0$ circumvents 
rank-deficient $\boldsymbol{V}_{\mathrm{A}\to \mathrm{B},t,t}$. Note that 
the positivity of the damping factors is not assumed since it is not required 
in state evolution.  

The main novelty for module~A in LM-OAMP is in the sufficient 
statistic~(\ref{statistic_mean_A}), the correction~(\ref{ext_mean_A}) for the 
posterior message to realize asymptotic Gaussianity, and the update rules 
(\ref{statistic_covariance_A}), (\ref{post_covariance_A}), 
(\ref{ext_covariance_A}), and (\ref{covariance_AB}) for the covariance 
messages. In particular, the sufficient statistic is a key technical tool 
to analyze the convergence of state evolution recursions for LM-OAMP 
rigorously. 

\begin{example}[LMMSE]
Consider the linear minimum mean-square error (LMMSE) filter
\begin{equation} \label{LMMSE}
\boldsymbol{W}_{t} 
= v_{\mathrm{B}\to \mathrm{A},t,t}^{\mathrm{suf}}\left(
 \sigma^{2}\boldsymbol{I}_{M} + v_{\mathrm{B}\to \mathrm{A},t,t}^{\mathrm{suf}}
 \boldsymbol{A}\boldsymbol{A}^{\mathrm{T}}
\right)^{-1}\boldsymbol{A}. 
\end{equation}
We know that the LMMSE filter minimizes the posterior variance 
$v_{\mathrm{A},t,t}^{\mathrm{post}}$ in (\ref{post_covariance_A}). Let $\eta(x)$ 
denote the $\eta$-transform~\cite{Tulino04} of the empirical eigenvalue 
distribution of $\boldsymbol{A}^{\mathrm{T}}\boldsymbol{A}$, given by 
\begin{equation} \label{eta_transform}
\eta(x) 
= \frac{1}{N}\mathrm{Tr}\left\{
 \left(
  \boldsymbol{I}_{N} + x\boldsymbol{A}^{\mathrm{T}}\boldsymbol{A} 
 \right)^{-1}
\right\}. 
\end{equation}
It is straightforward to confirm that the variance 
$v_{\mathrm{A},t,t}^{\mathrm{post}}$ in (\ref{post_covariance_A}) reduces to 
\begin{equation}
v_{\mathrm{A},t,t}^{\mathrm{post}} 
= \xi_{\mathrm{A},t}v_{\mathrm{B}\to \mathrm{A},t,t}^{\mathrm{suf}}, 
\end{equation}
where $\xi_{\mathrm{A},t}$ in (\ref{gamma_t}) is given by 
\begin{equation}
\xi_{\mathrm{A},t} 
= \eta\left(
 \frac{v_{\mathrm{B}\to \mathrm{A},t,t}^{\mathrm{suf}}}{\sigma^{2}} 
\right). 
\end{equation}
For $t'\neq t$, on the other hand, $v_{\mathrm{A},t',t}^{\mathrm{post}}$ reduces to 
\begin{IEEEeqnarray}{rl}
v_{\mathrm{A},t',t}^{\mathrm{post}} 
=& \gamma_{t',t}v_{\mathrm{B}\to \mathrm{A},t',t}^{\mathrm{suf}} 
+ \frac{v_{\mathrm{B}\to \mathrm{A},t',t'}^{\mathrm{suf}}
v_{\mathrm{B}\to \mathrm{A},t,t}^{\mathrm{suf}}}{v_{\mathrm{B}\to \mathrm{A},t',t'}^{\mathrm{suf}} 
- v_{\mathrm{B}\to \mathrm{A},t,t}^{\mathrm{suf}}} 
\nonumber \\
&\cdot\left\{
 \eta\left(
  \frac{v_{\mathrm{B}\to \mathrm{A},t,t}^{\mathrm{suf}}}{\sigma^{2}} 
 \right)
 - \eta\left(
  \frac{v_{\mathrm{B}\to \mathrm{A},t',t'}^{\mathrm{suf}}}{\sigma^{2}} 
 \right)
\right\},
\end{IEEEeqnarray}
where $\gamma_{t',t}$ in (\ref{gamma_t't}) is given by 
\begin{IEEEeqnarray}{rl}
\gamma_{t',t} 
=& \frac{v_{\mathrm{B}\to \mathrm{A},t',t'}^{\mathrm{suf}}}
{v_{\mathrm{B}\to \mathrm{A},t',t'}^{\mathrm{suf}} 
- v_{\mathrm{B}\to \mathrm{A},t,t}^{\mathrm{suf}}}
\eta\left(
 \frac{v_{\mathrm{B}\to \mathrm{A},t',t'}^{\mathrm{suf}}}{\sigma^{2}} 
\right) \nonumber \\
&- \frac{v_{\mathrm{B}\to \mathrm{A},t,t}^{\mathrm{suf}}}
{v_{\mathrm{B}\to \mathrm{A},t',t'}^{\mathrm{suf}} 
- v_{\mathrm{B}\to \mathrm{A},t,t}^{\mathrm{suf}}}
\eta\left(
 \frac{v_{\mathrm{B}\to \mathrm{A},t,t}^{\mathrm{suf}}}{\sigma^{2}} 
\right). 
\end{IEEEeqnarray}
\end{example}

\subsection{Module B (Nonlinear Estimation)}
Module~B is formulated in the same manner as in module~A. The only exception 
is element-wise nonlinear denoising while module~A has used the linear filter 
to mitigate the interference. Thus, the same explanation is omitted 
as in module~A. 

In the first step, module~B uses the preceding message matrix 
$\boldsymbol{X}_{\mathrm{A}\to \mathrm{B},t}$ and 
covariance $\{v_{\mathrm{A}\to \mathrm{B},\tau',\tau}: \tau', \tau\in\{0,\ldots,t\}\}$ 
in iteration~$t$ to compute a 
sufficient statistic $\boldsymbol{x}_{\mathrm{A}\to \mathrm{B},t}^{\mathrm{suf}}$ and 
the corresponding covariance 
$\{v_{\mathrm{A}\to \mathrm{B},t',t}^{\mathrm{suf}}\}_{t'=0}^{t}$, given by 
\begin{equation} \label{statistic_mean_B} 
\boldsymbol{x}_{\mathrm{A}\to \mathrm{B},t}^{\mathrm{suf}} 
= \frac{\boldsymbol{X}_{\mathrm{A}\to \mathrm{B},t}
\boldsymbol{V}_{\mathrm{A}\to \mathrm{B},t,t}^{-1}\boldsymbol{1}}
{\boldsymbol{1}^{\mathrm{T}}
\boldsymbol{V}_{\mathrm{A}\to \mathrm{B},t,t}^{-1}\boldsymbol{1}},
\end{equation}
\begin{equation}
v_{\mathrm{A}\to \mathrm{B},t',t}^{\mathrm{suf}}
= \frac{\boldsymbol{1}^{\mathrm{T}}\boldsymbol{V}_{\mathrm{A}\to \mathrm{B},t',t'}^{-1}
\boldsymbol{V}_{\mathrm{A}\to \mathrm{B},t',t}
\boldsymbol{V}_{\mathrm{A}\to \mathrm{B},t,t}^{-1}\boldsymbol{1}}
{\boldsymbol{1}^{\mathrm{T}}\boldsymbol{V}_{\mathrm{A}\to \mathrm{B},t',t'}^{-1}
\boldsymbol{1}
\boldsymbol{1}^{\mathrm{T}}\boldsymbol{V}_{\mathrm{A}\to \mathrm{B},t,t}^{-1}
\boldsymbol{1}}. 
\end{equation}

The second step is computation of the posterior mean 
$\boldsymbol{x}_{\mathrm{B},t+1}^{\mathrm{post}}\in\mathbb{R}^{N}$ 
and covariance\footnote{
If the true prior distribution of $\boldsymbol{x}$ is 
available, (\ref{post_covariance_B}) should be replaced with 
the posterior covariance $\langle 
C(\boldsymbol{x}_{\mathrm{A}\to \mathrm{B},t'}^{\mathrm{suf}},
\boldsymbol{x}_{\mathrm{A}\to \mathrm{B},t}^{\mathrm{suf}})
\rangle$ in (\ref{posterior_covariance}) 
while $v_{\mathrm{B},0,t+1}^{\mathrm{post}}$ should be set to the 
posterior variance $\langle C(\boldsymbol{x}_{\mathrm{A}\to \mathrm{B},t}^{\mathrm{suf}},
\boldsymbol{x}_{\mathrm{A}\to \mathrm{B},t}^{\mathrm{suf}})
\rangle$.   
} $\{v_{\mathrm{B},t'+1,t+1}^{\mathrm{post}}\}_{t'=0}^{t}$, which are defined 
with a scalar denoiser $f_{t}:\mathbb{R}\to\mathbb{R}$ as   
\begin{equation} \label{post_mean_B}
\boldsymbol{x}_{\mathrm{B},t+1}^{\mathrm{post}} 
= f_{t}(\boldsymbol{x}_{\mathrm{A}\to \mathrm{B},t}^{\mathrm{suf}}), 
\end{equation}
\begin{IEEEeqnarray}{r} 
v_{\mathrm{B},t'+1,t+1}^{\mathrm{post}}
= 1 + \frac{1}{N}f_{t'}^{\mathrm{T}}
(\boldsymbol{x}_{\mathrm{A}\to \mathrm{B},t'}^{\mathrm{suf}})
f_{t}(\boldsymbol{x}_{\mathrm{A}\to \mathrm{B},t}^{\mathrm{suf}}) 
\nonumber \\
+ v_{\mathrm{A}\to \mathrm{B},t',t}^{\mathrm{suf}}\xi_{\mathrm{B},t} 
- \frac{1}{N}f_{t}^{\mathrm{T}}(\boldsymbol{x}_{\mathrm{A}\to \mathrm{B},t}^{\mathrm{suf}}) 
\boldsymbol{x}_{\mathrm{A}\to \mathrm{B},t'}^{\mathrm{suf}}
\nonumber \\
+ v_{\mathrm{A}\to \mathrm{B},t',t}^{\mathrm{suf}}\xi_{\mathrm{B},t'} 
- \frac{1}{N}f_{t'}^{\mathrm{T}}
(\boldsymbol{x}_{\mathrm{A}\to \mathrm{B},t'}^{\mathrm{suf}}) 
\boldsymbol{x}_{\mathrm{A}\to \mathrm{B},t}^{\mathrm{suf}}, 
\label{post_covariance_B}
\end{IEEEeqnarray}
with 
\begin{equation} \label{xi_B_t}
\xi_{\mathrm{B},t} = \langle f_{t}'
(\boldsymbol{x}_{\mathrm{A}\to \mathrm{B},t}^{\mathrm{suf}}) \rangle.  
\end{equation}
The posterior mean message $\boldsymbol{x}_{\mathrm{B},t+1}^{\mathrm{post}}$ is 
used as an estimator of the signal vector $\boldsymbol{x}$. Since the 
sufficient statistic $\boldsymbol{x}_{\mathrm{A}\to \mathrm{B},t}^{\mathrm{suf}}$ 
depends on the preceding messages $\boldsymbol{X}_{\mathrm{A}\to \mathrm{B},t}$, 
the posterior message $\boldsymbol{x}_{\mathrm{B},t+1}^{\mathrm{post}}$ is a 
function of the preceding message matrix 
$\boldsymbol{X}_{\mathrm{B}\to \mathrm{A},t}$. In this sense, module~B utilizes 
LM denoising.  

The posterior covariance~(\ref{post_covariance_B}) is a consistent estimator 
of the covariance $N^{-1}\mathbb{E}[\{\boldsymbol{x}
- f_{t'}(\boldsymbol{x}_{\mathrm{A}\to \mathrm{B},t'}^{\mathrm{suf}})\}^{\mathrm{T}}
\{\boldsymbol{x} - f_{t}(\boldsymbol{x}_{\mathrm{A}\to \mathrm{B},t}^{\mathrm{suf}})\}]$ 
for the estimation errors. See Section~\ref{sec2_D} for the details. 

The third step is the Onsager correction to realize asymptotic Gaussianity. 
Module~B computes the extrinsic mean 
$\boldsymbol{x}_{\mathrm{B},t+1}^{\mathrm{ext}}\in\mathbb{R}^{N}$ and 
covariance $\{v_{\mathrm{B},t'+1,t+1}^{\mathrm{ext}}\}_{t'=0}^{t}$, 
\begin{equation} \label{ext_mean_B} 
\boldsymbol{x}_{\mathrm{B},t+1}^{\mathrm{ext}}
= \frac{\boldsymbol{x}_{\mathrm{B},t+1}^{\mathrm{post}} 
- \sum_{t'\in\mathcal{T}_{\mathrm{B},t}}\xi_{\mathrm{B},t',t}
\boldsymbol{x}_{\mathrm{A}\to \mathrm{B},t'}}{1 - \xi_{\mathrm{B},t}}, 
\end{equation}
\begin{equation} \label{ext_covariance_B}
v_{\mathrm{B},t'+1,t+1}^{\mathrm{ext}} 
= \frac{ v_{\mathrm{B},t'+1,t+1}^{\mathrm{post}}
- \xi_{\mathrm{B},t'}\xi_{\mathrm{B},t}v_{\mathrm{A}\to \mathrm{B},t',t}^{\mathrm{suf}}}
{(1-\xi_{\mathrm{B},t'})(1-\xi_{\mathrm{B},t})},
\end{equation}
with
\begin{equation} \label{xi_B} 
\xi_{\mathrm{B},t',t} 
= \xi_{\mathrm{B},t}
\frac{\boldsymbol{e}_{I_{\mathcal{T}_{\mathrm{B},t}}(t')}^{\mathrm{T}}
\boldsymbol{V}_{\mathrm{A}\to \mathrm{B},t,t}^{-1}\boldsymbol{1}}
{\boldsymbol{1}^{\mathrm{T}}\boldsymbol{V}_{\mathrm{A}\to \mathrm{B},t,t}^{-1}
\boldsymbol{1}} 
\end{equation}
for $t'\in\mathcal{T}_{\mathrm{B},t}$, where $\xi_{\mathrm{B},t}$ is given by 
(\ref{xi_B_t}).  

The last step is LM damping of the extrinsic messages. 
Module~B feeds the obtained messages  
$\boldsymbol{x}_{\mathrm{B}\to \mathrm{A},t+1}\in\mathbb{R}^{N}$ and 
$\{v_{\mathrm{B}\to \mathrm{A},t',t+1}\}_{t'=0}^{t+1}$ back to module~A. 
For $t', t\geq0$, 
\begin{equation} \label{mean_BA} 
\boldsymbol{x}_{\mathrm{B}\to \mathrm{A},t+1} 
= \sum_{\tau=0}^{t}\theta_{\mathrm{B},\tau,t}
\boldsymbol{x}_{\mathrm{B},\tau+1}^{\mathrm{ext}}, 
\end{equation}
\begin{equation} \label{covariance_BA}
v_{\mathrm{B}\to \mathrm{A},t'+1,t+1} 
= \sum_{\tau'=0}^{t'}\sum_{\tau=0}^{t}\theta_{\mathrm{B},\tau',t'}\theta_{\mathrm{B},\tau,t}
v_{\mathrm{B},\tau'+1,\tau+1}^{\mathrm{ext}}, 
\end{equation}
with
\begin{equation} \label{damping_B}
\sum_{\tau=0}^{t}\theta_{\mathrm{B},\tau,t} = 1, \quad \theta_{\mathrm{B},t,t}\neq0.  
\end{equation}
Otherwise, for $t\geq0$ we use 
\begin{equation}
v_{\mathrm{B}\to \mathrm{A},0,t+1} 
=\sum_{\tau=0}^{t}\frac{
\theta_{\mathrm{B},\tau,t}v_{\mathrm{B},0,\tau+1}^{\mathrm{post}}}{1 - \xi_{\mathrm{B},\tau}},
\end{equation}
with 
\begin{equation} 
v_{\mathrm{B},0,t+1}^{\mathrm{post}}
= 1 + v_{\mathrm{A}\to \mathrm{B},t,t}^{\mathrm{suf}}\xi_{\mathrm{B},t} 
- \frac{1}{N}f_{t}^{\mathrm{T}}(\boldsymbol{x}_{\mathrm{A}\to \mathrm{B},t}^{\mathrm{suf}}) 
\boldsymbol{x}_{\mathrm{A}\to \mathrm{B},t}^{\mathrm{suf}}.  
\end{equation}

\section{Theoretical Analysis} \label{sec4}
\subsection{State Evolution} 
We analyze the dynamics of LM-OAMP in the large system limit via 
state evolution~\cite{Takeuchi211}. A first step is a proof of 
asymptotic Gaussianity for the estimation errors in LM-OAMP. Asymptotic 
Gaussianity is proved by confirming that the error model for LM-OAMP is 
included into a general error model proposed in \cite{Takeuchi211}, which 
guarantees asymptotic Gaussianity. A second step is a derivation of state 
evolution recursions via asymptotic Gaussianity. 

We first postulate technical assumptions to justify 
asymptotic Gaussianity. 

\begin{assumption} \label{assumption_x}
The signal vector $\boldsymbol{x}$ has i.i.d.\ elements with zero mean, 
unit variance, and bounded $(2+\epsilon)$th moment for some $\epsilon>0$. 
\end{assumption}

Assumption~\ref{assumption_x} is an assumption to simplify state evolution 
analysis. We would need non-separable denoising if we postulated 
dependent signal elements. See \cite{Berthier19,Ma19,Fletcher19} 
for non-separable denoising. 

\begin{assumption} \label{assumption_A}
The sensing matrix $\boldsymbol{A}$ is right-orthogonally invariant: 
For any orthogonal matrix $\boldsymbol{\Phi}$ independent of $\boldsymbol{A}$, 
the equivalence in distribution 
$\boldsymbol{A}\boldsymbol{\Phi}\sim\boldsymbol{A}$ holds. 
More precisely, for the SVD 
$\boldsymbol{A}=\boldsymbol{U}\boldsymbol{\Sigma}
\boldsymbol{V}^{\mathrm{T}}$ the orthogonal matrix $\boldsymbol{V}$ is 
independent of $\boldsymbol{U}\boldsymbol{\Sigma}$ and 
Haar-distributed~\cite{Tulino04,Hiai00}. Furthermore, 
the empirical eigenvalue distribution of $\boldsymbol{A}^{\mathrm{T}}
\boldsymbol{A}$ converges almost surely to a compactly supported deterministic 
distribution with unit first moment in the large system limit. 
\end{assumption}

The unit-first-moment assumption is equivalent to the almost sure convergence 
$N^{-1}\mathrm{Tr}(\boldsymbol{A}^{\mathrm{T}}\boldsymbol{A})\ato1$ 
in the large system limit. The right-orthogonal invariance is an important 
assumption to justify asymptotic Gaussianity, for which a rough intuition 
is as follows: Let $\boldsymbol{m}\in\mathbb{R}^{N}$ denote a vector 
independent of $\boldsymbol{V}$. Since $\boldsymbol{V}$ is Haar-distributed, 
the orthogonal transform $\boldsymbol{V}\boldsymbol{m}$ is distributed 
as $(\|\boldsymbol{m}\|/\|\boldsymbol{z}\|)\boldsymbol{z}$ for a standard
Gaussian vector $\boldsymbol{z}$. If the amplitude 
$\|\boldsymbol{m}\|/\|\boldsymbol{z}\|$ converges in probability to a 
constant $a>0$ as $N\to\infty$, $\boldsymbol{V}\boldsymbol{m}$ is distributed 
as if it followed $\mathcal{N}(\boldsymbol{0},a^{2}\boldsymbol{I}_{N})$.  
  
It is a challenging issue in random matrix theory to weaken the 
right-orthogonal invariance to more practical assumptions, including 
discrete cosine transform (DCT) or Hadamard matrices with random 
permutation~\cite{Candes062}. See \cite{Anderson14,Male20,Dudeja22} 
for theoretical progress in this direction. 

\begin{assumption} \label{assumption_w}
The noise vector $\boldsymbol{w}$ is orthogonally invariant: 
For any orthogonal matrix $\boldsymbol{\Phi}$ independent of $\boldsymbol{w}$, 
the equivalence in distribution 
$\boldsymbol{\Phi}\boldsymbol{w}\sim\boldsymbol{w}$ holds. Furthermore, 
the almost sure convergence $M^{-1}\|\boldsymbol{w}\|^{2}\ato\sigma^{2}$ holds 
for the variance $\sigma^{2}>0$. The vector $\boldsymbol{w}$ has bounded 
$(2+\epsilon)$th moments for some $\epsilon>0$.  
\end{assumption}

The AWGN vector $\boldsymbol{w}\sim\mathcal{N}(\boldsymbol{0},
\sigma^{2}\boldsymbol{I}_{M})$ with variance $\sigma^{2}$ satisfies 
Assumption~\ref{assumption_w}. When the sensing matrix is left-orthogonally 
invariant, the orthogonal invariance in $\boldsymbol{w}$ can be induced 
from $\boldsymbol{A}$. 

\begin{assumption} \label{assumption_filter}
The linear filter matrix $\boldsymbol{W}_{t}$ in module~A has the same SVD 
structure as the sensing matrix $\boldsymbol{A}$, i.e.\ 
$\boldsymbol{W}_{t}=\boldsymbol{U}\tilde{\boldsymbol{W}}_{t}
\boldsymbol{V}^{\mathrm{T}}$ for $\boldsymbol{A}=\boldsymbol{U}\boldsymbol{\Sigma}
\boldsymbol{V}^{\mathrm{T}}$. Furthermore, the diagonal matrix 
$\tilde{\boldsymbol{W}}^{\mathrm{T}}\tilde{\boldsymbol{W}}$ is in the 
space spanned by $\{\boldsymbol{\Lambda}^{j}\}_{j=0}^{\infty}$ with 
$\boldsymbol{\Lambda}=\boldsymbol{\Sigma}^{\mathrm{T}}\boldsymbol{\Sigma}$.  
\end{assumption}

Assumption~\ref{assumption_filter} contains practically important linear 
filters, such as the MF $\boldsymbol{W}_{t}=\boldsymbol{A}$ 
and the LMMSE filter~(\ref{LMMSE}). 

\begin{assumption} \label{assumption_Lipschitz}
The scalar denoiser $f_{t}$ in module~B is Lipschitz-continuous and nonlinear. 
\end{assumption}

The Lipschitz-continuity in Assumption~\ref{assumption_Lipschitz} is a 
standard assumption in state evolution 
analysis~\cite{Bayati11,Rangan192,Takeuchi201,Takeuchi211} while the 
nonlinearity is required for the positive definiteness of 
$\boldsymbol{V}_{\mathrm{B}\to\mathrm{A},t,t}$. 
In module~A, Assumption~\ref{assumption_filter} and the compact support 
in Assumption~\ref{assumption_A} are substituted for 
Assumption~\ref{assumption_Lipschitz}. 

We are ready to present asymptotic Gaussianity for the estimation 
errors in LM-OAMP.

\begin{theorem}[Asymptotic Gaussianity] \label{THEOREM_GAUSSIANITY}
Suppose that Assumptions~\ref{assumption_x}--\ref{assumption_Lipschitz} 
are satisfied and consider any iteration $t$. Then, the following results 
hold for all iterations $\tau', \tau\in\{0,\ldots,t\}$. 
\begin{itemize}
\item The covariance $N^{-1}
(\boldsymbol{x}_{\mathrm{B}\to \mathrm{A},\tau'}-\boldsymbol{x})^{\mathrm{T}}
(\boldsymbol{x}_{\mathrm{B}\to \mathrm{A},\tau}-\boldsymbol{x})$ converges almost 
surely to some constant $\bar{v}_{\mathrm{B}\to \mathrm{A},\tau',\tau}$ in the large 
system limit. Furthermore, the $(t+1)\times(t+1)$ upper-left block of the  
matrix $\bar{\boldsymbol{V}}_{\mathrm{B}\to \mathrm{A}}$ with 
$[\bar{\boldsymbol{V}}_{\mathrm{B}\to \mathrm{A}}]_{\tau',\tau}
=\bar{v}_{\mathrm{B}\to \mathrm{A},\tau',\tau}$ 
is positive definite as long as $t$ is finite.

\item Let $\bar{\boldsymbol{V}}_{\mathrm{B}\to \mathrm{A},t',t}
\in\mathbb{R}^{|\mathcal{T}_{\mathrm{A},t'}|
\times|\mathcal{T}_{\mathrm{A},t}|}$ denote the matrix obtained by extracting 
the elements $\bar{v}_{\mathrm{B}\to \mathrm{A},\tau',\tau}$ for all 
$\tau'\in\mathcal{T}_{\mathrm{A},t'}$ 
and $\tau\in\mathcal{T}_{\mathrm{A},t}$ from the covariance matrix 
$\bar{\boldsymbol{V}}_{\mathrm{B}\to \mathrm{A}}$ defined in the first property.  
Then, the covariance $N^{-1}(\boldsymbol{x}_{\mathrm{A},\tau'}^{\mathrm{post}}
-\boldsymbol{x})^{\mathrm{T}}
(\boldsymbol{x}_{\mathrm{A},\tau}^{\mathrm{post}}-\boldsymbol{x})$ 
converges almost surely to some constant 
$\bar{v}_{\mathrm{A},\tau',\tau}^{\mathrm{post}}$ in the large system limit, given by 
\begin{IEEEeqnarray}{rl} 
\bar{v}_{\mathrm{A},\tau',\tau}^{\mathrm{post}} 
= \lim_{M=\delta N\to\infty}&\left\{
 \gamma_{\tau',\tau}\bar{v}_{\mathrm{B}\to \mathrm{A},\tau',\tau}^{\mathrm{suf}}
\right. \nonumber \\
&\left.
 + \frac{\sigma^{2}}{N}\mathrm{Tr}\left(
  \boldsymbol{W}_{\tau'}\boldsymbol{W}_{\tau}^{\mathrm{T}}
 \right)
\right\},
\label{SE_post_A} 
\end{IEEEeqnarray}
where $\gamma_{\tau',\tau}$ is given by (\ref{gamma_t't}), with 
\begin{equation} \label{SE_statistic_A_tmp} 
\bar{v}_{\mathrm{B}\to \mathrm{A},\tau',\tau}^{\mathrm{suf}} 
= \frac{\boldsymbol{1}^{\mathrm{T}}
\boldsymbol{V}_{\mathrm{B}\to \mathrm{A},\tau',\tau'}^{-1}
\bar{\boldsymbol{V}}_{\mathrm{B}\to \mathrm{A},\tau',\tau}
\boldsymbol{V}_{\mathrm{B}\to \mathrm{A},\tau,\tau}^{-1}
\boldsymbol{1}}
{\boldsymbol{1}^{\mathrm{T}}\boldsymbol{V}_{\mathrm{B}\to \mathrm{A},\tau',\tau'}^{-1}
\boldsymbol{1}\boldsymbol{1}^{\mathrm{T}}
\boldsymbol{V}_{\mathrm{B}\to \mathrm{A},\tau,\tau}^{-1}\boldsymbol{1}}.  
\end{equation}

\item The covariance $N^{-1}
(\boldsymbol{x}_{\mathrm{A}\to \mathrm{B},\tau'}-\boldsymbol{x})^{\mathrm{T}}
(\boldsymbol{x}_{\mathrm{A}\to \mathrm{B},\tau}-\boldsymbol{x})$ converges almost 
surely to some constant $\bar{v}_{\mathrm{A}\to \mathrm{B}, \tau',\tau}$ in the large 
system limit. Furthermore, the $(t+1)\times(t+1)$ upper-left block of the 
matrix $\bar{\boldsymbol{V}}_{\mathrm{A}\to \mathrm{B}}$ 
with $[\bar{\boldsymbol{V}}_{\mathrm{A}\to \mathrm{B}}]_{\tau',\tau}
=\bar{v}_{\mathrm{A}\to \mathrm{B},\tau',\tau}$ 
is positive definite as long as $t$ is finite.
 
\item Let $\bar{\boldsymbol{V}}_{\mathrm{A}\to \mathrm{B},t',t}
\in\mathbb{R}^{|\mathcal{T}_{\mathrm{B},t'}|
\times|\mathcal{T}_{\mathrm{B},t}|}$ denote the matrix defined by extracting 
the elements $\bar{v}_{\mathrm{A}\to \mathrm{B},\tau',\tau}$ for all 
$\tau'\in\mathcal{T}_{\mathrm{B},t'}$ 
and $\tau\in\mathcal{T}_{\mathrm{B},t}$ from the covariance matrix 
$\bar{\boldsymbol{V}}_{\mathrm{A}\to \mathrm{B}}$ defined in the third property. 
Then, the covariance $N^{-1}(\boldsymbol{x}_{\mathrm{B},\tau'+1}^{\mathrm{post}}
-\boldsymbol{x})^{\mathrm{T}}
(\boldsymbol{x}_{\mathrm{B},\tau+1}^{\mathrm{post}}-\boldsymbol{x})$ 
converges almost surely to some constant 
$\bar{v}_{\mathrm{B},\tau'+1,\tau+1}^{\mathrm{post}}$ in the large system limit,
\begin{IEEEeqnarray}{rl}
\bar{v}_{\mathrm{B},\tau'+1,\tau+1}^{\mathrm{post}} 
= \mathbb{E}[&\{f_{\tau'}(x_{1} + z_{\tau'}) - x_{1}\}
\nonumber \\
&\cdot\{f_{\tau}(x_{1} + z_{\tau}) - x_{1}\}], 
\label{SE_post_B}
\end{IEEEeqnarray}
where $\{z_{\tau'}, z_{\tau}\}$ are independent of the signal element $x_{1}$ 
and zero-mean Gaussian random variables with covariance 
$\mathbb{E}[z_{\tau'}z_{\tau}]
=\bar{v}_{\mathrm{A}\to \mathrm{B},\tau',\tau}^{\mathrm{suf}}$, given by  
\begin{equation} \label{SE_statistic_B_tmp}
\bar{v}_{\mathrm{A}\to \mathrm{B},\tau',\tau}^{\mathrm{suf}}
= \frac{\boldsymbol{1}^{\mathrm{T}}
\boldsymbol{V}_{\mathrm{A}\to \mathrm{B},\tau',\tau'}^{-1}
\bar{\boldsymbol{V}}_{\mathrm{A}\to \mathrm{B},\tau',\tau}
\boldsymbol{V}_{\mathrm{A}\to \mathrm{B},\tau,\tau}^{-1}\boldsymbol{1}}
{\boldsymbol{1}^{\mathrm{T}}\boldsymbol{V}_{\mathrm{A}\to \mathrm{B},\tau',\tau'}^{-1}
\boldsymbol{1}\boldsymbol{1}^{\mathrm{T}}
\boldsymbol{V}_{\mathrm{A}\to \mathrm{B},\tau,\tau}^{-1}\boldsymbol{1}}. 
\end{equation}
Furthermore, $\xi_{\mathrm{B},\tau}$ in (\ref{xi_B_t}) converges almost surely to 
$\bar{\xi}_{\mathrm{B},\tau}$ in the large system limit, 
\begin{equation} \label{xi_asym}
\bar{\xi}_{\mathrm{B},\tau}=\mathbb{E}[f_{\tau}'(x_{1}+z_{\tau})].
\end{equation} 
\end{itemize} 
\end{theorem}
\begin{IEEEproof}
See Appendix~\ref{proof_theorem_Gaussianity}. 
\end{IEEEproof}

The second and fourth results in Theorem~\ref{THEOREM_GAUSSIANITY} are the 
precise meaning of asymptotic Gaussianity for LM-OAMP while the asymptotic 
Gaussianity for the second result is not explicit, because of the linear 
filter in module~A. 
We cannot claim the joint Gaussianity for the estimation errors: 
the convergence of the joint distribution of 
$\{\boldsymbol{x}_{\mathrm{A}\to \mathrm{B},\tau}^{\mathrm{suf}}
-\boldsymbol{x}\}_{\tau=0}^{t}$ toward a joint Gaussian distribution with 
covariance $\mathbb{E}[(\boldsymbol{x}_{\mathrm{A}\to \mathrm{B},\tau}^{\mathrm{suf}}
-\boldsymbol{x})(\boldsymbol{x}_{\mathrm{A}\to \mathrm{B},\tau'}^{\mathrm{suf}}
-\boldsymbol{x})^{\mathrm{T}}]=\bar{v}_{\mathrm{A}\to\mathrm{B},\tau',\tau}^{\mathrm{suf}}
\boldsymbol{I}$. 
We only claim the almost sure convergence of 
$N^{-1}(\boldsymbol{x}_{\mathrm{B},\tau'+1}^{\mathrm{post}}
-\boldsymbol{x})^{\mathrm{T}}(\boldsymbol{x}_{\mathrm{B},\tau+1}^{\mathrm{post}}
-\boldsymbol{x})$ toward (\ref{SE_post_B}) given via the zero-mean Gaussian 
random variables $z_{\tau'}$ and $z_{\tau}$ with covariance 
$\mathbb{E}[z_{\tau'}z_{\tau}]=\bar{v}_{\mathrm{A}\to\mathrm{B},\tau',\tau}^{\mathrm{suf}}$. 

We next derive state evolution recursions by evaluating the covariance 
$\bar{v}_{\mathrm{A}\to \mathrm{B},\tau',\tau}$ and 
$\bar{v}_{\mathrm{B}\to \mathrm{A},\tau',\tau}$ in 
Theorem~\ref{THEOREM_GAUSSIANITY}. 

\begin{theorem}[State Evolution] \label{THEOREM_SE}
Suppose that Assumptions~\ref{assumption_x}--\ref{assumption_Lipschitz} 
are satisfied. 
Then, the state evolution recursions for module~A are given by  
\begin{equation} \label{SE_statistic_A}
\bar{v}_{\mathrm{B}\to \mathrm{A},t',t}^{\mathrm{suf}} 
= \frac{\boldsymbol{1}^{\mathrm{T}}
\bar{\boldsymbol{V}}_{\mathrm{B}\to \mathrm{A},t',t'}^{-1}
\bar{\boldsymbol{V}}_{\mathrm{B}\to \mathrm{A},t',t}
\bar{\boldsymbol{V}}_{\mathrm{B}\to \mathrm{A},t,t}^{-1}
\boldsymbol{1}}{\boldsymbol{1}^{\mathrm{T}}
\bar{\boldsymbol{V}}_{\mathrm{B}\to \mathrm{A},t',t'}^{-1}\boldsymbol{1}
\boldsymbol{1}^{\mathrm{T}}
\bar{\boldsymbol{V}}_{\mathrm{B}\to \mathrm{A},t,t}^{-1}\boldsymbol{1}},  
\end{equation}
\begin{equation} \label{gamma_t_asym}
\bar{\xi}_{\mathrm{A},t} 
= \lim_{M=\delta N\to\infty}\frac{1}{N}\mathrm{Tr}\left(
 \boldsymbol{I}_{N} - \boldsymbol{W}_{t}^{\mathrm{T}}\boldsymbol{A}
\right), 
\end{equation}
\begin{equation} \label{SE_A_ext}
\bar{v}_{\mathrm{A},t',t}^{\mathrm{ext}}
= \frac{\bar{v}_{\mathrm{A},t',t}^{\mathrm{post}} 
- \bar{\xi}_{\mathrm{A},t'}\bar{\xi}_{\mathrm{A},t}
\bar{v}_{\mathrm{B}\to \mathrm{A},t',t}^{\mathrm{suf}}}
{(1-\bar{\xi}_{\mathrm{A},t'})(1-\bar{\xi}_{\mathrm{A},t})},
\end{equation}
\begin{equation} 
\bar{v}_{\mathrm{A}\to \mathrm{B},t',t} 
= \sum_{\tau'=0}^{t'}\sum_{\tau=0}^{t}\theta_{\mathrm{A},\tau',t'}\theta_{\mathrm{A},\tau,t} 
\bar{v}_{\mathrm{A},\tau',\tau}^{\mathrm{ext}}, 
\label{SE_AB}
\end{equation}
with $\bar{v}_{\mathrm{B}\to \mathrm{A},0,0}=1$, 
where $\bar{v}_{\mathrm{A},t',t}^{\mathrm{post}}$ is defined in (\ref{SE_post_A}). 
In these expressions, $v_{\mathrm{B}\to\mathrm{A},t,t}^{\mathrm{suf}}$ in 
$\boldsymbol{W}_{t}$ is replaced with 
$\bar{v}_{\mathrm{B}\to\mathrm{A},t,t}^{\mathrm{suf}}$. 
On the other hand, the state evolution recursions for module~B are given by 
\begin{equation} \label{SE_statistic_B}
\bar{v}_{\mathrm{A}\to \mathrm{B},t',t}^{\mathrm{suf}}
= \frac{\boldsymbol{1}^{\mathrm{T}}
\bar{\boldsymbol{V}}_{\mathrm{A}\to \mathrm{B},t',t'}^{-1}
\bar{\boldsymbol{V}}_{\mathrm{A}\to \mathrm{B},t',t}
\bar{\boldsymbol{V}}_{\mathrm{A}\to \mathrm{B},t,t}^{-1}\boldsymbol{1}}
{\boldsymbol{1}^{\mathrm{T}}\bar{\boldsymbol{V}}_{\mathrm{A}\to \mathrm{B},t',t'}^{-1}
\boldsymbol{1}\boldsymbol{1}^{\mathrm{T}}
\bar{\boldsymbol{V}}_{\mathrm{A}\to \mathrm{B},t,t}^{-1}\boldsymbol{1}}, 
\end{equation}
\begin{equation} \label{SE_B_ext}
\bar{v}_{\mathrm{B},t'+1,t+1}^{\mathrm{ext}} 
= \frac{\bar{v}_{\mathrm{B},t'+1,t+1}^{\mathrm{post}}
 - \bar{\xi}_{\mathrm{B},t'}\bar{\xi}_{\mathrm{B},t}
\bar{v}_{\mathrm{A}\to \mathrm{B},t',t}^{\mathrm{suf}}}
{(1-\bar{\xi}_{\mathrm{B},t'})(1-\bar{\xi}_{\mathrm{B},t})},  
\end{equation}
\begin{equation}
\bar{v}_{\mathrm{B}\to \mathrm{A},t'+1,t+1} 
= \sum_{\tau'=0}^{t'}\sum_{\tau=0}^{t}\theta_{\mathrm{B},\tau',t'}\theta_{\mathrm{B},\tau,t}
\bar{v}_{\mathrm{B},\tau'+1,\tau+1}^{\mathrm{ext}} 
\label{SE_BA}
\end{equation}
for $t', t\geq0$, where $\bar{v}_{\mathrm{B},t'+1,t+1}^{\mathrm{post}}$ and 
$\bar{\xi}_{\mathrm{B},t}$ are defined as 
(\ref{SE_post_B}) and (\ref{xi_asym}), respectively. 
For $t'=-1$, (\ref{SE_BA}) is replaced with 
\begin{equation} \label{SE_BA_0}
\bar{v}_{\mathrm{B}\to \mathrm{A},0,t+1}
=\sum_{\tau=0}^{t}\frac{
\theta_{\mathrm{B},\tau,t}\bar{v}_{\mathrm{B},0,\tau+1}^{\mathrm{post}}}
{1 - \bar{\xi}_{\mathrm{B},\tau}}
\end{equation}
for $t\geq0$, with 
\begin{equation}
\bar{v}_{\mathrm{B},0,\tau+1}^{\mathrm{post}} 
= \mathbb{E}[x_{1}\{x_{1} - f_{\tau}(x_{1} + z_{\tau})\}],  
\end{equation}
where $z_{\tau}$ is independent of $x_{1}$ 
and a zero-mean Gaussian random variable with variance 
$\mathbb{E}[z_{\tau}^{2}]=\bar{v}_{\mathrm{A}\to \mathrm{B},\tau,\tau}^{\mathrm{suf}}$. 
\end{theorem}
\begin{IEEEproof}
See Appendix~\ref{proof_theorem_SE}.
\end{IEEEproof}

Theorem~\ref{THEOREM_SE} presents the state evolution recursions for 
LM-OAMP, which justify the update rules for the covariance messages in 
LM-OAMP. In particular, the MSE  
$N^{-1}\|\boldsymbol{x}_{\mathrm{B},t+1}^{\mathrm{post}}-\boldsymbol{x}\|^{2}$ 
converges almost surely to $\bar{v}_{\mathrm{B},t+1,t+1}^{\mathrm{post}}$ 
in the large system limit. 

\begin{example} \label{example2} 
As an example of Theorem~\ref{THEOREM_SE}, we derive state evolution 
recursions for damped OAMP. 
Let $\mathcal{T}_{\mathrm{A},t}=\mathcal{T}_{\mathrm{B},t}=\{t\}$ and consider 
damping $\theta_{\mathrm{A},0,t}=(1-\theta_{\mathrm{A}})^{t}$, 
$\theta_{\mathrm{A},\tau,t}=\theta_{\mathrm{A}}(1-\theta_{\mathrm{A}})^{t-\tau}$ for all 
$\tau>0$ in (\ref{mean_AB}) with some $\theta_{\mathrm{A}}\in[0, 1]$, and 
$\theta_{\mathrm{B},0,t}=(1-\theta_{\mathrm{B}})^{t}$, 
$\theta_{\mathrm{B},\tau,t}=\theta_{\mathrm{B}}(1-\theta_{\mathrm{B}})^{t-\tau}$ 
for all $\tau>0$ in 
(\ref{mean_BA}) with some $\theta_{\mathrm{B}}\in[0, 1]$~\cite{Rangan192}. 
From (\ref{mean_AB}) we have $\boldsymbol{x}_{\mathrm{A}\to \mathrm{B},0}
=\boldsymbol{x}_{\mathrm{A},0}^{\mathrm{ext}}$ and  
\begin{IEEEeqnarray}{rl}
\boldsymbol{x}_{\mathrm{A}\to \mathrm{B},t}
=& \theta_{\mathrm{A}}\sum_{\tau=1}^{t}(1-\theta_{\mathrm{A}})^{t-\tau}
\boldsymbol{x}_{\mathrm{A},\tau}^{\mathrm{ext}}
+ (1-\theta_{\mathrm{A}})^{t}\boldsymbol{x}_{\mathrm{A},0}^{\mathrm{ext}} 
\nonumber \\
=& \theta_{\mathrm{A}}\boldsymbol{x}_{\mathrm{A},t}^{\mathrm{ext}}
+ (1-\theta_{\mathrm{A}})\boldsymbol{x}_{\mathrm{A}\to \mathrm{B},t-1}  
\end{IEEEeqnarray}
for $t>0$. 
Similarly, from (\ref{mean_BA}) we have $\boldsymbol{x}_{\mathrm{B}\to \mathrm{A},1}
=\boldsymbol{x}_{\mathrm{B}\to \mathrm{A},1}^{\mathrm{ext}}$ and for $t>0$  
\begin{equation}
\boldsymbol{x}_{\mathrm{B}\to \mathrm{A},t+1} 
= \theta_{\mathrm{B}}\boldsymbol{x}_{\mathrm{B},t+1}^{\mathrm{ext}}
+ (1-\theta_{\mathrm{B}})\boldsymbol{x}_{\mathrm{B}\to \mathrm{A},t}. 
\end{equation}

The state evolution recursion for module~A in damped OAMP is given by 
\begin{IEEEeqnarray}{rl} 
\bar{v}_{\mathrm{A},t',t}^{\mathrm{ext}} 
= \frac{\bar{v}_{\mathrm{A},t',t}^{\mathrm{post}} 
- \bar{\xi}_{\mathrm{A},t'}\bar{\xi}_{\mathrm{A},t}\bar{v}_{\mathrm{B}\to \mathrm{A},t',t}}
{(1-\bar{\xi}_{\mathrm{A},t'})(1-\bar{\xi}_{\mathrm{A},t})},
\end{IEEEeqnarray}
where $\bar{v}_{\mathrm{A},t',t}^{\mathrm{post}}$ is defined in (\ref{SE_post_A}) 
with $\bar{v}_{\mathrm{B}\to \mathrm{A},t',t}^{\mathrm{suf}}
=\bar{v}_{\mathrm{B}\to \mathrm{A},t',t}$. Furthermore, 
we have $\bar{v}_{\mathrm{A}\to \mathrm{B},0,0}=\bar{v}_{\mathrm{A},0,0}^{\mathrm{ext}}$ and 
\begin{equation}
\bar{v}_{\mathrm{A}\to \mathrm{B},0,t} 
= \theta_{\mathrm{A}}\bar{v}_{\mathrm{A},0,t}^{\mathrm{ext}} 
+ (1-\theta_{\mathrm{A}})\bar{c}_{\mathrm{A},0,t-1}, 
\end{equation}
\begin{equation} \label{covariance_AB_dOAMP} 
\bar{v}_{\mathrm{A}\to \mathrm{B},t',t} 
= \theta_{\mathrm{A}}\bar{c}_{\mathrm{A},t',t} 
+ (1-\theta_{\mathrm{A}})\bar{v}_{\mathrm{A}\to \mathrm{B},t'-1,t}, 
\end{equation}
for all $t', t>0$, with  
\begin{IEEEeqnarray}{l}
\bar{c}_{\mathrm{A},t',t} 
= \lim_{M=\delta N\to\infty}\frac{1}{N}
(\boldsymbol{x}_{\mathrm{A},t'}^{\mathrm{ext}} - \boldsymbol{x})^{\mathrm{T}}
(\boldsymbol{x}_{\mathrm{A}\to \mathrm{B},t} - \boldsymbol{x})
\nonumber \\
\aeq \theta_{\mathrm{A}} \bar{v}_{\mathrm{A},t',t}^{\mathrm{ext}} 
+ (1-\theta_{\mathrm{A}})\bar{c}_{\mathrm{A},t',t-1}, 
\; \bar{c}_{\mathrm{A},t',0} = \bar{v}_{\mathrm{A},t',0}^{\mathrm{ext}}. 
\end{IEEEeqnarray}

The state evolution recursions for module~B reduce to 
\begin{equation}
\bar{v}_{\mathrm{B},0,t+1}^{\mathrm{ext}} 
= \frac{\bar{v}_{\mathrm{B},0,t+1}^{\mathrm{post}}}{1 - \bar{\xi}_{\mathrm{B},t}}, 
\end{equation}
\begin{equation}
\bar{v}_{\mathrm{B},t'+1,t+1}^{\mathrm{ext}}
= \frac{\bar{v}_{\mathrm{B},t'+1,t+1}^{\mathrm{post}}
 - \bar{\xi}_{\mathrm{B},t'}\bar{\xi}_{\mathrm{B},t}
\bar{v}_{\mathrm{A}\to \mathrm{B},t',t}}
{(1-\bar{\xi}_{\mathrm{B},t'})(1-\bar{\xi}_{\mathrm{B},t})}
\end{equation}
for $t'\geq0$, where $\bar{v}_{\mathrm{B},t'+1,t+1}^{\mathrm{post}}$ is given by 
(\ref{SE_post_B}) with $\bar{v}_{\mathrm{A}\to \mathrm{B},\tau',\tau}^{\mathrm{suf}}
=\bar{v}_{\mathrm{A}\to \mathrm{B},\tau',\tau}$. Furthermore, we have 
$\bar{v}_{\mathrm{B}\to \mathrm{A},0,0}=1$, $\bar{v}_{\mathrm{B}\to \mathrm{A},0,1} 
= \bar{v}_{\mathrm{B},0,1}^{\mathrm{ext}}$, 
$\bar{v}_{\mathrm{B}\to \mathrm{A},1,1}=\bar{v}_{\mathrm{B},1,1}^{\mathrm{ext}}$, and   
\begin{equation} 
\bar{v}_{\mathrm{B}\to \mathrm{A},0,t+1}
= \theta_{\mathrm{B}}\bar{v}_{\mathrm{B},0,t+1}^{\mathrm{ext}} 
+ (1-\theta_{\mathrm{B}})\bar{v}_{\mathrm{B}\to \mathrm{A},0,t},  
\end{equation}
\begin{equation}
\bar{v}_{\mathrm{B}\to \mathrm{A},1,t+1}
= \theta_{\mathrm{B}}\bar{v}_{\mathrm{B},1,t+1}^{\mathrm{ext}} 
+ (1-\theta_{\mathrm{B}})\bar{c}_{\mathrm{B},1,t},
\end{equation}    
\begin{equation} \label{covariance_BA_dOAMP} 
\bar{v}_{\mathrm{B}\to \mathrm{A},t'+1,t+1} 
= \theta_{\mathrm{B}}\bar{c}_{\mathrm{B},t'+1,t+1} 
+ (1-\theta_{\mathrm{B}})\bar{v}_{\mathrm{B}\to \mathrm{A},t',t+1}
\end{equation}
for all $t', t>0$, with  
\begin{IEEEeqnarray}{l}
\bar{c}_{\mathrm{B},t',t+1} 
= \lim_{M=\delta N\to\infty}\frac{1}{N}
(\boldsymbol{x}_{\mathrm{B},t'}^{\mathrm{ext}} - \boldsymbol{x})^{\mathrm{T}}
(\boldsymbol{x}_{\mathrm{B}\to \mathrm{A},t+1} - \boldsymbol{x})
\nonumber \\
\aeq \theta_{\mathrm{B}} \bar{v}_{\mathrm{B},t',t+1}^{\mathrm{ext}} 
+ (1-\theta_{\mathrm{B}})\bar{c}_{\mathrm{B},t',t}, 
\quad \bar{c}_{\mathrm{B},t',1} = \bar{v}_{\mathrm{B},t',1}^{\mathrm{ext}}. 
\end{IEEEeqnarray}
\end{example}

These results imply that the correct state evolution 
recursions for damped OAMP are the two-dimensional discrete systems with 
respect to the covariance messages. Equivalent results were 
derived for the MF $\boldsymbol{W}_{t}=\boldsymbol{A}$ and zero-mean 
i.i.d.\ Gaussian sensing matrices~\cite{Mimura19}. 

The following heuristic damping for the variance messages  
was used:  
\begin{equation}
\frac{1}{v_{\mathrm{A}\to \mathrm{B},t,t}} 
= \theta_{\mathrm{A}}\left(
 \frac{1}{v_{\mathrm{A},t,t}^{\mathrm{post}}} - \frac{1}{v_{\mathrm{B}\to \mathrm{A},t,t}}
\right) 
+ \frac{1-\theta_{\mathrm{A}}}{v_{\mathrm{A}\to \mathrm{B},t-1,t-1}}, 
\end{equation}
\begin{equation}
\frac{1}{v_{\mathrm{B}\to \mathrm{A},t+1,t+1}} 
= \theta_{\mathrm{B}}\left(
 \frac{1}{v_{\mathrm{B},t+1,t+1}^{\mathrm{post}}} 
 - \frac{1}{v_{\mathrm{A}\to \mathrm{B},t,t}}
\right) 
+ \frac{1-\theta_{\mathrm{B}}}{v_{\mathrm{B}\to \mathrm{A},t,t}}
\end{equation}
in precision domain~\cite{Rangan192} or 
\begin{IEEEeqnarray}{rl}
v_{\mathrm{A}\to \mathrm{B},t,t}
=& \theta_{\mathrm{A}}\left(
 \frac{1}{v_{\mathrm{A},t,t}^{\mathrm{post}}} - \frac{1}{v_{\mathrm{B}\to \mathrm{A},t,t}}
\right)^{-1}  \nonumber \\
&+ (1-\theta_{\mathrm{A}})v_{\mathrm{A}\to \mathrm{B},t-1,t-1}, 
\label{covariance_AB_hdOAMP}
\end{IEEEeqnarray}
\begin{IEEEeqnarray}{rl}
v_{\mathrm{B}\to \mathrm{A},t+1,t+1}
=& \theta_{\mathrm{B}}\left(
 \frac{1}{v_{\mathrm{B},t+1,t+1}^{\mathrm{post}}} - \frac{1}{v_{\mathrm{B}\to \mathrm{A},t,t}}
\right)^{-1} \nonumber \\
&+ (1-\theta_{\mathrm{B}})v_{\mathrm{A}\to \mathrm{B},t,t}
\label{covariance_BA_hdOAMP}
\end{IEEEeqnarray}
in variance domain. 
The heuristic damping cannot provide accurate estimates for the MSE in 
each module while it results in a one-dimensional discrete system 
to describe the variance messages in OAMP. 

\section{Optimization} \label{sec5}
\subsection{Damping} 
LM-OAMP has several design parameters such as the subsets of indices 
$\mathcal{T}_{\mathrm{A},t}$, $\mathcal{T}_{\mathrm{B},t}$, the linear filter 
$\boldsymbol{W}_{t}$, damping factors $\{\theta_{\mathrm{A},\tau,t}\}$, 
$\{\theta_{\mathrm{B},\tau,t}\}$, and the denoiser $f_{t}$. Since this paper 
focuses on the Bayes-optimal LM-OAMP, the LMMSE filter~(\ref{LMMSE}) and 
the Bayes-optimal denoiser are used. Thus, the remaining design parameters 
are $\mathcal{T}_{\mathrm{A},t}$, $\mathcal{T}_{\mathrm{B},t}$, 
$\{\theta_{\mathrm{A},\tau,t}\}$, and $\{\theta_{\mathrm{B},\tau,t}\}$. 

We use Theorem~\ref{THEOREM_SE} to optimize the damping factors 
$\{\theta_{\mathrm{A},\tau,t}\}$ and $\{\theta_{\mathrm{B},\tau,t}\}$. A reasonable 
criterion is the minimization of the variance 
$\bar{v}_{\mathrm{B}\to \mathrm{A},t,t}^{\mathrm{suf}}$ and 
$\bar{v}_{\mathrm{A}\to \mathrm{B},t,t}^{\mathrm{suf}}$ for the sufficient statistics 
given in (\ref{SE_statistic_A}) and (\ref{SE_statistic_B}). 
This criterion results in greedy minimization of the MSEs 
$\bar{v}_{\mathrm{A},t,t}^{\mathrm{post}}$ and $\bar{v}_{\mathrm{B},t+1,t+1}^{\mathrm{post}}$ 
for the posterior estimators. 

When all preceding messages are used, i.e.\ $\mathcal{T}_{\mathrm{A},t}
=\mathcal{T}_{\mathrm{B},t}=\{0,\ldots,t\}$, we have 
\begin{equation} \label{SE_statistic_all}
\bar{v}_{\mathrm{B}\to \mathrm{A},t,t}^{\mathrm{suf}} 
= \frac{1}{\boldsymbol{1}^{\mathrm{T}}
\bar{\boldsymbol{V}}_{\mathrm{B}\to \mathrm{A},t,t}^{-1}
\boldsymbol{1}}, \quad
\bar{v}_{\mathrm{A}\to \mathrm{B},t,t}^{\mathrm{suf}} 
= \frac{1}{\boldsymbol{1}^{\mathrm{T}}
\bar{\boldsymbol{V}}_{\mathrm{A}\to \mathrm{B},t,t}^{-1}
\boldsymbol{1}}. 
\end{equation}
In this case, the damping factors can be optimized via the following lemma:  
\begin{lemma} \label{LEMMA_DAMPING}
For any $T$, let $\boldsymbol{C}_{T}\in\mathbb{R}^{(T+1)\times(T+1)}$ denote 
a positive-definite symmetric matrix with $[\boldsymbol{C}_{T}]_{\tau',\tau}
=c_{\tau',\tau}$. 
Suppose that $\boldsymbol{V}_{T}\in\mathbb{R}^{(T+1)\times (T+1)}$ has 
the $(t', t)$ element $v_{t',t}$ for $t', t\in\{0,\ldots,T\}$, 
given by 
\begin{equation} \label{v_t't}
v_{t',t} = \sum_{\tau'=0}^{t'}\sum_{\tau=0}^{t}c_{\tau',\tau}
\theta_{\tau',t'}\theta_{\tau,t}, 
\end{equation}
where $\{\theta_{\tau,t}\}_{\tau=0}^{t}$ satisfy the normalization condition, 
\begin{equation}  \label{constraint}
\sum_{\tau=0}^{t}\theta_{\tau,t} = 1. 
\end{equation}
Define the upper triangular matrix $\boldsymbol{\Theta}_{T}\in
\mathbb{R}^{(T+1)\times(T+1)}$ as 
\begin{equation}
\boldsymbol{\Theta}_{T} 
= \begin{bmatrix}
\theta_{0,0} & \cdots & \theta_{0,T} \\
& \ddots & \vdots \\
\boldsymbol{O} & & \theta_{T,T} 
\end{bmatrix}. 
\end{equation}
If $\boldsymbol{\Theta}_{T}$ has full rank, 
then the following identity holds:  
\begin{equation}
\boldsymbol{1}^{\mathrm{T}}\boldsymbol{V}_{T}^{-1}\boldsymbol{1}
= \boldsymbol{1}^{\mathrm{T}}\boldsymbol{C}_{T}^{-1}\boldsymbol{1}. 
\end{equation}
\end{lemma}
\begin{IEEEproof}
By definition, $\boldsymbol{V}_{T}$ is represented as 
$\boldsymbol{V}_{T} = \boldsymbol{\Theta}_{T}^{\mathrm{T}}
\boldsymbol{C}_{T}\boldsymbol{\Theta}_{T}$. Since both $\boldsymbol{V}_{T}$ and 
$\boldsymbol{\Theta}_{T}$ are invertible, we have 
$\boldsymbol{1}^{\mathrm{T}}\boldsymbol{V}_{T}^{-1}\boldsymbol{1}
=\boldsymbol{1}^{\mathrm{T}}\boldsymbol{\Theta}_{T}^{-1}
\boldsymbol{C}_{T}^{-1}(\boldsymbol{\Theta}_{T}^{-1})^{\mathrm{T}}\boldsymbol{1}$. 
The normalization~(\ref{constraint}) implies $\boldsymbol{1}^{\mathrm{T}}
\boldsymbol{\Theta}_{T}=\boldsymbol{1}^{\mathrm{T}}$, so that 
$\boldsymbol{1}^{\mathrm{T}}=\boldsymbol{1}^{\mathrm{T}}
\boldsymbol{\Theta}_{T}\boldsymbol{\Theta}_{T}^{-1}
=\boldsymbol{1}^{\mathrm{T}}\boldsymbol{\Theta}_{T}^{-1}$ holds. 
Combining these results, we arrive at Lemma~\ref{LEMMA_DAMPING}. 
\end{IEEEproof}

Using Lemma~\ref{LEMMA_DAMPING} for (\ref{SE_AB}) and (\ref{SE_BA}), we find 
that the damping factors $\{\theta_{\mathrm{A},\tau,t}\}$ and 
$\{\theta_{\mathrm{B},\tau,t}\}$ are independent of the 
variance~(\ref{SE_statistic_all}) for the sufficient statistics if 
$\mathcal{T}_{\mathrm{A},t}=\mathcal{T}_{\mathrm{B},t}
=\{0,\ldots,t\}$ are used. This observation is intuitively trivial since 
$\boldsymbol{x}_{\mathrm{B}\to \mathrm{A},t}^{\mathrm{suf}}$ and 
$\boldsymbol{x}_{\mathrm{A}\to \mathrm{B},t}^{\mathrm{suf}}$ in 
(\ref{statistic_mean_A}) and 
(\ref{statistic_mean_B}) are sufficient statistics. Thus, we assume no 
damping $\theta_{\mathrm{A},\tau,t}=\theta_{\mathrm{B},\tau,t}=\delta_{\tau,t}$ when 
all preceding information 
$\mathcal{T}_{\mathrm{A},t}=\mathcal{T}_{\mathrm{B},t}=\{0,\ldots,t\}$ are used.

\subsection{Bayes-Optimal LM-OAMP} 
We consider Bayes-optimal LM-OAMP. 
In general, the state evolution recursions in Theorem~\ref{THEOREM_SE} 
are two-dimensional discrete systems. Interestingly, the systems reduce to 
one-dimensional systems when all preceding information  
$\mathcal{T}_{\mathrm{A},t}=\mathcal{T}_{\mathrm{B},t}=\{0,\ldots,t\}$, the LMMSE 
filter~(\ref{LMMSE}), and the Bayes-optimal denoiser are used 
in each iteration. 

The Bayes-optimal denoiser is defined as the denoiser $f_{\tau}$ that 
minimizes the MSE $\bar{v}_{\mathrm{B},\tau+1,\tau+1}^{\mathrm{post}}$ in
 (\ref{SE_post_B}). 
By definition, the Bayes-optimal denoiser is equal to the posterior 
mean estimator $f_{\tau}=\mathbb{E}[x_{1}|x_{1}+z_{\tau}]$ with $z_{\tau}$ 
defined in Theorem~\ref{THEOREM_GAUSSIANITY}. 

To present the state evolution recursions for the Bayes-optimal LM-OAMP, 
we review state evolution recursions for conventional Bayes-optimal 
OAMP~\cite{Rangan192,Takeuchi201}, which are one-dimensional systems with 
respect to two variance parameters $\bar{v}_{\mathrm{A}\to \mathrm{B},t}$ 
and $\bar{v}_{\mathrm{B}\to \mathrm{A},t}$,
\begin{equation} \label{SE_AB_Bayes}
\bar{v}_{\mathrm{A}\to \mathrm{B},t} 
= \left(
 \frac{1}{\bar{\xi}_{\mathrm{A},t}\bar{v}_{\mathrm{B}\to \mathrm{A},t}}
 - \frac{1}{\bar{v}_{\mathrm{B}\to \mathrm{A},t}}
\right)^{-1}, 
\end{equation}
\begin{equation} \label{SE_BA_Bayes}
\bar{v}_{\mathrm{B}\to \mathrm{A},t+1} 
= \left(
 \frac{1}{\bar{v}_{\mathrm{B},t+1}^{\mathrm{post}}}
 - \frac{1}{\bar{v}_{\mathrm{A}\to \mathrm{B},t}}
\right)^{-1}, 
\end{equation}
with $\bar{v}_{\mathrm{B}\to \mathrm{A},0}=1$. In (\ref{SE_AB_Bayes}), 
$\bar{\xi}_{\mathrm{A},t}$ is given by 
\begin{IEEEeqnarray}{rl}
\bar{\xi}_{\mathrm{A},t} 
= 1 - \lim_{M=\delta N\to\infty}&\frac{\bar{v}_{\mathrm{B}\to \mathrm{A},t}}{N}
\mathrm{Tr}\left\{
 \boldsymbol{A}\boldsymbol{A}^{\mathrm{T}}
\right.
\nonumber \\
&\left.
 \cdot\left(
  \sigma^{2}\boldsymbol{I}_{M} + \bar{v}_{\mathrm{B}\to \mathrm{A},t}\boldsymbol{A}
  \boldsymbol{A}^{\mathrm{T}}
 \right)^{-1}
\right\}. \label{gamma_t_Bayes}
\end{IEEEeqnarray}
Furthermore, $\bar{v}_{\mathrm{B},t+1}^{\mathrm{post}}
=\mathbb{E}[\{\mathbb{E}[x_{1}|x_{1}+z_{t}]-x_{1}\}^{2}]$ in (\ref{SE_BA_Bayes}) 
corresponds to the MSE for Bayes-optimal OAMP in the large system limit, 
with $z_{t}$ denoting a zero-mean independent Gaussian 
random variable with variance $\bar{v}_{\mathrm{A}\to \mathrm{B},t}$. 

\begin{theorem}[Bayes-Optimal LM-OAMP] \label{THEOREM_SE_BAYES}
Suppose that Assumptions~\ref{assumption_x}--\ref{assumption_Lipschitz} 
are satisfied. 
Consider $\mathcal{T}_{\mathrm{A},t}=\mathcal{T}_{\mathrm{B},t}=\{0,\ldots,t\}$, 
$\theta_{\mathrm{A},\tau,t}=\theta_{\mathrm{B},\tau,t}=\delta_{\tau,t}$, 
the LMMSE filter~(\ref{LMMSE}), and the Bayes-optimal denoiser. Then, 
\begin{itemize}
\item The state evolution recursions for the Bayes-optimal LM-OAMP with 
the initialization $\bar{v}_{\mathrm{B}\to \mathrm{A},0,0}=1$ satisfy 
$\bar{v}_{\mathrm{A}\to \mathrm{B},t',t}=\bar{v}_{\mathrm{A}\to \mathrm{B},t}$ 
and $\bar{v}_{\mathrm{B}\to \mathrm{A},t'+1,t+1}=\bar{v}_{\mathrm{B}\to \mathrm{A},t+1}$    
for all $t'\leq t$, which are defined via the state evolution 
recursions~(\ref{SE_AB_Bayes}) and (\ref{SE_BA_Bayes}) for Bayes-optimal OAMP. 
\item
The state $(\bar{v}_{\mathrm{A}\to \mathrm{B},t',t}, 
\bar{v}_{\mathrm{B}\to \mathrm{A},t'+1,t+1})$ converges to a FP 
as $t$ tends to infinity for all $t'\leq t$. 
\end{itemize}
\end{theorem}
\begin{IEEEproof}
See Appendix~\ref{proof_theorem_SE_Bayes}. 
\end{IEEEproof}

Theorem~\ref{THEOREM_SE_BAYES} implies that the Bayes-optimal LM-OAMP converges 
for general signal prior and asymptotic eigenvalue distributions of 
$\boldsymbol{A}^{\mathrm{T}}\boldsymbol{A}$, while conventional MP algorithms 
require convergence analysis for individual 
problems~\cite{Liu16,Gerbelot20,Mondelli21}. 

The state evolution recursions for the Bayes-optimal 
LM-OAMP are essentially one-dimensional: $\bar{v}_{\mathrm{A}\to \mathrm{B},t',t}$ 
and $\bar{v}_{\mathrm{B}\to \mathrm{A},t'+1,t+1}$ are well defined via the 
one-dimensional recursions (\ref{SE_AB_Bayes}) and (\ref{SE_BA_Bayes}) for 
Bayes-optimal OAMP~\cite{Rangan192,Takeuchi201}. As a by-product, 
we arrive at the convergence of conventional Bayes-optimal OAMP.  

\begin{corollary}[Bayes-Optimal OAMP] \label{COROLLARY_OPTIMALITY}
Suppose that Assumptions~\ref{assumption_x}--\ref{assumption_Lipschitz} 
are satisfied. Then, the state evolution recursions~(\ref{SE_AB_Bayes}) and 
(\ref{SE_BA_Bayes}) for conventional Bayes-optimal OAMP converge to a FP. 
Furthermore, the FP characterizes the Bayes-optimal performance 
if it is unique.  
\end{corollary}
\begin{IEEEproof}
Since the Bayes-optimality of the unique FP is due to \cite{Ma17}, we only 
prove the convergence of the state evolution recursions for Bayes-optimal 
OAMP. 

The latter property in Theorem~\ref{THEOREM_SE_BAYES} implies 
the convergence of $\bar{v}_{\mathrm{A}\to \mathrm{B},t',t}$ and 
$\bar{v}_{\mathrm{B}\to \mathrm{A},t',t}$ for the Bayes-optimal LM-OAMP to a FP. 
Furthermore, the former property in Theorem~\ref{THEOREM_SE_BAYES} claims 
that the diagonal elements 
$\bar{v}_{\mathrm{A}\to \mathrm{B},t,t}$ and $\bar{v}_{\mathrm{B}\to \mathrm{A},t,t}$ 
are respectively equal to $\bar{v}_{\mathrm{A}\to \mathrm{B},t}$ and 
$\bar{v}_{\mathrm{B}\to \mathrm{A},t}$ in the state evolution 
recursions~(\ref{SE_AB_Bayes}) and 
(\ref{SE_BA_Bayes}) for Bayes-optimal OAMP~\cite{Rangan192,Takeuchi201}. 
These observations imply the convergence of the state evolution recursions 
for Bayes-optimal OAMP to a FP. 
\end{IEEEproof}

Theorem~\ref{THEOREM_SE_BAYES} implies that the dynamics of the MSE 
performance for the Bayes-optimal LM-OAMP is equal to that for Bayes-optimal 
OAMP. Interestingly, we can confirm an equivalence between the Bayes-optimal 
LM-OAMP and Bayes-optimal OAMP.  

\begin{proposition} \label{proposition_equivalence}
Consider $\mathcal{T}_{\mathrm{A},t}=\mathcal{T}_{\mathrm{B},t}=\{0,\ldots,t\}$, 
$\theta_{\mathrm{A},\tau,t}=\theta_{\mathrm{B},\tau,t}=\delta_{\tau,t}$, 
the LMMSE filter~(\ref{LMMSE}), and the Bayes-optimal denoiser. If 
the covariance matrices $\boldsymbol{V}_{\mathrm{A}\to\mathrm{B},t,t}$ and 
$\boldsymbol{V}_{\mathrm{B}\to\mathrm{A},t,t}$ given via 
(\ref{covariance_AB}) and (\ref{covariance_BA}) are positive definite 
for all $t$, 
then the Bayes-optimal LM-OAMP is equivalent to Bayes-optimal OAMP.  
\end{proposition}
\begin{IEEEproof}
Repeating the proof of the former property in Theorem~\ref{THEOREM_SE_BAYES}, 
we can prove the identities $v_{\mathrm{A}\to \mathrm{B},t',t}
=v_{\mathrm{A}\to \mathrm{B},t,t}$ and $v_{\mathrm{B}\to \mathrm{A},t'+1,t+1}
=v_{\mathrm{B}\to \mathrm{A},t+1,t+1}$ for all $t'\leq t$.     
From the positive definiteness of $\boldsymbol{V}_{\mathrm{A}\to\mathrm{B},t,t}$ and 
$\boldsymbol{V}_{\mathrm{B}\to\mathrm{A},t,t}$, we can use the last property in 
Lemma~\ref{LEMMA_MONOTONICITY} to find that (\ref{statistic_mean_A}) and 
(\ref{statistic_mean_B}) reduce to 
$\boldsymbol{x}_{\mathrm{B}\to \mathrm{A},t}^{\mathrm{suf}} 
=\boldsymbol{x}_{\mathrm{B}\to \mathrm{A},t}$ and 
$\boldsymbol{x}_{\mathrm{A}\to \mathrm{B},t}^{\mathrm{suf}} 
=\boldsymbol{x}_{\mathrm{A}\to \mathrm{B},t}$, respectively, and that 
$\xi_{\mathrm{A},t',t}=\xi_{\mathrm{B},t',t}=0$ in (\ref{xi_A}) and (\ref{xi_B}) 
holds for all $t'\neq t$. 

The former observation implies that one can skip the computation of the 
sufficient statistics in the first step of the Bayes-optimal LM-OAMP. 
The latter observation indicates that the extrinsic mean messages 
in (\ref{ext_mean_A}) and (\ref{ext_mean_B}) are equivalent to those 
in Bayes-optimal OAMP~\cite{Ma17}. 
Thus, the Bayes-optimal LM-OAMP is equivalent to Bayes-optimal OAMP.   
\end{IEEEproof} 

Theorem~\ref{THEOREM_GAUSSIANITY} justifies the positive-definiteness 
assumption in Proposition~\ref{proposition_equivalence} in the large system 
limit. Intuitively, Proposition~\ref{proposition_equivalence} holds since 
$\boldsymbol{x}_{\mathrm{A}\to \mathrm{B},t}$ and 
$\boldsymbol{x}_{\mathrm{B}\to \mathrm{A},t}$ are sufficient statistics 
for estimation of the signal vector given all preceding messages in each 
module. The Bayes-optimal LM-OAMP may be 
regarded as a theoretical tool to prove that 
$\boldsymbol{x}_{\mathrm{A}\to \mathrm{B},t}$ and 
$\boldsymbol{x}_{\mathrm{B}\to \mathrm{A},t}$ are sufficient statistics. 
These observations imply that conventional 
Bayes-optimal OAMP is the best option among all possible LM-MP algorithms 
in terms of the reconstruction performance.

\section{Numerical Results} \label{sec6}
\subsection{Simulation Conditions}
In all numerical results, we assume an i.i.d.\ Bernoulli-Gaussian (BG) 
signal vector $\boldsymbol{x}$ and artificial sensing matrices 
$\boldsymbol{A}$. Each signal element $x_{n}$ is independently 
sampled from the Gaussian distribution $\mathcal{N}(0,\rho^{-1})$ 
with probability $\rho\in[0,1]$. Otherwise, $x_{n}$ takes zero. 
See \cite[Appendix F]{Takeuchi211} for properties of the BG prior. 

In artificial sensing matrices with condition number~$\kappa>1$, 
$\boldsymbol{A}$ is assumed to have non-zero singular values 
$\sigma_{0}\geq\cdots\geq\sigma_{M-1}>0$ satisfying the condition number 
$\kappa=\sigma_{0}/\sigma_{M-1}$, $\sigma_{m}/\sigma_{m-1}
=\kappa^{-1/(M-1)}$, and $\sigma_{0}^{2}=N(1-\kappa^{-2/(M-1)})
/(1-\kappa^{-2M/(M-1)})$. As shown in \cite[Eq.~(65)]{Takeuchi211}, 
the $\eta$-transform $\eta$ in (\ref{eta_transform}) 
converges almost surely to  
\begin{equation} \label{eta_transform_artificial}
\eta(x) 
\ato 1 - \frac{1}{C}\ln\left(
 \frac{\kappa^{2}-1+\kappa^{2}Cx}{{\kappa^{2}-1+Cx}} 
\right) 
\end{equation}
in the large system limit, with $C=2\delta^{-1}\ln\kappa$. The 
$\eta$-transform~(\ref{eta_transform_artificial}) is used in solving state 
evolution recursions. 

To reduce the computational complexity, we assume the SVD structure 
$\boldsymbol{A}=\boldsymbol{\Sigma}\boldsymbol{V}^{\mathrm{T}}$, in which 
$\boldsymbol{V}$ denotes a Hadamard matrix with random row permutation. 
Such orthogonal matrices $\boldsymbol{V}$ may be regarded as a practical 
alternative of Haar orthogonal matrices~\cite{Anderson14,Male20,Dudeja22}. 

\subsection{Implementation} 
We consider the LMMSE filter~(\ref{LMMSE}) in module~A and the Bayes-optimal 
denoiser $f_{t}(s_{t})=\mathbb{E}[x_{1}|s_{t}]$ with $s_{t}=x_{1}+z_{t}$ in 
module~B, in which $\{z_{\tau}\}$ are defined in 
Theorem~\ref{THEOREM_GAUSSIANITY}. 
In computing the covariance message $v_{\mathrm{B},t'+1,t+1}^{\mathrm{post}}$, 
we use the posterior covariance $v_{\mathrm{B},t'+1,t+1}^{\mathrm{post}}
=\langle C(\boldsymbol{x}_{\mathrm{A}\to\mathrm{B},t'}^{\mathrm{suf}},
\boldsymbol{x}_{\mathrm{A}\to\mathrm{B},t}^{\mathrm{suf}})\rangle$ with 
$C(s_{t'},s_{t})=\mathbb{E}[\{x_{1}-f_{t'}(s_{t'})\}\{x_{1}-f_{t}(s_{t})\}
|s_{t'}, s_{t}]$, instead of (\ref{post_covariance_B}). 
See \cite[Appendix F]{Takeuchi211} for the details. 

Let $\boldsymbol{V}_{\mathrm{B},t,t}^{\mathrm{ext}}
\in\mathbb{R}^{|\mathcal{T}_{\mathrm{B},t}|\times|\mathcal{T}_{\mathrm{B},t}|}$ denote the 
covariance matrix with $[\boldsymbol{V}_{\mathrm{B},t,t}^{\mathrm{ext}}]_{\tau',\tau}
=v_{\mathrm{B},\tau',\tau}^{\mathrm{ext}}$ given in (\ref{ext_covariance_B}). 
The matrix $\boldsymbol{V}_{\mathrm{B},t,t}^{\mathrm{ext}}$ is guaranteed to be 
positive definite in the large system limit. For finite-sized systems, however, 
it might not be positive definite for some~$t$. To circumvent this issue, 
it is recommended to replace $v_{\mathrm{B},\tau',\tau}^{\mathrm{ext}}$ and 
$v_{\mathrm{B},\tau,\tau'}^{\mathrm{ext}}$ with $v_{\mathrm{B},\tau,\tau}^{\mathrm{ext}}$ if 
$v_{\mathrm{B},\tau',\tau'}^{\mathrm{ext}}v_{\mathrm{B},\tau,\tau}^{\mathrm{ext}} 
- (v_{\mathrm{B},\tau',\tau}^{\mathrm{ext}})^{2}<\epsilon$ holds for small 
$\epsilon>0$, which was set to $\epsilon=10^{-6}$ in numerical simulations. 

As discussed in Section~\ref{sec2}, this replacement implies that 
the AWGN observation $s_{\tau}=x_{1}+z_{\tau}$ in 
Theorem~\ref{THEOREM_GAUSSIANITY} becomes 
a sufficient statistic for estimation of $x_{1}$ based on both $s_{\tau'}$ 
and $s_{\tau}$ for $\tau'<\tau$. In other words, it means that $s_{\tau'}$ 
is not used since it is correlated with $s_{\tau}$ strongly. 

\subsection{Numerical Simulations}
Damped OAMP with the correct state 
evolution~(\ref{covariance_AB_dOAMP}) and 
(\ref{covariance_BA_dOAMP})---called LM-OAMP---is compared to damped OAMP 
with the heuristic damping (\ref{covariance_AB_hdOAMP}) and 
(\ref{covariance_BA_hdOAMP})~\cite{Rangan192}---called OAMP. 
See Example~\ref{example2} for the details. 

\begin{figure}[t]
\begin{center}
\includegraphics[width=\hsize]{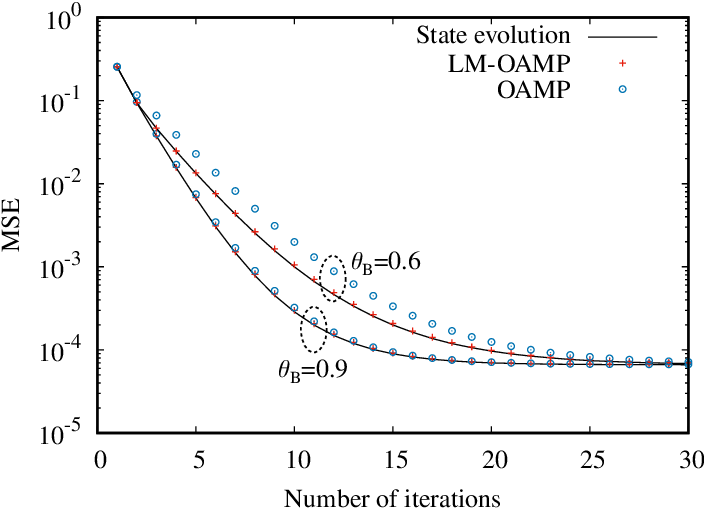}
\caption{
MSE versus the number of iterations for $M=2^{12}$, $N=2^{13}$, 
BG signals with signal density~$\rho=0.1$, condition number~$\kappa=10^{3}$, 
$\theta_{\mathrm{A}}=1$, and SNR~$1/\sigma^{2}=40$~dB. 
}
\label{fig2} 
\end{center}
\end{figure}

We first verify the correctness of the state evolution results via 
numerical simulations for large systems $N=2^{13}$. 
Figure~\ref{fig2} shows the MSEs versus the number of iterations for 
LM-OAMP with damping only in module B, i.e.\ $\theta_{\mathrm{A}}=1$ and 
$\theta_{\mathrm{B}}\in(0, 1)$. The state evolution results predict the MSEs 
of LM-OAMP accurately while OAMP is not in agreement with the state evolution 
results for small $\theta_{\mathrm{B}}$. 

\begin{figure}[t]
\begin{center}
\includegraphics[width=\hsize]{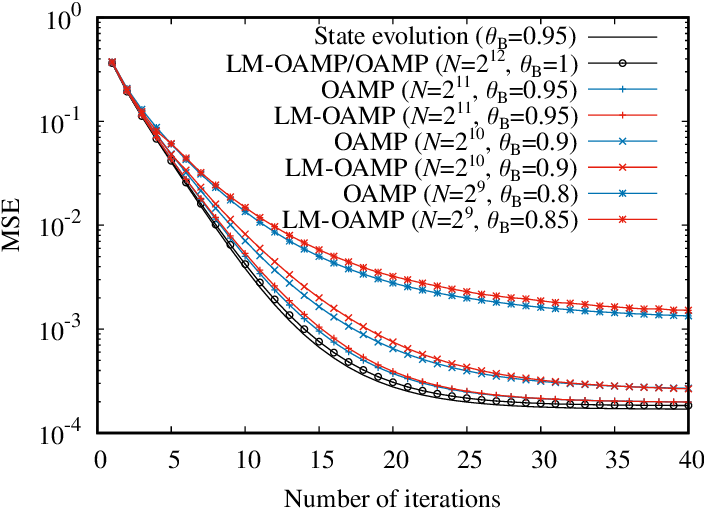}
\caption{
MSE versus the number of iterations for compression ratio $\delta=0.5$, 
BG signals with signal density~$\rho=0.1$, condition number~$\kappa=10^{4}$, 
$\theta_{\mathrm{A}}=1$, and SNR~$1/\sigma^{2}=40$~dB. 
}
\label{fig3} 
\end{center}
\end{figure}

We next compare LM-OAMP with OAMP for smaller systems than in Fig.~\ref{fig2}. 
As shown in Fig.~\ref{fig3}, LM-OAMP is slightly inferior to 
OAMP especially for small systems. This is surprising since 
LM-OAMP uses the consistent state evolution in the large system limit. 

This result is intuitively understood as follows: LM-OAMP uses the covariance 
messages in all preceding iterations to compute the state evolution prediction 
in the current iteration. As a result, all prediction errors in the preceding 
iterations are accumulated for finite-sized systems, so that LM-OAMP should 
have larger prediction errors than conventional OAMP without LM processing. 
Figure~\ref{fig3} shows that this prediction errors due to finite size effects 
are more dominant than the inconsistency of the heuristic damping 
in the large system limit. 

\section{Conclusions} \label{sec7}
This paper has proposed LM-MP, constructed by adding computation 
of sufficient statistics for estimation of the signal vector given all 
preceding messages to the beginning of each module in conventional MP. 
The proposed LM-MP is a modification of conventional MP in two points. 

In one point, the Bayes-optimal LM-MP provides a proof strategy to guarantee 
the convergence of state evolution recursions for conventional Bayes-optimal 
MP systematically. In the proof strategy, one-dimensional state evolution 
recursions for the Bayes-optimal MP are embedded into two-dimensional state 
evolution recursions for the Bayes-optimal LM-MP, which are 
systematically guaranteed to converge. As a result, one can prove the 
convergence of the state evolution recursions for the Bayes-optimal MP. While 
it has been applied to Bayes-optimal OAMP~\cite{Ma17,Rangan192} in this paper, 
the proposed LM-MP strategy is general and applicable to the convergence 
analysis of the other Bayes-optimal MP algorithms. 

The other point is that LM-MP may improve the convergence properties of 
conventional MP. If state evolution recursions for MP with Bayes-optimal 
denoiser are not embedded into those for the corresponding LM-MP, 
the LM-MP strategy indicates that the conventional MP has room for 
improvement: The constructed LM-MP improves the convergence 
properties of the conventional MP for large systems. In this sense, 
MAMP~\cite{Liu21} can be regarded as a modification of an MP algorithm  
without LM damping.  

Possible directions for future work are summarized to conclude this paper. 
A first issue is another criterion for optimality such as the maximum a 
posteriori (MAP) scenario while this paper has focused on the minimum MSE 
(MMSE) scenario. It is an interesting direction whether the LM-MP strategy 
can be generalized to the other criteria for optimality. 

The other issue is an improvement of the state evolution prediction for LM-MP 
in finite-sized systems. State evolution recursions for LM-MP are more 
sensitive to finite size effects than for conventional MP while they are 
consistent in the large system limit. It is important to construct covariance 
estimators that are consistent in the large system limit and accurate for 
finite-sized systems. Such covariance estimators should increase practical 
value of LM-MP for finite-sized systems while the numerical simulations in 
this paper have shown the inferiority of damped LM-OAMP to OAMP with 
heuristic damping for finite-sized systems.

\appendices
\section{Proofs}
\subsection{Proof of Proposition~\ref{proposition_denoiser}}
\label{appen_AA}
We follow \cite[Appendix B]{Ma17} to prove 
Proposition~\ref{proposition_denoiser}. 
We use the definition $f_{\mathrm{opt}}(S_{t})=\mathbb{E}[X|S_{t}]$ 
to obtain 
\begin{IEEEeqnarray}{rl}
\mathbb{E}[(\psi_{t} - X)^{2}]
=& \mathbb{E}[(\psi_{t} - f_{\mathrm{opt}} + f_{\mathrm{opt}} - X)^{2}] 
\nonumber \\
=& \mathbb{E}[(\psi_{t} - f_{\mathrm{opt}})^{2}] 
+ \mathbb{E}[(f_{\mathrm{opt}} - X)^{2}].  
\end{IEEEeqnarray}
Thus, the minimization of $\mathbb{E}[(\psi_{t} - X)^{2}]$ is equivalent 
to that of $\mathbb{E}[(\psi_{t} - f_{\mathrm{opt}})^{2}]$. 

To evaluate $\mathbb{E}[(\psi_{t} - f_{\mathrm{opt}})^{2}]$, we repeat 
the same argument for the extrinsic denoiser $f_{t}^{\mathrm{ext}}$ to obtain 
\begin{IEEEeqnarray}{rl}
&\mathbb{E}[(\psi_{t} - f_{\mathrm{opt}})^{2}]
= \mathbb{E}[(\psi_{t} - f_{t}^{\mathrm{ext}} + f_{t}^{\mathrm{ext}} 
- f_{\mathrm{opt}})^{2}] 
\nonumber \\
=& \mathbb{E}[(\psi_{t} - f_{t}^{\mathrm{ext}})^{2}] 
+ 2\mathbb{E}[(\psi_{t} - f_{t}^{\mathrm{ext}})
(f_{t}^{\mathrm{ext}} - f_{\mathrm{opt}})]
\nonumber \\
&+ \mathbb{E}[(f_{t}^{\mathrm{ext}} - f_{\mathrm{opt}})^{2}],
\end{IEEEeqnarray}
which implies that $\psi_{t}=f_{t}^{\mathrm{ext}}$ minimizes 
$\mathbb{E}[(\psi_{t} - f_{\mathrm{opt}})^{2}]$ if the second term is equal 
to zero for all extrinsic denoisers $\psi_{t}$. 

To complete the proof of Proposition~\ref{proposition_denoiser}, it is 
sufficient to prove $\mathbb{E}[\psi_{t}(f_{t}^{\mathrm{ext}} - f_{\mathrm{opt}})]=0$ 
for any Lipschitz-continuous extrinsic denoiser $\psi_{t}$ with 
$\mathbb{E}[\partial\psi_{t}/\partial Y_{\tau}]=0$. 
From the definition of $f_{t}^{\mathrm{ext}}$ in (\ref{extrinsic_denoiser}), 
we have 
\begin{equation} 
f_{t}^{\mathrm{ext}} - f_{\mathrm{opt}} 
= \frac{\sum_{\tau\in\mathcal{T}_{t}}\xi_{\tau,t}(f_{\mathrm{opt}} - X - W_{\tau})}
{1-\sum_{\tau\in\mathcal{T}_{t}}\xi_{\tau,t}}. 
\end{equation}
Since $\{W_{\tau}: \tau\in\mathcal{T}_{t}\}$ are zero-mean Gaussian, we can use 
\cite[Lemma~2]{Takeuchi211} for Lipschitz-continuous $\psi_{t}$ to obtain 
\begin{equation}
\mathbb{E}[W_{\tau}\psi_{t}]
= \sum_{t'\in\mathcal{T}_{t}}\mathbb{E}[W_{\tau}W_{t'}]
\mathbb{E}\left[
 \frac{\partial\psi_{t}}{\partial Y_{t'}}
\right]
= 0, 
\end{equation}
where the last follows from the assumption 
$\mathbb{E}[\partial\psi_{t}/\partial Y_{t'}]=0$. Thus, we have  
\begin{equation}
\mathbb{E}[\psi_{t}(f_{t}^{\mathrm{ext}} - f_{\mathrm{opt}})]
= \frac{\sum_{\tau\in\mathcal{T}_{t}}\xi_{\tau,t}
\mathbb{E}[\psi_{t}(f_{\mathrm{opt}} - X)]}
{1-\sum_{\tau\in\mathcal{T}_{t}}\xi_{\tau,t}}
= 0,
\end{equation}
where the last equality follows from $\mathbb{E}[\psi_{t}X]
= \mathbb{E}[\psi_{t}f_{\mathrm{opt}}]$ due to the definition 
$f_{\mathrm{opt}}(S_{t})=\mathbb{E}[X|S_{t}]=\mathbb{E}[X|\boldsymbol{Y}_{t}]$. 
Thus, Proposition~\ref{proposition_denoiser} holds. 

\subsection{Proof of Lemma~\ref{LEMMA_CONSISTENT}}
\label{proof_LEMMA_CONSISTENT} 
Since $\{S_{t',n}, S_{t,n}\}_{n=1}^{N}$ are independent samples, we use 
the strong law of large numbers to obtain 
\begin{equation}
\lim_{N\to\infty}\frac{1}{N}\sum_{n=1}^{N}\hat{C}(S_{t',n},S_{t,n})\aeq 
\mathbb{E}[\hat{C}(S_{t'},S_{t})]. 
\end{equation}
Thus, it is sufficient to prove the unbiasedness of $\hat{C}$, i.e.\ 
\begin{equation}
\mathbb{E}[\hat{C}(S_{t'},S_{t})]
= \mathbb{E}[\{X-f_{t'}(S_{t'})\}\{X-f_{t}(S_{t})\}]. 
\end{equation}

We first evaluate the expectation 
$\mathbb{E}[S_{t'}f_{t}(S_{t})]$. Substituting $S_{t'}=X+\tilde{W}_{t'}$ and 
using Stein's lemma~\cite{Stein72} for Lipschitz-continuous $f_{t}$, we have 
\begin{equation}
\mathbb{E}[S_{t'}f_{t}(S_{t})]
= \mathbb{E}[Xf_{t}(S_{t})]
+ \mathbb{E}[\tilde{W}_{t'}\tilde{W}_{t}]\mathbb{E}[f_{t}'(S_{t})]. 
\end{equation}
Applying this result to (\ref{consistent_estimator}), we arrive at 
\begin{IEEEeqnarray}{rl}
\mathbb{E}[\hat{C}(S_{t'},S_{t})]
=& \mathbb{E}[X^{2}] + \mathbb{E}[f_{t'}(S_{t'})f_{t}(S_{t})]
\nonumber \\
&- \mathbb{E}[Xf_{t}(S_{t})] - \mathbb{E}[Xf_{t'}(S_{t'})]
\nonumber \\
=& \mathbb{E}[\{X-f_{t'}(S_{t'})\}\{X-f_{t}(S_{t})\}].
\end{IEEEeqnarray}
Thus, Lemma~\ref{LEMMA_CONSISTENT} holds.

\subsection{Proof of Lemma~\ref{LEMMA_BAYES}}
\label{proof_LEMMA_BAYES}
We prove the former property. Since $\{Y_{\tau}\}_{\tau=0}^{t}$ contains 
$\{Y_{\tau}\}_{\tau=0}^{t'}$ for $t'<t$, the optimality of the posterior 
mean estimator $f_{\mathrm{opt}}(S_{t})$ implies 
$\mathbb{E}[\{X-f_{\mathrm{opt}}(S_{t})\}^{2}]
\leq\mathbb{E}[\{X-f_{\mathrm{opt}}(S_{t'})\}^{2}]$. 
The monotonicity $\mathbb{E}[\tilde{W}_{t'}^{2}]\geq
\mathbb{E}[\tilde{W}_{t}^{2}]$ follows 
from the monotonicity of the MSE $\mathbb{E}[\{X-f_{\mathrm{opt}}(S_{t})\}^{2}]$ 
with respect to the variance $\mathbb{E}[\tilde{W}_{t}^{2}]$. 

We next prove the latter property. When 
$\mathbb{E}[\tilde{W}_{t'}\tilde{W}_{t}]=\mathbb{E}[\tilde{W}_{t}^{2}]$ holds, 
we have the following representation for the sufficient statistic: 
\begin{equation} \label{representation_S}
S_{t'} = S_{t} + Z_{t'}, \quad S_{t} = X + \tilde{W}_{t},
\end{equation}
where $Z_{t'}$ is a zero-mean Gaussian random variable with variance 
$\mathbb{E}[\tilde{W}_{t'}^{2}]-\mathbb{E}[\tilde{W}_{t}^{2}]\geq0$ and 
independent of $X$ and $\tilde{W}_{t}$. 
The representation~(\ref{representation_S}) can be verified from 
\begin{equation}
\mathbb{E}[(\tilde{W}_{t}+Z_{t'})^{2}] = \mathbb{E}[\tilde{W}_{t'}^{2}], 
\quad \mathbb{E}[(\tilde{W}_{t}+Z_{t'})\tilde{W}_{t}] 
= \mathbb{E}[\tilde{W}_{t}^{2}].  
\end{equation} 

The representation~(\ref{representation_S}) implies that $S_{t}$ is a 
sufficient statistic for estimation of $X$ given both $S_{t'}$ and $S_{t}$. 
Thus, we have $\mathbb{E}[X|S_{t'},S_{t}]=\mathbb{E}[X|S_{t}]
=f_{\mathrm{opt}}(S_{t})$. Applying this identity to the definition of 
$C(S_{t'},S_{t})$ in (\ref{posterior_covariance}), we arrive at 
\begin{IEEEeqnarray}{rl}
&C(S_{t'},S_{t}) - C(S_{t},S_{t})
\nonumber \\
=& \{f_{\mathrm{opt}}(S_{t}) - f_{\mathrm{opt}}(S_{t'})\}\left\{
 \mathbb{E}[X | S_{t'},S_{t}] - f_{\mathrm{opt}}(S_{t})
\right\} \nonumber \\
=& 0.
\end{IEEEeqnarray}
Thus, the latter property holds. 

\subsection{Proof of Lemma~\ref{LEMMA_MONOTONICITY}}
\label{proof_LEMMA_MONOTONICITY}
We first prove the first property. It is straightforward to confirm 
\begin{equation} \label{determinant} 
\det\boldsymbol{\Sigma}_{t} 
= [\boldsymbol{\Sigma}_{t}]_{t,t}
\prod_{\tau=0}^{t-1}\Delta\Sigma_{\tau,t},
\end{equation}
with $\Delta\Sigma_{\tau,t}=[\boldsymbol{\Sigma}_{t}]_{\tau,\tau} 
- [\boldsymbol{\Sigma}_{t}]_{\tau+1,\tau+1}$. Since 
$\{\boldsymbol{\Sigma}_{\tau}\}_{\tau=1}^{t}$ have been assumed to be positive 
definite, we have the positivity 
$\{\det\boldsymbol{\Sigma}_{\tau}>0\}_{\tau=1}^{t}$. Using (\ref{determinant})  
yields the first property $\Delta\Sigma_{\tau,t}>0$ for all $\tau$.  

Let us prove (\ref{determinant}). 
Since $[\boldsymbol{\Sigma}_{t}]_{\tau',\tau}
=[\boldsymbol{\Sigma}_{t}]_{\tau,\tau'}=[\boldsymbol{\Sigma}_{t}]_{\tau,\tau}$ 
holds for all $\tau'<\tau$, 
we subtract the $(\tau+1)$th column in $\boldsymbol{\Sigma}_{t}$ from 
the $\tau$th column for all $\tau\in\{0,\ldots,t-1\}$ to obtain the 
upper triangular expression for the determinant,  
\begin{equation}
\det\boldsymbol{\Sigma}_{t} 
= \begin{vmatrix}
\Delta\Sigma_{0,t} & \cdots & \Delta\Sigma_{t-1,t} & 
[\boldsymbol{\Sigma}_{t}]_{t,t} \\
\boldsymbol{O} & \ddots & \vdots & \vdots \\
\vdots & \ddots & \Delta\Sigma_{t-1,t} & \vdots \\
\boldsymbol{O} & \cdots & \boldsymbol{O} & 
[\boldsymbol{\Sigma}_{t}]_{t,t}
\end{vmatrix},
\end{equation}
which implies (\ref{determinant}). Thus, the first property holds. 

We next prove the second property. Using the assumption 
$[\boldsymbol{\Sigma}_{t}]_{\tau',\tau}=[\boldsymbol{\Sigma}_{t}]_{\tau,\tau}$ 
and the first property $[\boldsymbol{\Sigma}_{t}]_{\tau',\tau'}>
[\boldsymbol{\Sigma}_{t}]_{\tau,\tau}$ for all $\tau'<\tau$, we have the 
following representation for $\{Y_{\tau}\}$:  
\begin{equation}
Y_{t} = X + W_{t}, 
\quad 
Y_{\tau-1} = Y_{\tau} + V_{\tau-1}
\end{equation}
for $\tau\in\{1,\ldots,t\}$, where $\{V_{\tau-1}\}$ are independent zero-mean 
Gaussian random variables with 
$\mathbb{E}[V_{\tau-1}^{2}]=\Delta\Sigma_{\tau-1,t}>0$. This representation 
implies that $Y_{t}$ is a sufficient statistic for estimation of $X$ based on 
$\{Y_{\tau}\}_{\tau=0}^{t}$. Thus, we arrive at 
$\mathbb{E}[\{X - f_{\mathrm{opt}}(S_{t})\}^{2}]
=\mathbb{E}[\{X - f_{\mathrm{opt}}(Y_{t})\}^{2}]$, which implies  
$\mathbb{E}[\tilde{W}_{t}^{2}] = \mathbb{E}[W_{t}^{2}]$. 

Finally, the last property follows immediately from the definitions of $S_{t}$ 
and $\xi_{\tau,t}$ in (\ref{sufficient_statistic}) and (\ref{xi}), 
and the identity 
$\boldsymbol{\Sigma}_{t}^{-1}\boldsymbol{1}=([\boldsymbol{\Sigma}_{t}]_{t,t})^{-1}
\boldsymbol{e}_{t}$. The identity $S_{t}=Y_{t}$ provides an alternative 
proof of the second property: 
$\mathbb{E}[\tilde{W}_{t}^{2}]=\mathbb{E}[(S_{t}-X)^{2}]
=\mathbb{E}[(Y_{t}-X)^{2}]=\mathbb{E}[W_{t}^{2}]$.

The identity $\boldsymbol{\Sigma}_{t}^{-1}\boldsymbol{1}
=([\boldsymbol{\Sigma}_{t}]_{t,t})^{-1}\boldsymbol{e}_{t}$ can be confirmed 
as follows: From the assumption in Lemma~\ref{LEMMA_MONOTONICITY}, 
$\boldsymbol{\Sigma}_{t}$ has the last column $[\boldsymbol{\Sigma}_{t}]_{t,t}
\boldsymbol{1}$. Thus, we have $\boldsymbol{\Sigma}_{t}\boldsymbol{e}_{t}
=[\boldsymbol{\Sigma}_{t}]_{t,t}\boldsymbol{1}$, which is equivalent to 
the identity $\boldsymbol{\Sigma}_{t}^{-1}\boldsymbol{1}
=([\boldsymbol{\Sigma}_{t}]_{t,t})^{-1}\boldsymbol{e}_{t}$. 

\section{Proof of Theorem~\ref{THEOREM_GAUSSIANITY}}
\label{proof_theorem_Gaussianity}
Asymptotic Gaussianity has been proved for a general error model proposed 
in \cite{Takeuchi211}. Thus, we first prove that the error model 
for LM-OAMP is included into the general error model. 

\begin{lemma} \label{lemma_error_model}
Suppose that Assumption~\ref{assumption_filter} holds. 
Let $\boldsymbol{h}_{t}=\boldsymbol{x}_{\mathrm{A}\to \mathrm{B},t}-\boldsymbol{x}$ 
denote the estimation error before denoising in iteration~$t$. Define an error 
vector $\boldsymbol{q}_{t+1}$ associated with the estimation error after 
denoising by   
\begin{equation} \label{q}
\boldsymbol{q}_{t+1} 
= \sum_{\tau=0}^{t}\frac{\theta_{\mathrm{B},\tau,t}}{1 - \xi_{\mathrm{B},\tau}}
(\boldsymbol{x}_{\mathrm{B},\tau+1}^{\mathrm{post}}-\boldsymbol{x}).  
\end{equation}
Then, the error model for LM-OAMP is included into the general error model 
in \cite{Takeuchi211} and given by 
\begin{equation} \label{b}
\boldsymbol{b}_{t} = \boldsymbol{V}^{\mathrm{T}}\tilde{\boldsymbol{q}}_{t},
\end{equation}
\begin{equation} \label{m}
\boldsymbol{m}_{t} 
= \sum_{\tau=0}^{t}\frac{\theta_{\mathrm{A},\tau,t}}{1-\xi_{\mathrm{A},\tau}}
\boldsymbol{V}^{\mathrm{T}}(\boldsymbol{x}_{\mathrm{A},\tau}^{\mathrm{post}} 
- \boldsymbol{x}),
\end{equation}
\begin{equation} \label{m_tilde}
\tilde{\boldsymbol{m}}_{t} 
= \sum_{\tau=0}^{t}\frac{\theta_{\mathrm{A},\tau,t}
\{\boldsymbol{V}^{\mathrm{T}}(\boldsymbol{x}_{\mathrm{A},\tau}^{\mathrm{post}} 
- \boldsymbol{x}) - \sum_{t'\in\mathcal{T}_{\mathrm{A},\tau}}\xi_{\mathrm{A},t',\tau}
\boldsymbol{b}_{t'}\}}{1-\xi_{\mathrm{A},\tau}}, 
\end{equation}
\begin{IEEEeqnarray}{rl}
\boldsymbol{V}^{\mathrm{T}}(\boldsymbol{x}_{\mathrm{A},t}^{\mathrm{post}} 
- \boldsymbol{x})
=& \left(
 \boldsymbol{I}_{N} - \tilde{\boldsymbol{W}}_{t}^{\mathrm{T}}\boldsymbol{\Sigma}
\right)
\frac{\boldsymbol{B}_{t}\boldsymbol{V}_{\mathrm{B}\to \mathrm{A},t,t}^{-1}
\boldsymbol{1}}
{\boldsymbol{1}^{\mathrm{T}}\boldsymbol{V}_{\mathrm{B}\to \mathrm{A},t,t}^{-1}
\boldsymbol{1}} \nonumber \\
&+ \tilde{\boldsymbol{W}}_{t}^{\mathrm{T}}\boldsymbol{U}^{\mathrm{T}}\boldsymbol{w},  
\label{post_mean_A_tmp}
\end{IEEEeqnarray}
\begin{equation}
\boldsymbol{h}_{t}=\boldsymbol{V}\tilde{\boldsymbol{m}}_{t}, 
\end{equation}
\begin{equation} \label{q_tilde} 
\tilde{\boldsymbol{q}}_{t+1}
=\sum_{\tau=0}^{t}\frac{
\theta_{\mathrm{B},\tau,t}
(\boldsymbol{x}_{\mathrm{B},\tau+1}^{\mathrm{post}}-\boldsymbol{x}
 - \sum_{t'\in\mathcal{T}_{\mathrm{B},\tau}}\xi_{\mathrm{B},t',\tau}\boldsymbol{h}_{t'})}
{1 - \xi_{\mathrm{B},\tau}},  
\end{equation}
with $\tilde{\boldsymbol{q}}_{0}=-\boldsymbol{x}$, where 
$\boldsymbol{B}_{t}\in\mathbb{R}^{N\times|\mathcal{T}_{\mathrm{A},t}|}$ consists of 
$\{\boldsymbol{b}_{\tau}:\tau\in\mathcal{T}_{\mathrm{A},t}\}$. In particular, 
$\tilde{\boldsymbol{q}}_{t}
=\boldsymbol{x}_{\mathrm{B}\to \mathrm{A},t}-\boldsymbol{x}$ holds. 
\end{lemma} 
\begin{IEEEproof}
We first prove  
\begin{IEEEeqnarray}{rl} \label{q_tilde_def}
\tilde{\boldsymbol{q}}_{t+1}
= \boldsymbol{q}_{t+1} 
- \sum_{t'=0}^{t}\left(
 \frac{1}{N}\sum_{n=1}^{N}
 \frac{\partial q_{n,t+1}}{\partial h_{n,t'}}
\right)\boldsymbol{h}_{t'}, 
\end{IEEEeqnarray}
which is equivalent to the definition in \cite[Eq.~(9)]{Takeuchi211}. 
From the definition of $\boldsymbol{q}_{t+1}$ in (\ref{q}), we have 
\begin{equation}
\frac{1}{N}\sum_{n=1}^{N}\frac{\partial q_{n,t+1}}{\partial h_{n,t'}}
= \frac{1}{N}\sum_{n=1}^{N}\sum_{\tau=0}^{t}
\frac{\theta_{\mathrm{B},\tau,t}}{1 - \xi_{\mathrm{B},\tau}}
\frac{\partial x_{\mathrm{B},n,\tau+1}^{\mathrm{post}}}{\partial h_{n,t'}}. 
\end{equation}
We define the error matrix $\boldsymbol{H}_{t}
=\boldsymbol{X}_{\mathrm{A}\to \mathrm{B},t} 
- \boldsymbol{1}^{\mathrm{T}}\otimes\boldsymbol{x}$ for the preceding messages 
before denoising, so that $\boldsymbol{H}_{t}$ consists of the columns 
$\{\boldsymbol{h}_{t'}: t'\in\mathcal{T}_{\mathrm{B},t}\}$. 
Using the definition of $\boldsymbol{x}_{\mathrm{B},t+1}^{\mathrm{post}}$ 
in (\ref{post_mean_B}) with (\ref{statistic_mean_B}), 
we find that $\boldsymbol{x}_{\mathrm{B},t+1}^{\mathrm{post}}$ is 
a function of $\{\boldsymbol{h}_{t'}: t'\in\mathcal{T}_{\mathrm{B},t}\}$ 
since the sufficient statistic~(\ref{statistic_mean_B}) is a function of 
$\boldsymbol{X}_{\mathrm{A}\to \mathrm{B},t}=\boldsymbol{1}^{\mathrm{T}}
\otimes\boldsymbol{x}+\boldsymbol{H}_{t}$. Thus, we obtain 
\begin{IEEEeqnarray}{rl}
\frac{1}{N}&\sum_{n=1}^{N}
\frac{\partial x_{\mathrm{B},n,\tau+1}^{\mathrm{post}}}{\partial h_{n,t'}}
= \frac{1}{N}\sum_{n=1}^{N}f_{\tau}'(x_{\mathrm{A}\to \mathrm{B},n,\tau}^{\mathrm{suf}}) 
\nonumber \\
&\cdot\frac{\partial}{\partial h_{n,t'}}
\frac{([\boldsymbol{1}^{\mathrm{T}}\otimes\boldsymbol{x}
+\boldsymbol{H}_{\tau}]_{,n})^{\mathrm{T}}
\boldsymbol{V}_{\mathrm{B}\to \mathrm{A},\tau,\tau}^{-1}\boldsymbol{1}}
{\boldsymbol{1}^{\mathrm{T}}\boldsymbol{V}_{\mathrm{B}\to \mathrm{A},\tau,\tau}^{-1}
\boldsymbol{1}}
\nonumber \\
=& \langle f_{\tau}'(\boldsymbol{x}_{\mathrm{A}\to \mathrm{B},\tau}^{\mathrm{suf}}) 
\rangle 
\frac{\boldsymbol{e}_{I_{\mathcal{T}_{\mathrm{B},\tau}}(t')}^{\mathrm{T}}
\boldsymbol{V}_{\mathrm{B}\to \mathrm{A},\tau,\tau}^{-1}\boldsymbol{1}}
{\boldsymbol{1}^{\mathrm{T}}\boldsymbol{V}_{\mathrm{B}\to \mathrm{A},\tau,\tau}^{-1}
\boldsymbol{1}} 
= \xi_{\mathrm{B},t',\tau} \label{xi_B_tmp}
\end{IEEEeqnarray}
for $t'\in\mathcal{T}_{\mathrm{B},\tau}$, with $\xi_{\mathrm{B},t',\tau}$ defined in 
(\ref{xi_B}), where $[\cdots]_{,n}$ denotes the $n$th row of $\cdots$. 
Substituting these results and the definition of $\boldsymbol{q}_{t+1}$ in 
(\ref{q}) into (\ref{q_tilde_def}), we arrive at the equivalence 
between (\ref{q_tilde}) and (\ref{q_tilde_def}). 

We next prove $\tilde{\boldsymbol{q}}_{t+1}
=\boldsymbol{x}_{\mathrm{B}\to \mathrm{A},t+1}
-\boldsymbol{x}$. Using the definition of 
$\boldsymbol{x}_{\mathrm{B},t}^{\mathrm{ext}}$ 
in (\ref{ext_mean_B}) and 
$\boldsymbol{h}_{t}=\boldsymbol{x}_{\mathrm{A}\to \mathrm{B},t}-\boldsymbol{x}$ 
yields  
\begin{equation} 
\boldsymbol{x}_{\mathrm{B},t+1}^{\mathrm{ext}}-\boldsymbol{x}
= \frac{\boldsymbol{x}_{\mathrm{B},t+1}^{\mathrm{post}}-\boldsymbol{x}
- \sum_{t'\in\mathcal{T}_{\mathrm{B},t}}\xi_{\mathrm{B},t',t}\boldsymbol{h}_{t'}}
{1 - \xi_{\mathrm{B},t}},
\end{equation}
where we have used $\sum_{t'\in\mathcal{T}_{\mathrm{B},t}}\xi_{\mathrm{B},t',t}
=\xi_{\mathrm{B},t}$. 
Applying this expression to (\ref{q_tilde}), we have 
\begin{equation}
\tilde{\boldsymbol{q}}_{t+1}
=\sum_{\tau=0}^{t}\theta_{\mathrm{B},\tau,t}
(\boldsymbol{x}_{\mathrm{B},\tau+1}^{\mathrm{ext}}-\boldsymbol{x}).   
\end{equation}
Using the definition of $\boldsymbol{x}_{\mathrm{B}\to \mathrm{A},t+1}$ in 
(\ref{mean_BA}) and the normalization~(\ref{damping_B}), we arrive at 
$\tilde{\boldsymbol{q}}_{t+1}=\boldsymbol{x}_{\mathrm{B}\to \mathrm{A},t+1}
-\boldsymbol{x}$. 

We shall derive (\ref{post_mean_A_tmp}). 
Applying $\tilde{\boldsymbol{q}}_{t}=\boldsymbol{x}_{\mathrm{B}\to \mathrm{A},t}
-\boldsymbol{x}$ 
to the sufficient statistic~(\ref{statistic_mean_A}) yields  
\begin{equation} \label{statistic_mean_A_tmp}
\boldsymbol{x}_{\mathrm{B}\to \mathrm{A},t}^{\mathrm{suf}} 
= \boldsymbol{x}
+ \frac{\tilde{\boldsymbol{Q}}_{t}
\boldsymbol{V}_{\mathrm{B}\to \mathrm{A},t,t}^{-1}\boldsymbol{1}}
{\boldsymbol{1}^{\mathrm{T}}\boldsymbol{V}_{\mathrm{B}\to \mathrm{A},t,t}^{-1}
\boldsymbol{1}},  
\end{equation}
with $\tilde{\boldsymbol{Q}}_{t}=\boldsymbol{X}_{\mathrm{B}\to \mathrm{A},t}
-(\boldsymbol{1}^{\mathrm{T}}\otimes\boldsymbol{x})$. Substituting 
this result and the measurement model~(\ref{model}) into the definition 
of $\boldsymbol{x}_{\mathrm{A},t}^{\mathrm{post}}$ in (\ref{post_mean_A}), we have 
\begin{equation}  \label{post_mean_A_tmp2}
\boldsymbol{x}_{\mathrm{A},t}^{\mathrm{post}} - \boldsymbol{x}
= \left(
 \boldsymbol{I}_{N} - \boldsymbol{W}_{t}^{\mathrm{T}}\boldsymbol{A}
\right)
\frac{\tilde{\boldsymbol{Q}}_{t}
\boldsymbol{V}_{\mathrm{B}\to \mathrm{A},t,t}^{-1}\boldsymbol{1}}
{\boldsymbol{1}^{\mathrm{T}}\boldsymbol{V}_{\mathrm{B}\to \mathrm{A},t,t}^{-1}
\boldsymbol{1}} 
+ \boldsymbol{W}_{t}^{\mathrm{T}}\boldsymbol{w}. 
\end{equation}
For the SVD $\boldsymbol{A}
=\boldsymbol{U}\boldsymbol{\Sigma}\boldsymbol{V}^{\mathrm{T}}$, 
applying the SVD $\boldsymbol{W}_{t}=\boldsymbol{U}\tilde{\boldsymbol{W}}_{t}
\boldsymbol{V}^{\mathrm{T}}$ in Assumption~\ref{assumption_filter} to 
(\ref{post_mean_A_tmp2}) and using the definition of $\boldsymbol{b}_{t}$ 
in (\ref{b}), we arrive at (\ref{post_mean_A_tmp}) 
with $\boldsymbol{B}_{t}=\boldsymbol{V}^{\mathrm{T}}
\tilde{\boldsymbol{Q}}_{t}$. 

Let us prove 
\begin{equation} \label{m_tilde_def}
\tilde{\boldsymbol{m}}_{t} 
= \boldsymbol{m}_{t} 
- \sum_{t'=0}^{t}\left(
 \frac{1}{N}\sum_{n=1}^{N}\frac{\partial m_{n,t}}{\partial b_{n,t'}}
\right)\boldsymbol{b}_{t'},   
\end{equation}
which is equivalent to the definition in \cite[Eq.~(11)]{Takeuchi211}. 
Using the expression for $\boldsymbol{m}_{t}$ in (\ref{m}) with 
(\ref{post_mean_A_tmp}), we repeat the derivation of (\ref{xi_B_tmp}) to have 
\begin{equation}
\frac{1}{N}\sum_{n=1}^{N}\frac{\partial m_{n,t}}{\partial b_{n,t'}} 
= \sum_{\tau=0}^{t}\frac{\theta_{\mathrm{A},\tau,t}\xi_{\mathrm{A},t',\tau}}
{1-\xi_{\mathrm{A},\tau}}   
\end{equation}
for $t'\in\mathcal{T}_{\mathrm{A},t}$, with $\xi_{\mathrm{A},t',t}$ defined in 
(\ref{xi_A}), where we have used the identity $\mathrm{Tr}(\boldsymbol{I}_{N} 
- \tilde{\boldsymbol{W}}_{t}^{\mathrm{T}}\boldsymbol{\Sigma})
= \mathrm{Tr}(\boldsymbol{I}_{N} - \boldsymbol{W}_{t}^{\mathrm{T}}\boldsymbol{A})$. 
Substituting this expression and the definition of $\boldsymbol{m}_{t}$ 
in (\ref{m}) into (\ref{m_tilde_def}), we arrive at the equivalence 
between (\ref{m_tilde}) and (\ref{m_tilde_def}). 

We have so far proved that the error model~(\ref{b})--(\ref{q_tilde}) is 
included into the general error model in \cite{Takeuchi211}. 
To complete the proof of Lemma~\ref{lemma_error_model}, 
we prove $\boldsymbol{h}_{t}=\boldsymbol{V}\tilde{\boldsymbol{m}}_{t}$. 
From the definition of $\tilde{\boldsymbol{m}}_{t}$ in (\ref{m_tilde}), 
we have 
\begin{equation}
\boldsymbol{V}\tilde{\boldsymbol{m}}_{t} 
= \sum_{\tau=0}^{t}\theta_{\mathrm{A},\tau,t}\frac{
\boldsymbol{x}_{\mathrm{A},\tau}^{\mathrm{post}} - \boldsymbol{x} 
- \sum_{t'\in\mathcal{T}_{\mathrm{A},\tau}}\xi_{\mathrm{A},t',\tau}\tilde{\boldsymbol{q}}_{t'}}
{1-\xi_{\mathrm{A},\tau}},  
\end{equation}
with $\tilde{\boldsymbol{q}}_{t}=\boldsymbol{V}\boldsymbol{b}_{t}$. 
Using the definition of $\boldsymbol{x}_{\mathrm{A},t}^{\mathrm{ext}}$ in 
(\ref{ext_mean_A}) yields 
\begin{equation}
\boldsymbol{x}_{\mathrm{A},t}^{\mathrm{ext}} - \boldsymbol{x} 
= \frac{\boldsymbol{x}_{\mathrm{A},t}^{\mathrm{post}}-\boldsymbol{x} 
- \sum_{t'\in\mathcal{T}_{\mathrm{A},t}}\xi_{\mathrm{A},t',t}\tilde{\boldsymbol{q}}_{t'}}
{1 - \xi_{\mathrm{A},t}},
\end{equation}
with $\tilde{\boldsymbol{q}}_{t}=\boldsymbol{x}_{\mathrm{B}\to \mathrm{A},t}
-\boldsymbol{x}$, 
where we have used the identity $\sum_{t'\in\mathcal{T}_{\mathrm{A},t}}
\xi_{\mathrm{A},t',t}=\xi_{\mathrm{A},t}$. 
Thus, we have 
\begin{equation}
\boldsymbol{V}\tilde{\boldsymbol{m}}_{t}
= \sum_{\tau=0}^{t}\theta_{\mathrm{A},\tau,t}
(\boldsymbol{x}_{\mathrm{A},\tau}^{\mathrm{ext}} - \boldsymbol{x})
= \boldsymbol{x}_{\mathrm{A}\to \mathrm{B},t}-\boldsymbol{x}
= \boldsymbol{h}_{t}, 
\end{equation}
where the second equality follows from the definition of 
$\boldsymbol{x}_{\mathrm{A}\to \mathrm{B},t}$ in (\ref{mean_AB}) and 
the normalization~(\ref{damping_A}). 
Thus, Lemma~\ref{lemma_error_model} holds. 
\end{IEEEproof} 

Lemma~\ref{lemma_error_model} and 
Assumptions~\ref{assumption_x}--\ref{assumption_Lipschitz} allow us to 
use \cite[Theorem 6]{Takeuchi211} to prove Theorem~\ref{THEOREM_GAUSSIANITY}. 
The first property in Theorem~\ref{THEOREM_GAUSSIANITY} follows from 
\cite[Theorem~6 (B4) and Eq.~(107)]{Takeuchi211}. Similarly, the third 
property follows from \cite[Theorem~6 (A4) and Eq.~(98)]{Takeuchi211}. 
 
We next prove the second property. 
Using (\ref{post_mean_A_tmp}) and \cite[Eq.~(94)]{Takeuchi211} yields
\begin{IEEEeqnarray}{rl}
&\frac{1}{N}(\boldsymbol{x}_{\mathrm{A},\tau'}^{\mathrm{post}}
-\boldsymbol{x})^{\mathrm{T}}
(\boldsymbol{x}_{\mathrm{A},\tau}^{\mathrm{post}}-\boldsymbol{x}) 
\nonumber \\
\aeq& \frac{1}{N}\frac{\boldsymbol{1}^{\mathrm{T}}
\boldsymbol{V}_{\mathrm{B}\to \mathrm{A},\tau',\tau'}^{-1}}
{\boldsymbol{1}^{\mathrm{T}}\boldsymbol{V}_{\mathrm{B}\to \mathrm{A},\tau',\tau'}^{-1}
\boldsymbol{1}}
\mathbb{E}\left[
 \boldsymbol{B}_{\tau'}^{\mathrm{T}}\boldsymbol{D}_{\tau',\tau}
 \boldsymbol{B}_{\tau}
\right]
\frac{\boldsymbol{V}_{\mathrm{B}\to \mathrm{A},\tau,\tau}^{-1}\boldsymbol{1}}
{\boldsymbol{1}^{\mathrm{T}}\boldsymbol{V}_{\mathrm{B}\to \mathrm{A},\tau,\tau}^{-1}
\boldsymbol{1}} \nonumber \\
&+ N^{-1}\mathbb{E}\left[
 \boldsymbol{w}^{\mathrm{T}}\boldsymbol{U}\tilde{\boldsymbol{W}}_{\tau'}
 \tilde{\boldsymbol{W}}_{\tau}^{\mathrm{T}}\boldsymbol{U}^{\mathrm{T}}\boldsymbol{w}
\right]
+ o(1) \label{SE_post_A_tmp}
\end{IEEEeqnarray}
in the large system limit, with 
\begin{equation}
\boldsymbol{D}_{\tau',\tau}
= \left(
 \boldsymbol{I}_{N} - \tilde{\boldsymbol{W}}_{\tau'}^{\mathrm{T}}
 \boldsymbol{\Sigma}
\right)^{\mathrm{T}}
\left(
 \boldsymbol{I}_{N} - \tilde{\boldsymbol{W}}_{\tau}^{\mathrm{T}}\boldsymbol{\Sigma}
\right), 
\end{equation} 
where $\{\boldsymbol{B}_{\tau}\}$ follow zero-mean Gaussian 
random matrices with covariance 
$\mathbb{E}[\boldsymbol{b}_{\tau}\boldsymbol{b}_{\tau'}^{\mathrm{T}}]
=\bar{v}_{\mathrm{B}\to \mathrm{A},\tau',\tau}\boldsymbol{I}_{N}$ while 
$\boldsymbol{U}^{\mathrm{T}}\boldsymbol{w}$ follows $\mathcal{N}(\boldsymbol{0},
\sigma^{2}\boldsymbol{I}_{M})$. 
Evaluating the expectation in (\ref{SE_post_A_tmp}) with the 
identities $\mathrm{Tr}(\boldsymbol{D}_{\tau',\tau})
=\mathrm{Tr}\{(\boldsymbol{I}_{N} - \boldsymbol{W}_{\tau'}^{\mathrm{T}}
\boldsymbol{A})^{\mathrm{T}}
(\boldsymbol{I}_{N} - \boldsymbol{W}_{\tau}^{\mathrm{T}}\boldsymbol{A})\}$ and 
$\mathrm{Tr}(\tilde{\boldsymbol{W}}_{\tau'}
\tilde{\boldsymbol{W}}_{\tau}^{\mathrm{T}})
=\mathrm{Tr}(\boldsymbol{W}_{\tau'}\boldsymbol{W}_{\tau}^{\mathrm{T}})$, 
we find that the right-hand side (RHS) is equal to (\ref{SE_post_A}) 
in the large system limit. Thus, the second property in 
Theorem~\ref{THEOREM_GAUSSIANITY} holds. 

Finally, we prove the last property. Repeating the derivation of 
(\ref{statistic_mean_A_tmp}) for the sufficient 
statistic~(\ref{statistic_mean_B}) yields 
\begin{equation} \label{statistic_mean_B_tmp}
\boldsymbol{x}_{\mathrm{A}\to \mathrm{B},t}^{\mathrm{suf}} 
= \boldsymbol{x}
+ \boldsymbol{z}_{t}, 
\quad
\boldsymbol{z}_{t}
= \frac{\boldsymbol{H}_{t}\boldsymbol{V}_{\mathrm{A}\to \mathrm{B},t,t}^{-1}
\boldsymbol{1}}
{\boldsymbol{1}^{\mathrm{T}}\boldsymbol{V}_{\mathrm{A}\to \mathrm{B},t,t}^{-1}
\boldsymbol{1}}.
\end{equation}
We use the definition of $\boldsymbol{x}_{\mathrm{B},\tau+1}$ in 
(\ref{post_mean_B}) and \cite[Eq.~(103)]{Takeuchi211} to obtain 
\begin{IEEEeqnarray}{rl}
&\frac{1}{N}(\boldsymbol{x}_{\mathrm{B},\tau'+1}^{\mathrm{post}}
-\boldsymbol{x})^{\mathrm{T}}
(\boldsymbol{x}_{\mathrm{B},\tau+1}^{\mathrm{post}}-\boldsymbol{x})
\nonumber \\
\aeq& \frac{1}{N}\sum_{n=1}^{N}\mathbb{E}\left[
 \{f_{\tau'}(x_{n}+z_{n,\tau'}) - x_{n}\}\{f_{\tau}(x_{n}+z_{n,\tau}) - x_{n}\}
\right] \nonumber \\
&+ o(1), 
\end{IEEEeqnarray}
where $\{\boldsymbol{H}_{\tau}\}$ in (\ref{statistic_mean_B_tmp}) follow 
zero-mean Gaussian random matrices with covariance 
$\mathbb{E}[\boldsymbol{h}_{\tau}\boldsymbol{h}_{\tau'}^{\mathrm{T}}]
=\bar{v}_{\mathrm{A}\to \mathrm{B},\tau',\tau}\boldsymbol{I}_{N}$. 
Evaluating the correlation $\mathbb{E}[z_{n,\tau'}z_{n,\tau}]$ with 
(\ref{correlation_W}), we arrive at (\ref{SE_post_B}).

The almost sure convergence of $\xi_{\mathrm{B},\tau}$ in (\ref{xi_B})
to $\bar{\xi}_{\mathrm{B},\tau}$ follows from 
Assumption~\ref{assumption_Lipschitz} and 
\cite[Eq.~(104)]{Takeuchi211}. 
Thus, Theorem~\ref{THEOREM_GAUSSIANITY} holds.

\section{Proof of Theorem~\ref{THEOREM_SE}}
\label{proof_theorem_SE} 
We first evaluate the covariance $\bar{v}_{\mathrm{A}\to \mathrm{B},t',t}$.  
Lemma~\ref{lemma_error_model} implies 
\begin{equation}
\bar{v}_{\mathrm{A}\to \mathrm{B},t',t} 
= \lim_{M=\delta N\to\infty}\frac{1}{N}
\tilde{\boldsymbol{m}}_{t'}^{\mathrm{T}}\tilde{\boldsymbol{m}}_{t},
\end{equation} 
where $\tilde{\boldsymbol{m}}_{t}$ is given by (\ref{m_tilde}). 
By definition, we have 
\begin{equation}
\bar{v}_{\mathrm{A}\to \mathrm{B},t',t} 
\aeq \sum_{\tau'=0}^{t'}\sum_{\tau=0}^{t}
\theta_{\mathrm{A},\tau',t'}\theta_{\mathrm{A},\tau,t}
\bar{v}_{\mathrm{A},\tau',\tau}^{\mathrm{ext}}
+ o(1),
\end{equation}
which is equal to (\ref{SE_AB}), with 
\begin{IEEEeqnarray}{rl}
&(1-\xi_{\mathrm{A},t'})(1-\xi_{\mathrm{A},t})\bar{v}_{\mathrm{A},t',t}^{\mathrm{ext}} 
\nonumber \\
=& \bar{v}_{\mathrm{A},t',t}^{\mathrm{post}} 
- \sum_{\tau'\in\mathcal{T}_{\mathrm{A},t'}}\xi_{\mathrm{A},\tau',t'}
\frac{\boldsymbol{b}_{\tau'}^{\mathrm{T}}\boldsymbol{V}^{\mathrm{T}}
(\boldsymbol{x}_{\mathrm{A},t}^{\mathrm{post}} - \boldsymbol{x})}{N}
\nonumber \\
&- \sum_{\tau\in\mathcal{T}_{\mathrm{A},t}}\xi_{\mathrm{A},\tau,t}
\frac{(\boldsymbol{x}_{\mathrm{A},t'}^{\mathrm{post}} - \boldsymbol{x})^{\mathrm{T}}
\boldsymbol{V}\boldsymbol{b}_{\tau}}{N} 
\nonumber \\
&+ \sum_{\tau'\in\mathcal{T}_{\mathrm{A},t'}}\sum_{\tau\in\mathcal{T}_{\mathrm{A},t}}
\xi_{\mathrm{A},\tau',t'}\xi_{\mathrm{A},\tau,t}
\bar{v}_{\mathrm{B}\to \mathrm{A},\tau',\tau}, \label{SE_A_ext_tmp}
\end{IEEEeqnarray}
where $N^{-1}\boldsymbol{b}_{\tau'}^{\mathrm{T}}
\boldsymbol{b}_{\tau}=N^{-1}\tilde{\boldsymbol{q}}_{\tau'}^{\mathrm{T}}
\tilde{\boldsymbol{q}}_{\tau}\ato\bar{v}_{\mathrm{B}\to \mathrm{A},\tau',\tau}$ has 
been used. We apply (\ref{gamma_t}), (\ref{post_mean_A_tmp}), and 
\cite[Eq.~(31)]{Takeuchi211} to obtain 
\begin{equation}
\frac{1}{N}\boldsymbol{b}_{\tau'}^{\mathrm{T}}\boldsymbol{V}^{\mathrm{T}}
(\boldsymbol{x}_{\mathrm{A},t}^{\mathrm{post}} - \boldsymbol{x}) 
\aeq \sum_{\tau\in\mathcal{T}_{\mathrm{A},t}}\xi_{\mathrm{A},\tau,t}
\bar{v}_{\mathrm{B}\to \mathrm{A},\tau',\tau}
+ o(1), 
\end{equation}
where we have used (\ref{xi_A}). 
Applying this result to (\ref{SE_A_ext_tmp}) yields 
\begin{IEEEeqnarray}{l}
(1-\xi_{\mathrm{A},t'})(1-\xi_{\mathrm{A},t})\bar{v}_{\mathrm{A},t',t}^{\mathrm{ext}} 
\aeq \bar{v}_{\mathrm{A},t',t}^{\mathrm{post}} \nonumber \\
- \sum_{\tau'\in\mathcal{T}_{\mathrm{A},t'}}\sum_{\tau\in\mathcal{T}_{\mathrm{A},t}}
\xi_{\mathrm{A},\tau',t'}\xi_{\mathrm{A},\tau,t}\bar{v}_{\mathrm{B}\to \mathrm{A},\tau',\tau} 
+ o(1)
\nonumber \\
= \bar{v}_{\mathrm{A},t',t}^{\mathrm{post}}
- \bar{\xi}_{\mathrm{A},t'}\bar{\xi}_{\mathrm{A},t}
\bar{v}_{\mathrm{B}\to \mathrm{A},t',t}^{\mathrm{suf}} + o(1), 
\label{SE_A_ext_tmp2}
\end{IEEEeqnarray}
where the second equality follows from (\ref{xi_A}), 
(\ref{SE_statistic_A_tmp}), and (\ref{gamma_t_asym}). 
Thus, we have arrived at (\ref{SE_A_ext}) with 
$\bar{v}_{\mathrm{B}\to \mathrm{A},t',t}^{\mathrm{suf}}$ given by 
(\ref{SE_statistic_A_tmp}), instead of (\ref{SE_statistic_A}). 

We next evaluate $\bar{v}_{\mathrm{B}\to \mathrm{A},t'+1,t+1}$ for $t', t\geq0$. 
Lemma~\ref{lemma_error_model} implies 
\begin{equation}
\bar{v}_{\mathrm{B}\to \mathrm{A},t'+1,t+1}
= \frac{1}{N}\tilde{\boldsymbol{q}}_{t'+1}^{\mathrm{T}}
\tilde{\boldsymbol{q}}_{t+1}, 
\end{equation}
where $\tilde{\boldsymbol{q}}_{t+1}$ is given by (\ref{q_tilde}). 
We use the definition of $\xi_{\mathrm{B},t',t}$ in (\ref{xi_B}) and 
Theorem~\ref{THEOREM_GAUSSIANITY} to obtain 
\begin{equation}
\bar{v}_{\mathrm{B}\to \mathrm{A},t'+1,t+1} 
\aeq \sum_{\tau'=0}^{t'}\sum_{\tau=0}^{t}
\theta_{\mathrm{B},\tau',t'}\theta_{\mathrm{B},\tau,t}
\bar{v}_{\mathrm{B},\tau'+1,\tau+1}^{\mathrm{ext}} + o(1),  
\end{equation}
which is equal to (\ref{SE_BA}), with 
\begin{IEEEeqnarray}{rl}
&(1-\bar{\xi}_{\mathrm{B},t'})(1-\bar{\xi}_{\mathrm{B},t})
\bar{v}_{\mathrm{B},t'+1,t+1}^{\mathrm{ext}}
\nonumber \\
=& \bar{v}_{\mathrm{B},t'+1,t+1}^{\mathrm{post}}
- \frac{1}{N}\sum_{\tau\in\mathcal{T}_{\mathrm{B},t}}\xi_{\mathrm{B},\tau,t}
(\boldsymbol{x}_{\mathrm{B},t'+1}^{\mathrm{post}}-\boldsymbol{x})^{\mathrm{T}}
\boldsymbol{h}_{\tau}
\nonumber \\
&- \frac{1}{N}\sum_{\tau'\in\mathcal{T}_{\mathrm{B},t'}}\xi_{\mathrm{B},\tau',t'}
\boldsymbol{h}_{\tau'}^{\mathrm{T}}
(\boldsymbol{x}_{\mathrm{B},t+1}^{\mathrm{post}}-\boldsymbol{x})
\nonumber \\
&+ \frac{1}{N}\sum_{\tau'\in\mathcal{T}_{\mathrm{B},t'}}\sum_{\tau\in\mathcal{T}_{\mathrm{B},t}}
\xi_{\mathrm{B},\tau',t'}\xi_{\mathrm{B},\tau,t}
\boldsymbol{h}_{\tau'}^{\mathrm{T}}\boldsymbol{h}_{\tau}. 
\label{SE_B_ext_tmp}
\end{IEEEeqnarray}
Utilizing \cite[Eq.~(30)]{Takeuchi211} for 
$\boldsymbol{x}_{\mathrm{B},t+1}^{\mathrm{post}}$ given in (\ref{post_mean_B}) with 
(\ref{statistic_mean_B}), and applying (\ref{xi_B}) and  
Theorem~\ref{THEOREM_GAUSSIANITY}, we have 
\begin{equation}
\frac{1}{N}\boldsymbol{h}_{\tau'}^{\mathrm{T}}
(\boldsymbol{x}_{\mathrm{B},t+1}^{\mathrm{post}}-\boldsymbol{x})
\aeq \sum_{\tau\in\mathcal{T}_{\mathrm{B},t}}
\xi_{\mathrm{B},t',t}\bar{v}_{\mathrm{A}\to \mathrm{B},\tau',\tau}
+ o(1).
\end{equation}
Substituting this result into (\ref{SE_B_ext_tmp}) and repeating the 
derivation of the second equality in (\ref{SE_A_ext_tmp2}), 
we arrive at (\ref{SE_B_ext}) with 
$\bar{v}_{\mathrm{A}\to \mathrm{B},t',t}^{\mathrm{suf}}$ 
given by (\ref{SE_statistic_B_tmp}), instead of (\ref{SE_statistic_B}). 

Finally, we evaluate $\bar{v}_{\mathrm{B}\to \mathrm{A},0,t+1}=N^{-1}
\tilde{\boldsymbol{q}}_{0}^{\mathrm{T}}\tilde{\boldsymbol{q}}_{t+1}$. 
By definition, $\bar{v}_{\mathrm{B}\to \mathrm{A},0,0}=1$ is trivial. 
Using $\tilde{\boldsymbol{q}}_{0}=-\boldsymbol{x}$, the definition of 
$\tilde{\boldsymbol{q}}_{t+1}$ in (\ref{q_tilde}), and 
Theorem~\ref{THEOREM_GAUSSIANITY}, we arrive at 
(\ref{SE_BA_0}) for $t\geq0$. 

To complete the proof of Theorem~\ref{THEOREM_SE}, 
we derive (\ref{SE_statistic_A}) and (\ref{SE_statistic_B}).   
Since we have defined the covariance messages that are consistent to the 
state evolution recursions in the large system limit, 
we have $\boldsymbol{V}_{\mathrm{A}\to \mathrm{B},\tau,\tau}\ato
\bar{\boldsymbol{V}}_{\mathrm{A}\to \mathrm{B},\tau,\tau}$ and 
$\boldsymbol{V}_{\mathrm{B}\to \mathrm{A},\tau,\tau}\ato
\bar{\boldsymbol{V}}_{\mathrm{B}\to \mathrm{A},\tau,\tau}$. Thus, 
(\ref{SE_statistic_A_tmp}) and (\ref{SE_statistic_B_tmp}) reduce to 
(\ref{SE_statistic_A}) and (\ref{SE_statistic_B}), respectively. 
This justifies the replacement of 
$v_{\mathrm{B}\to\mathrm{A},t,t}^{\mathrm{suf}}$ in $\boldsymbol{W}_{t}$ 
with $\bar{v}_{\mathrm{B}\to\mathrm{A},t,t}^{\mathrm{suf}}$. 

\section{Poof of Theorem~\ref{THEOREM_SE_BAYES}}
\label{proof_theorem_SE_Bayes}
We first prove the identity $\bar{v}_{\mathrm{A}\to \mathrm{B},t',t}
=\bar{v}_{\mathrm{A}\to \mathrm{B},t}$ for all $t'\leq t$.  
When all preceding information 
$\mathcal{T}_{\mathrm{A},t}=\mathcal{T}_{\mathrm{B},t}=\{0,\ldots,t\}$ is used for 
all $t$, $\bar{v}_{\mathrm{B}\to \mathrm{A},t',t}^{\mathrm{suf}}$ in 
(\ref{SE_statistic_A}) for $t'\leq t$ reduces to 
\begin{IEEEeqnarray}{rl}
\bar{v}_{\mathrm{B}\to \mathrm{A},t',t}^{\mathrm{suf}} 
=& \frac{\boldsymbol{1}^{\mathrm{T}}
\bar{\boldsymbol{V}}_{\mathrm{B}\to \mathrm{A},t',t'}^{-1}
(\boldsymbol{I}_{t'}, \boldsymbol{O})
\bar{\boldsymbol{V}}_{\mathrm{B}\to \mathrm{A},t,t}
\bar{\boldsymbol{V}}_{\mathrm{B}\to \mathrm{A},t,t}^{-1}
\boldsymbol{1}}{\boldsymbol{1}^{\mathrm{T}}
\bar{\boldsymbol{V}}_{\mathrm{B}\to \mathrm{A},t',t'}^{-1}\boldsymbol{1}
\boldsymbol{1}^{\mathrm{T}}
\bar{\boldsymbol{V}}_{\mathrm{B}\to \mathrm{A},t,t}^{-1}\boldsymbol{1}}
\nonumber \\
=& \frac{1}{\boldsymbol{1}^{\mathrm{T}}
\bar{\boldsymbol{V}}_{\mathrm{B}\to \mathrm{A},t,t}^{-1}\boldsymbol{1}},  
\label{SE_statistic_A_Bayes_tmp}
\end{IEEEeqnarray}
as shown in (\ref{correlation_W_reduction}). 
Using this result, the definition of $\gamma_{t',t}$ in (\ref{gamma_t't}), 
the LMMSE filter~(\ref{LMMSE}) with $v_{\mathrm{B}\to \mathrm{A},t,t}^{\mathrm{suf}}$ 
replaced by $\bar{v}_{\mathrm{B}\to \mathrm{A},t,t}^{\mathrm{suf}}$,  
we find that $\bar{v}_{\mathrm{A},t',t}^{\mathrm{post}}$ in (\ref{SE_post_A}) 
for $t'\leq t$ reduces to   
\begin{IEEEeqnarray}{l}
\bar{v}_{\mathrm{A},t',t}^{\mathrm{post}}
\aeq o(1) + \frac{\sigma^{2}}{N} 
\mathrm{Tr}\left(
 \boldsymbol{\Xi}_{t'}\boldsymbol{A}\boldsymbol{A}^{\mathrm{T}}
 \boldsymbol{\Xi}_{t}
\right) \nonumber \\
+ \frac{\bar{v}_{\mathrm{B}\to \mathrm{A},t,t}^{\mathrm{suf}}}{N}\mathrm{Tr}\left\{
 \left(
  \boldsymbol{I}_{N} - \boldsymbol{A}^{\mathrm{T}}\boldsymbol{\Xi}_{t'}
  \boldsymbol{A}
 \right)
 \left(
  \boldsymbol{I}_{N} - \boldsymbol{A}^{\mathrm{T}}\boldsymbol{\Xi}_{t}
  \boldsymbol{A}
 \right)
\right\} 
 \nonumber \\
= \frac{\bar{v}_{\mathrm{B}\to \mathrm{A},t,t}^{\mathrm{suf}}}{N}\mathrm{Tr}\left(
 \boldsymbol{I}_{N} - \boldsymbol{A}^{T}\boldsymbol{\Xi}_{t}\boldsymbol{A}
\right) + o(1) \nonumber \\
= \bar{\xi}_{\mathrm{A},t}\bar{v}_{\mathrm{B}\to \mathrm{A},t,t}^{\mathrm{suf}} + o(1), 
\label{SE_post_A_Bayes_tmp}
\end{IEEEeqnarray}
with 
\begin{equation}
\boldsymbol{\Xi}_{t} 
= \bar{v}_{\mathrm{B}\to \mathrm{A},t,t}^{\mathrm{suf}}\left(
 \sigma^{2}\boldsymbol{I}_{M} 
 + \bar{v}_{\mathrm{B}\to \mathrm{A},t,t}^{\mathrm{suf}}\boldsymbol{A}
 \boldsymbol{A}^{\mathrm{T}}
\right)^{-1},  
\end{equation}
where the last equality follows from (\ref{gamma_t_asym}). This expression 
is independent of $t'$. Thus, we have 
$\bar{v}_{\mathrm{A},t',t}^{\mathrm{post}}=\bar{v}_{\mathrm{A},t,t}^{\mathrm{post}}$.  
Substituting $\bar{\xi}_{\mathrm{A},t}=\bar{v}_{\mathrm{A},t,t}^{\mathrm{post}}/
\bar{v}_{\mathrm{B}\to \mathrm{A},t,t}^{\mathrm{suf}}$ into (\ref{SE_A_ext}), 
we find that, for all $t'\leq t$, $\bar{v}_{\mathrm{A}\to \mathrm{B},t',t}$ 
in (\ref{SE_AB}) is equal to $\bar{v}_{\mathrm{A}\to \mathrm{B},t}$ given in 
(\ref{SE_AB_Bayes}) with $\bar{v}_{\mathrm{B}\to \mathrm{A},t,t}$ replaced by 
$\bar{v}_{\mathrm{B}\to \mathrm{A},t,t}^{\mathrm{suf}}$. 

We next prove the identity $\bar{v}_{\mathrm{B}\to \mathrm{A},t'+1,t+1}
=\bar{v}_{\mathrm{B}\to \mathrm{A},t+1}$ for all $t'\leq t$. 
Repeating the same derivation for (\ref{SE_statistic_B}) as in 
(\ref{SE_statistic_A_Bayes_tmp}), we find 
\begin{equation} \label{SE_statistic_B_Bayes_tmp}  
\bar{v}_{\mathrm{A}\to \mathrm{B},t',t}^{\mathrm{suf}}
= \frac{1}{\boldsymbol{1}^{\mathrm{T}}
\bar{\boldsymbol{V}}_{\mathrm{A}\to \mathrm{B},t,t}^{-1}
\boldsymbol{1}} 
= \bar{v}_{\mathrm{A}\to \mathrm{B},t,t}^{\mathrm{suf}}
\end{equation}
for all $t'\leq t$. 
Using the second property in Lemma~\ref{LEMMA_BAYES} yields 
$\bar{v}_{\mathrm{B},t',t}^{\mathrm{post}}
=\bar{v}_{\mathrm{B},t,t}^{\mathrm{post}}$ for all $t'\leq t$. Furthermore, we have 
the general relationship $\bar{\xi}_{\mathrm{B},t}
=\bar{v}_{\mathrm{B},t+1,t+1}^{\mathrm{post}}
/\bar{v}_{\mathrm{A}\to \mathrm{B},t,t}^{\mathrm{suf}}$~\cite[Lemma 2]{Takeuchi201} 
between the 
Bayes-optimal denoiser and the MMSE. Applying these results to 
(\ref{SE_B_ext}), we find that, for all $t'\leq t$, 
$\bar{v}_{\mathrm{B}\to \mathrm{A},t'+1,t+1}$ in (\ref{SE_BA}) is equal to 
$\bar{v}_{\mathrm{B}\to \mathrm{A},t+1}$ given in (\ref{SE_BA_Bayes})
with $\bar{v}_{\mathrm{A}\to \mathrm{B},t,t}$ replaced by 
$\bar{v}_{\mathrm{A}\to \mathrm{B},t,t}^{\mathrm{suf}}$. 

To complete the proof of the former property in 
Theorem~\ref{THEOREM_SE_BAYES}, 
we need to prove the identities 
$\bar{v}_{\mathrm{A}\to \mathrm{B},t,t}^{\mathrm{suf}}=\bar{v}_{\mathrm{A}\to \mathrm{B},t,t}$ 
and $\bar{v}_{\mathrm{B}\to \mathrm{A},t,t}^{\mathrm{suf}}
=\bar{v}_{\mathrm{B}\to \mathrm{A},t,t}$.  
Without loss of generality, we focus on the former  
$\bar{v}_{\mathrm{A}\to \mathrm{B},t,t}^{\mathrm{suf}}
=\bar{v}_{\mathrm{A}\to \mathrm{B},t,t}$.   

We have already proved 
$\bar{v}_{\mathrm{A}\to \mathrm{B},t',t}=\bar{v}_{\mathrm{A}\to \mathrm{B},t,t}$ 
for all $t'\leq t$. Furthermore, Theorem~\ref{THEOREM_GAUSSIANITY} implies 
that $\{\bar{\boldsymbol{V}}_{\mathrm{A}\to \mathrm{B},\tau,\tau}\}_{\tau=1}^{t}$ are 
positive definite. Thus, we can use the second property in 
Lemma~\ref{LEMMA_MONOTONICITY} to arrive at 
$\bar{v}_{\mathrm{A}\to \mathrm{B},t,t}^{\mathrm{suf}}
=\bar{v}_{\mathrm{A}\to \mathrm{B},t,t}$, so that the former property 
in Theorem~\ref{THEOREM_SE_BAYES} holds.

Let us prove the latter property in Theorem~\ref{THEOREM_SE_BAYES}. 
Since $\bar{v}_{\mathrm{A}\to \mathrm{B},t,t}^{\mathrm{suf}}
=\bar{v}_{\mathrm{A}\to \mathrm{B},t,t}$ and 
$\bar{v}_{\mathrm{B}\to \mathrm{A},t,t}^{\mathrm{suf}}
=\bar{v}_{\mathrm{B}\to \mathrm{A},t,t}$ hold, 
the first property in Lemma~\ref{LEMMA_BAYES} implies 
that $\{\bar{v}_{\mathrm{A}\to \mathrm{B},t,t}\geq0\}$ and 
$\{\bar{v}_{\mathrm{B}\to \mathrm{A},t,t}\geq0\}$ are 
monotonically non-increasing sequences with respect to $t$. 
Also, it is possible to prove that $\{\bar{v}_{\mathrm{A}\to \mathrm{B},t,t}\geq0\}$ 
and $\{\bar{v}_{\mathrm{B}\to \mathrm{A},t,t}\geq0\}$ are strictly deceasing 
sequences, by using the first property in Lemma~\ref{LEMMA_MONOTONICITY}. 
Thus, there are the limits $\lim_{t\to\infty}\bar{v}_{\mathrm{A}\to \mathrm{B},t,t}
=\bar{v}_{\mathrm{A}\to \mathrm{B}}$ and 
$\lim_{t\to\infty}\bar{v}_{\mathrm{B}\to \mathrm{A},t,t}
=\bar{v}_{\mathrm{B}\to \mathrm{A}}$, which imply the convergence of 
$\bar{v}_{\mathrm{B},t+1,t+1}^{\mathrm{post}}$ in (\ref{SE_post_B}) to some 
constant $\bar{v}_{\mathrm{B}}^{\mathrm{post}}$. 
From (\ref{xi_asym}) and (\ref{gamma_t_Bayes}), 
$\bar{\xi}_{\mathrm{B},t}$ and $\bar{\xi}_{\mathrm{A},t}$ converge to some constants 
$\bar{\xi}_{\mathrm{B}}$ and $\bar{\xi}_{\mathrm{A}}$ as $t\to\infty$. 
These observations imply that $\bar{v}_{\mathrm{A}\to \mathrm{B},t',t}
=\bar{v}_{\mathrm{A}\to \mathrm{B},t}$ and 
$\bar{v}_{\mathrm{B}\to \mathrm{A},t',t}=\bar{v}_{\mathrm{B}\to \mathrm{A},t}$ given 
in (\ref{SE_AB_Bayes}) and (\ref{SE_BA_Bayes}) converge to a FP as 
$t$ tends to infinity for all $t'\leq t$.

\section*{Acknowledgment}
The author thanks the anonymous reviewers for their suggestions that have 
improved the quality of the manuscript greatly.

\ifCLASSOPTIONcaptionsoff
  \newpage
\fi



\bibliographystyle{IEEEtran}
\bibliography{IEEEabrv,kt-it2021}
%

%







\end{document}